\documentclass[prd,showpacs,nofootinbib,preprintnumbers]{revtex4}

\usepackage{latexsym}
\usepackage{amssymb}
\usepackage{amsmath}
\usepackage{slashbox}
\usepackage{subfigure}
\usepackage{array}
\usepackage[dvips]{epsfig}

 \numberwithin{equation}{section}

\newcommand{\comment}[1]{}

\newcommand{\hsp}{\hspace{0.1em}}
\newcommand{\nhsp}{\hspace{-0.1em}}
\newcommand{\nsmhsp}{\hspace{-0.05em}}

\newcommand{\lph}{\vphantom{\frac{\frac{d}{}}{d^d}}}
\newcommand{\smph}{\vphantom{ d^0_0}}
\newcommand{\vp}{\vphantom{\frac{a}{a}}}
\newcommand{\vph}{\vphantom{\frac{d}{d}}}
\newcommand{\tc}{\centering\arraybackslash}

\renewcommand{\theequation}{\arabic{section}.\arabic{equation}}
\renewcommand{\geq}{\geqslant}

\newcommand{\vk}{\varkappa_D}
\newcommand{\ga}{\gamma}

\newcommand{\om}{\omega}

\newcommand{\ka}{\kappa}
\newcommand{\pa}{\partial}
\newcommand{\s}{\sigma}

\newcommand{\be}{\begin{equation}}
\newcommand{\ee}{\end{equation}}
\newcommand{\ba}{\begin{align}}
\newcommand{\ea}{\end{align}}
\newcommand{\bea}{\begin{eqnarray}}
\newcommand{\eea}{\end{eqnarray}}
\newcommand{\bw}{\begin{widetext}}
\newcommand{\ew}{\end{widetext}}

\newcommand{\ffi}{{\varphi}}

\newcommand{\e}{{\rm e}}

\newcommand{\nn}{\nonumber}
\newcommand{\zt}{\dot{z}}
\newcommand{\zp}{\left.\zt^{\smhsp\prime}\right.}
\newcommand{\ztt}{\ddot{z}}
\newcommand{\zpp}{\left.\ztt^{\smhsp\prime}\right.}

\newcommand{\cd}{\smhsp,\smhsp}
\newcommand{\un}{\, ^1 \!}
\newcommand{\de}{\, ^2 \!}

\newcommand{\pp}{\ldots}

\newcommand{\ds}{\displaystyle}
\newcommand{\nul}{\, ^0 }
\newcommand{\cm}{\mathrm{cm}}
\newcommand{\rad}{\mathrm{rad}}

\newcommand{\coa}{\xi}

\newcommand{\od}{\omega}

\newcommand{\dm}{(d\smhsp)}

\newcommand{\mco}{\mathcal{O}\smhsp}

\newcommand{\dx}{d^{\hsp D}\nhsp x}
\newcommand{\dk}{d^{\hsp D}\nhsp k}
\newcommand{\dq}{d^{\hsp D}\nhsp q}
\newcommand{\ddp}{d^{\hsp D}\nhsp p}

\newcommand{\smhsp}{\hspace{0.06em}}

\begin{document}
\hfill CCTP-2013-18

\title{Vector Bremsstrahlung by Ultrarelativistic Collisions in Higher Dimensions}
\author{Yiannis Constantinou$^a$ and Pavel Spirin$^b$
\thanks{\tt E-mail: gkonst@physics.uoc.gr,
salotop@list.ru}} \affiliation{
 \mbox{$^a$Crete Center for Theoretical Physics, University of Crete, 71003, Heraklion,
 Greece;}\\
 \mbox{$^b$Department of Theoretical
Physics, Moscow State University, 119899, Moscow, Russian
Federation.}}

\date{\today}

\begin{abstract}
A classical computation of vector bremsstrahlung in
ultrarelativistic gravitational-force collisions of massive point
particles is presented in an arbitrary number $d$ of extra
dimensions. Our method adapts the post-linear formalism of General
Relativity to the multidimensional case. The total emitted energy,
as well as its angular and frequency distribution and
characteristic values, are discussed in detail.

For an electromagnetic mediation propagated in the bulk, the
emitted energy $E_{\rm em}$ of scattering with  impact parameter
$b$ has magnitude $E_{\rm em} \sim e^4\smhsp{e'}^2 \gamma^{d+2}\!/
(m^2\smhsp b^{\hsp 3d+3} )$, with dominant frequency $\omega_{\rm
em} \sim \gamma^2\!/\smhsp b$. For the gravitational force the
charge emits via vector field, propagated in the bulk,  energy
$E_{\rm rad} \sim  [G_D m'{e} ]^2 \gamma^{d+2}\!/\hsp b^{\hsp
3d+3}$ for $d\geqslant 2$, with dominant frequency  $\omega  \sim
\gamma^2\!/\smhsp b$; and   energy $E_{\rm rad} \sim \left[G_5 m'
{e_5}\right]^2\,\gamma^{3}\ln \gamma\smhsp /\hsp b^{\hsp 6}$ for
$d=1$, with most of the energy  coming from a wide frequency
region $\omega \in [\mco(\gamma/\smhsp b),\mco(\gamma^2\!/\smhsp
b)] $. For the UED model with extra space volume $V=(2\pi R)^d$
the emitted energy is $E_{\rm UED}\sim (b^{\hsp d}/V)^2 E_{\rad}$.
Finally, for the ADD model, including four dimensions, the
electromagnetic field living on 3-brane, loses on emission the
energy $E_{\rm ADD} \sim \left[G_D m'
{e}\right]^2\,\gamma^{3}\!/(V\smhsp b^{\hsp 2d+3})$, with
characteristic frequency $\omega_{\rm ADD} \sim \gamma/\smhsp b$.

The contribution of the low frequency part of the radiation (soft
photons) to the total radiated energy is shown to be negligible
for all values of $d$. The domain of validity of the classical
result is discussed. The result is analyzed from the viewpoint of
the de\hsp{}Witt\hsp{}--\hsp{}Brehme\hsp{}--\hsp{}Hobbs equation
(and corresponding equations in higher dimensions). The different
frequency domains and their competition mentioned above, may be
explained as coming from different terms in this equation. Thus
the whole emission process may be naturally split in two
sub-processes with drastically different spectral and temporal
characteristics.
\end{abstract}\maketitle

\tableofcontents

\thispagestyle{empty}
\section{Introduction}\label{intr}


The first experiments of the Large Hadron Collider (LHC) at CERN
have shown that creation of Black holes is much less than
predicted by theorists. When the discovery of new physics at LHC
associated with supersymmetry at low energies fails, the models of
TeV-scale gravity become of particular interest. The LHC can be
used to test models with Large Extra Dimensions (LEDs) and set
bounds on their parameters \cite{BH,recent}. Initially proposed as
an alternative to supersymmetry in solving the hierarchy problem,
such models are motivated by string theory and open new
interesting directions in cosmology. Inspired by earlier ideas of
the Universe as a topological defect in higher-dimensional
space-time and the TeV-scale supersymmetry breaking in heterotic
string theory associated with compactification \cite{ablt}, they
appeared in several proposals.

A conceptually and technically simple one is the Arkani-Hamed,
Dimopoulos and Dvali (ADD) scenario \cite{ADD1}, with the Standard
Model particles living in the four-dimensional space-time and
gravity propagating in the $D$-dimensional bulk with the $d=D-4$
flat dimensions compactified on a torus. Gravity is strong with a
corresponding Planck mass  $M_{\rm Pl}^*$ at the (presumably) TeV
scale.

Other LED scenaria include the warped compactification
Randall-Sundrum (RS) models \cite{RS}, which are based on an
identification of the physical four-dimensional space-time with a
3-brane embedded into a five-dimensional bulk endowed with the
cosmological constant, in which case the fifth dimension may be
infinite. The model known as "Universal Extra Dimensions"
\cite{UED} (UED) allows \emph{all} fields to propagate through the
bulk.


The common feature of all these models is the existence of a large
(in Planck units) length  $L_{\rm Pl}$, which may appear either
via a compactification radius, or via inverse powers of curvature
of the infinite bulk. If the quantum gravity scale happens to be
of order of TeV, the LHC, expected to reach center of mass
energies one order of magnitude higher, will be able to study
information about gravity at ultraplanckian energies \cite{LHC}.
The gravitational radius associated with the center of mass
collision energy increases with energy, and in the transplanckian
regime becomes larger than the Planck length, indicating that
gravity behaves \emph{classically} at least for some region of
momentum transfers \cite{'tHooft:1987rb}. Thus the transplanckian
gravity is believed to be adequately described by the classical
Einstein equations \cite{GiRaWeTrans}. This, presumably, allows
one to make reliable theoretical predictions of gravitational
effects without entering into the complications related to quantum
gravity.


\comment{Black Hole production is arguably the most exciting
inelastic process in the context of the TeV-gravity. Apart from
creation of black holes, another inelastic gravitational process
is radiation. Bremsstrahlung itself represents the natural process
as probe to test extra dimensions, since or the ultrarelativistic
particles the difference between dimensions can be easily revealed
by the power of Lorentz factor \mbox{$\gamma\!\gg\! 1$} of
collision.}

Black Hole production is arguably the most exciting inelastic
process in the context of the TeV-gravity. Apart from the creation
of black holes, another inelastic gravitational process is
radiation. Bremsstrahlung itself represents the natural   process
to test the existence of extra dimensions and probe them.
Colliding ultrarelativistic particles will radiate and the number
of dimensions can easily be determined by the dependence of the
radiated energy from the Lorentz factor \mbox{$\gamma\!\gg\! 1$}
of collision.

Bremsstrahlung is characterized by the only one length parameter
of experiment -- the impact parameter $b$. To keep gravity
classical, it is expected to be much greater than the
Schwarzschild radius $r_S$, associated with the energy
$\mathcal{E} \simeq \sqrt{s}$, where $s$ stands for the Mandelstam
$s$-variable:
\begin{equation}\label{restr_A}
  b\gg \gamma^{1/(d+1)} r_S \sim \left(\vk^2 \gamma \sqrt{s}
\right)^{1/(d+1)} .
\end{equation}

However, the calculation of classical ultraplanckian
\emph{gravitational} bremsstrahlung in the context of the ADD
model \cite{GKST4} predicts strong enhancement of radiation losses
as compared to theories without extra dimensions already for large
values of the impact parameter. These extreme losses possibly
originate from the large number of light Kaluza-Klein (KK) modes
\cite{other,Miro}. Our estimate shows that transplanckian
collisions should be heavily damped by radiation,  and classical
radiation reaction has to be taken into account in the study of
gravitational collapse and BH production in colliders.


 On the other hand, the theory of electromagnetic radiation (both
classical and quantum) has been developed to much greater extent
than gravitational radiation. The same applies to the
corresponding \emph{detectors of the emitted waves}. Thus it is
natural to include vector bremsstrahlung among the realistic
inelastic problems, where the force causing the acceleration may
be either gravitational or non-gravitational.

Nevertheless the problem of radiation reaction is far from solved,
even in electrodynamics. Inspired by the pioneering  work of Dirac
\cite{Dirac}, it was developed by Rohrlich and Teitelboim in flat
space-time \cite{Rohrlich,Teitelboim}, adapted by de\,Witt and
Brehme for curved background \cite{DeWitt:1960fc} and generalized
to curved background in higher dimensions in \cite{GaSp}.

Some attempts to include radiation reaction in QED have been made
during the last thirty years \cite{deWitt,Kriv,Higuchi}. However,
the number of physical cases where these attempts have succeeded
in producing a closed form result, is quite modest
\cite{DeWitt-deWitt}.

Thus electromagnetic bremsstrahlung in an external gravitational
field (generated by the partner particle) represents a process of
particular theoretical interest in the context of another
application of tail appearance coming from the non-local part of
the Green's function in curved background.

It actualizes the purposes of this paper. Furthermore, the
synchrotron radiation shows that within some region of parameters,
the electromagnetic field can be also treated classically,
accurately matching the result of quantum electrodynamics.

Thereby, in addition, to make the scheme self-consistent, one has
to demand also the classicality of the particles' trajectory and
\emph{classicality} of the electrodynamics.\comment{: namely,
introducing the 'classical radius' of a massive charge $e$ in
$D\equiv 4+d$ dimensions
$$r_{\rm cl}\sim \left(\frac{e^2}{m}\right)^{1/(d+1)}\,,$$
one requires
$$b\gg r_{\rm cl}\,.$$}


Perturbation theory over the gravitational constant $\vk$ will be
of usage in the computation presented here. Given as a
zeroth-order solution,  Minkowski space-time will be used as an
effective background for the wave propagation. The significance of
such a choice is highlighted by the following facts: (i) it
ensures the asymptotically flat space-time, (ii) one considers
tensors and their variations as tensors in flat space with simple
raise/lowering indices and (iii) it allows the freedom to use
Fourier-transforms.

 Thus one considers the Minkowski space-time as the background,
while the direct nature of  modes (Kaluza-Klein modes for toroidal
extra dimensions or curvature-mediated modes in cosmological
models with no compactification, like RS2) should be taken into
account as a correction due to the  curvature.  Depending on the
choice of model, the vector field can either propagate through the
bulk, or not, even though the charges are confined on the 3-brane.
Thus we generically consider Minkowski space-time as the
background with \emph{arbitrary} dimensionality $D \geqslant 4$,
while all interesting cases can be obtained as limiting cases of
the generic calculation.


This work continues a series of papers
\cite{GKST1,GKST2,GKST3,GKST4}: pure gravitational transplanckian
bremsstrahlung is considered in  \cite{GKST4}, the classical
scalar bremsstrahlung in \cite{GKST2}, while \cite{GKST3} is
devoted to the scalar emission in the gravity-mediated
bremsstrahlung. Mathematically, in the ADD model the Minkowski
limit appears as the reduction of summation over KK-modes into the
integration, as long as the restriction on the large size of extra
dimensions holds. Therefore, one has to  assume
\begin{equation}\label{restr_B}
    b\ll R
\end{equation}
to have large number of KK-quanta, for each model to be applied
to.


Most of the previous works on classical bremsstrahlung were
concerned with gravitational radiation: for reviews see
\cite{GKST4} and references therein, and \cite{Tasso} among the
most recent.

Among the previous works in four dimensions on the electromagnetic
radiation caused by gravitational force, one emphasizes the papers
by Peters \cite{Peters}, by Matzner and Nutku \cite{Matzner} and
the work by Gal'tsov, Grats and Matyukhin \cite{ggm}. In
\cite{Peters} the post-linear formalism is used in the coordinate
space  for Schwarzschild background, considering bremsstrahlung
near the vicinity of black hole.

Some qualitative arguments and estimates are given in
\cite{Khrip}. In \cite{Matzner} the equivalent-photons method was
adapted for gravitons. This approach was criticized in \cite{ggm},
who found that this method is of limited range when the frequency
range is decreased $\gamma$ times, and  thereby inappropriate.

In \cite{ggm} the iteration scheme accompanied by the perturbation
theory is used   -- as well as in the present work, while
mathematical techniques are different: contour integration in
\cite{ggm} versus expansion of Macdonald functions here. The
similar features are: (i) the damping of radiation amplitude at
high frequencies $\omega \sim \gamma^2\!/\smhsp b$ (at Lab frame),
(ii) the significant frequency $\omega \sim \gamma/\smhsp b$,
coming from the partial cancelation of local and non-local
currents, and (iii) the final power of Lorentz factor:
$$E_{\rm rad} \sim \frac{(G e \hsp m')^2}{b^{\hsp 3}}\,\gamma^3\,.$$
 The difference is related with the erroneous neglect of the local
 current (which turns out to be significant) at the dominant frequency $\omega
\sim \gamma/\smhsp b$ in \cite{ggm}, whereas it has the same
magnitude as the non-local part which is retained. Because of
this, the total
  coefficient is determined with an error, as well as
the small- and medium-frequency behavior. Thus our answer in four
dimensions corrects the overall coefficient obtained in
\cite{ggm}, and generalizes it to the higher dimensions.
Furthermore, we show that in higher dimensions the
higher-frequency regime $$ \omega \sim \gamma^2\!/\smhsp b$$
dominates over the domain $ \omega \sim \gamma/\smhsp b$, due to
the volume factors in the momentum space.


 Taking into account some similar
features appearing in these works \cite{GKST3,GKST4}, we minimize
the derivations and refer to the previously derived ones, when it
is possible. Meanwhile we would like to emphasize the features not
observed in previous works: conservation of source (validity of
the gauge condition), influence of self-action, the bremsstrahlung
of two charges, the length of the emitted wave formation
(coherence length), etc.

In order to distinguish vector radiation by gravitational
scattering from pure electromagnetic bremsstrahlung (which is
expected to represent much larger effect due to the values of
couplings in 4D), we charge only one
  particle  \emph{in the most} of the paper,
while a subsection in the Discussion section is devoted to the
radiation effects coming from the scattering of two charges.

 The
paper is organized as follows: the model, approximation method and
formulae  necessary for subsequent computation of the emitted
energy, including the polarization vectors, are described in the
Section \ref{setup}. The local and non-local amplitudes, their
combination and the amplitude damping   at high frequencies (the
destructive interference effect) are derived in Section
\ref{radiation_amplitudes}. Section \ref{emitted_energy} is
devoted to the computation of total emitted energy. Some
additional aspects (zero-frequency limits) are discussed.
Particular attention is paid to  the emission in the ADD model.
Possible cut-offs, the comparison of electromagnetic
bremsstrahlung by  gravitational and non-gravitational forces, the
conclusions and prospects are presented in the Discussion section.
Finally, some necessary formulae for computation and the simple
proof of the destructive interference phenomenon in the vector
case, dealing with just the integration-by-parts technique, are
given in three Appendices.

\section{The model}\label{setup}

We compute here a  classical spin-one bremsstrahlung in
ultra-relativistic gravity-mediated scattering of two massive
point particles $m$ and $m'$. The space-time is assumed to be
$M_{1,D-1}$ with coordinates $x^M$, $M=0,1,\ldots, D-1$, with the
mostly minus signature $(+, -, \pp , -)$. The units we use are
$c=\hbar=1$.

Particles are localized on the observable 3-brane  and interact
via the gravitational field $g_{MN}$, which propagates in the
whole space-time $M_{1,D-1}$. We also assume the existence of a
massless bulk vector field $A^M$, which interacts with $m$, but
not with $m'$. Thus only $m$ has an electromagnetic charge $e$.

\subsection{Setup and Equations of motion}
 The action of the model is symbolically of
the form
\begin{equation} S\equiv S_g+S_A+S_m+S_{mA}+S_{m'}\,,
\nonumber
\end{equation}and explicitly, in an obvious correspondence,  in the
reparametrization-invariant form
\begin{align}\label{action}
 S=-\nhsp\int\nhsp \dx \sqrt{|g|}\nhsp
\left[ \frac{R}{\vk^2} +
\frac{1}{4}\,g^{MN}g^{RS}F_{MR}F_{NS}\right]- \int \left[m \sqrt{
g_{MN} \dot{z}^M \dot{z}^N}   - e A_M \dot{z}^M \,\right] d\tau
-\int m^\prime\sqrt{ g_{MN} \dot{z}'^M \dot{z}'^N}  \, d\tau'
\end{align}
with $\vk^2 \equiv 16\smhsp \pi \hsp G_D$ where $G_D$ stands for
the $D$-dimensional Newton's constant. $F_{MN}$ is the field
strength defined as usual: $F_{MN}=\nabla_M A_N-\nabla_N A_M$
\footnote{We do not deal with massless particles. Thus the
Polyakov form of the mechanical action is not required.}. Our
convention for the Riemann tensor is $R^B{}_{\!NRS}\equiv
\Gamma^B_{NS , R} - \Gamma^B_{NR , S} + \Gamma^A_{NS}
\Gamma^B_{AR} - \Gamma^A_{NR} \Gamma^B_{AS}$, with
$\Gamma^A_{NR}=(1/2)\, g^{AB}(g_{BR, N}+g_{N B , R}-g_{NR, B})$.
Finally, the Ricci tensor and curvature scalar are defined as
$R_{MN}\equiv \delta^B_A\, R^A{}_{\!MBN}$ and $R\equiv g^{MN}
\,R_{MN}$, respectively.

In the sequel we deal with the affine parameter of the both
particles' worldline, so $g_{MN} \dot{z}^M \dot{z}^N =g_{MN}
\dot{z}'^M \dot{z}'^N =1$. Thus we consider only that class of the
worldline reparametrizations, which maintains the natural (affine)
parametrization of the trajectory.

Variation of (\ref{action}) with respect to $z^M$ and ${z'}^M$
gives the particles' equations of motion in the covariant form
\begin{align}\label{eom_part}
m \, D\zt^M=e\, F^{MN}\zt_N\,, \qquad\qquad  D'{}\zt'^M=0\, ,
\end{align}
where the covariant derivative is defined as
\begin{align}\label{cov_diff}
D  \pi^M \equiv \frac{\partial
\pi^M}{\partial\tau}+\Gamma^{M}_{RS} \,\pi^R \zt^S\,.
\end{align}

Variation over $A^M$ leads to
\begin{align}\label{emden0}
\nabla_{N}F^{MN}=- J^{M} \, ,\qquad\qquad  J^{M}(x)=e\int
\dot{z}^{M}(\tau)\,\frac{\delta^D\!\left(x-z(\tau)\vp\right)}{\sqrt{|g|}}\:
d \tau\,.
\end{align}
Finally, varying the action with respect to the metric $g_{MN}$,
one obtains the Einstein equations
\begin{align}\label{emden2}
 R^{MN}-\frac{1}{2}\, g^{MN}\hsp R=\frac{\vk^2 }{2}\,T^{MN} \, ,
\end{align}
where $T^{MN}$ is a total matter of the system-at-hand.

In order to resolve the equations of motion we use perturbation
theory with respect to the gravitational coupling and the
electromagnetic coupling.

As   was argued in the Introduction, one expands the metric as a
perturbation on the Minkowski background:
$$g_{MN}=\eta_{MN}+\vk h_{MN}
$$ and then finds the solution of equations of motion in each order
iteratively. Respectively, all tensors are to be considered as
tensors in flat space-time, as well as raising/lowering of their
indices.

\subsection{Approximation method}

We intend to use an approximation technique that relies on the
fact that the deviation from the Minkowski metric is small i.e.
$\vk h_{MN} \ll 1$. In particular, we have to evaluate $\vk
h_{MN}$ at the location of the charge, i.e. considering $m'$ as
the source of an external gravitational field. In what follows:
 \begin{equation}\label{restr_0}
 b\gg r_g \, , \qquad\qquad {r'_g}^{d+1}=   \frac{8\hsp \Gamma
\left( \frac{d+3}{2}\right)}{ {\pi}^{(d+1)/2}(d+2)}\;  G_D\smhsp
 m' \,.
 \end{equation}
The possible restrictions due to the charge do not affect the
perturbative approximation we use and their discussion is
postponed to the Discussion section.

 As mentioned above we will be
solving the equations of motion iteratively. Therefore all fields
and kinematical quantities are to be expanded as follows:
 \begin{equation}
 \label{snickers}
\phi = \nul \phi + \un
 \phi + \de \phi+ \pp \,,
 \end{equation}
where $\phi$ can be $h_{MN}$,  $T^{MN}$, $A_{M}$, $z^M$ and
${z'}^M$ as well as their derivatives. Thus the left superscript
is used to denote the order of iteration.

Next, to perform the iterations, it is more useful to work with a
flat-derivative interpretation of the EoM (\ref{emden0}):
\begin{align}\label{emden0a}
\frac{1}{\sqrt{|g|}}\,\left(\sqrt{|g|}\,g^{ML}g^{NR}F_{LR}\right)_{\!,N}=-
J^{M} \, , \qquad\qquad F_{MN}=\partial_M A_N-\partial_N A_M
\end{align}
and to  rewrite it, introducing ''new'' current\footnote{It
represents the vector density with respect to the total metric,
but each term of expansion of it will represent the vector in flat
background.}
 $\tilde{J}^{M}$:
\begin{align}\label{emden0b}
\partial_N\left(\sqrt{|g|}\,g^{ML}g^{NR}F_{LR}\right)=-\tilde{J}^{M}
 \, ,\qquad\qquad \tilde{J}^{M}(x)=e\int
\dot{z}^{M}(\tau)\,\delta^D\!\left(x-z(\tau)\smph\right)\, d
\tau\,.
\end{align}
Finally, one has to explicitly manifest the matter sources of the
generic equations to vary them in the sequel: the mechanical
energy-momentum tensor of two particles and the stress-tensor of
the bulk vector field are given by corresponding action variation
over the total metric $g_{MN}$ and read (in the gauge $g_{MN}\zt^M
\zt^N=1$)
\begin{align}\label{totalTm}
 T^{MN}_{\rm m}= m\int \frac{\zt^M \zt^N
\delta^D\!\left(x-z(\tau)\smph \right)}{\sqrt{-g}}\,d\tau \,
\qquad \qquad T^{MN}_{\rm m'}= m' \int \frac{\zt'^M \zt'^N
\delta^D\!\left(x-z'(\tau')\smph \right)}{\sqrt{-g}}\,d\tau',
\end{align}
and
\begin{align}
\label{totalTf} T^{MN}_{\rm em}= F^{ML}{F_L}^N+
\frac{1}{4}\,g^{MN}F_{LP}F^{LP} ,\,
\end{align}
respectively\footnote{Raising/lowering of indices here is
performed using the total metric, $g_{MN}$. Parallel displacement
bi-vectors $\bar{g}^{M}{}_{\!\nhsp \nhsp A}(x,z)$ are assumed in
(\ref{emden0},\ref{totalTm}) and omitted, due to the coincidence
limit $\delta^D(x-z)$.}.

 \vspace{0.5em}

\textbf{Zeroth order.} To zeroth order one expects the flat space
with no fields in it:
$$\nul h_{MN}=0\, , \qquad\qquad \nul A^M=0\, .$$
In what follows,  to this order both particles move freely:
$$\nul \ztt^M=\nul \!\!\zpp^M=0 $$
 with constant velocities $\nul\zt^M \equiv u^M$ and $\nul\!\!\left.\zt'\right.^M \equiv
{u'}^M$.

Furthermore we will   be working in the Lorentz frame where the
uncharged particle $m'$ is at rest (at zeroth order): in addition,
we set the origin of coordinate system to coincide with its
zeroth-order location.
\begin{align}\label{zprnul}
{u'}^M=(1,0, \pp\, 0)\, , \qquad\qquad  \nul
{z'}^{M}={u'}^M\,\tau'\,.
\end{align}
The charged particle $m$ is ultra-relativistic and moves along the
 3-brane with high-speed $v\lesssim 1$ and  large Lorentz factor $\gamma=(1-v^2)^{-1/2}
\gg 1$. We choose the spatial direction of zeroth-order motion as
the $z-$axis, while the vector of closest proximity $b^{\smhsp M}$
between the two particles is chosen to coincide with the $x-$axis.
Finally we choose the time of scattering to be zero. In what
follows
\begin{align}\label{znul}
u^M=\gamma(1,0,0,v, 0 \pp\, 0)\, , \qquad\qquad  \nul z^{M}=\nul
u^M\, \tau+b^M\,, \qquad\qquad b^{\smhsp M}=(0,b,0,\pp, 0) \,.
\end{align}
Thus $\gamma=u \cdot u'$ represents the Lorentz factor of
collision, $b>0$ represents the impact parameter of this
scattering, while  both $u^M$ and $ b^{\smhsp M}$ lie on the brane
and are mutually orthogonal.

Finally, vectorial and tensorial sources coming  from equations
(\ref{emden0b}) and  (\ref{totalTm}) are  given by
\begin{align}\label{vecq}
\nul\tilde{J}^{M}(x)=e\,u^M \int \delta^D \!\left(x-\nul z(\tau)
\right) \,d\tau
\end{align}
and
\begin{align}\label{Tzero}
\nul T^{MN}=m\,u^{M} u^{N}\int  \delta^D\!\left(x-\nul
z(\tau)\right)\, d\tau\, ,\qquad\qquad \nul {T'}^{MN}= m\,{u'}^{M}
{u'}^{N}\int \delta^D\!\left(x-\nul z'(\tau) \right) \,  d\tau\, ,
\end{align}
respectively, while $\nul T^{MN}_{\rm em}=0\hsp$.

\vspace{0.5em}

\textbf{First order.} The zeroth-order sources produce
corresponding first-order fields. Namely, from the Einstein
equations (\ref{emden2}) one expects to get the equation for $\un
h_{MN}$.

Consecutively computing the first-order
variations\footnote{Notice, here $h_{MN}$ represents the entire
tower of its iterations. In these notations with right superscript
we follow Weinberg \cite{Weinberg}.}
\begin{align}\label{elemvars}
 & g^{(1)}_{MN}=h_{MN} &\qquad\qquad&  g^{(1) MN}=-h^{MN}
  \nn\\& \Gamma^{(1)R}_{MN}=(h^{R}_{M \cd N}+h^{R}_{N\cd
M}-h_{MN}{}^{ \cd R})/2
 &\qquad\qquad&
  \Gamma^{(1)M}_{NR}\eta^{N
R}  =0\nn \\ &   R_{MN}^{(1)}=\frac{1}{2}\left(\vp h^{R}_{N\cd
MR}+h^{R}_{M\cd NR}-\Box\, h_{MN}-h_{\cd MN}\right)  &\qquad\qquad&    R^{(1)}=-\Box\, h+h_{MN}{}^{ \cd \nhsp MN}-R^{(1)}_{MN}h^{MN}  \nn\\
&
G_{MN}^{(1)}=\frac{1}{2}\left(-\Box\,\psi_{MN}-\eta_{MN}\xi_{L}^{\;
\cd L}+\xi_{M \cd N}+\xi_{N \cd M}\right)
   &\qquad\qquad& \xi_{M} \equiv \partial^{N}\psi_{MN} \,
\end{align}
 with $\Box \equiv \eta^{MN}\partial_M\partial_N$,
one introduces
\begin{align}\label{psidef}
\psi^{MN}=h^{MN} -\eta^{MN} \frac{h}{D-2}\, , \qquad\qquad h\equiv
h_P^P
\end{align}
and  sets the flat de\,Donder gauge
\begin{align}\label{psiguage}
\partial_N\psi^{MN}=0\, , \qquad\qquad \partial_N h^{MN}=\frac{1}{2}\,
h^{\hsp ,M}\,,
\end{align}
which leads to
\begin{align}\label{elemvars2}
 R_{MN}^{(1)}=- \frac{1}{2}\, \Box\,
h_{MN}\, ,\qquad \qquad R^{(1)}=- \frac{1}{2} \, \Box\, h \, ,
\qquad\qquad G_{MN}^{(1)}=-\frac{1}{2}\,\Box\,\psi_{MN}\; .
\end{align}
We note that the gauge fixation (\ref{psiguage}) implies
\begin{align}\label{psiguage2}
\partial_N\,{}^{k}\psi^{MN}=0\, , \qquad\qquad {}^{k}\psi^{MN}\equiv{}^{k}h^{MN}
-\frac{{}^{k} h_{LP}\hsp \eta^{LP}}{D-2}\, \eta^{MN}\,.
\end{align}
 Eventually, substituting $h_{MN}=\un
h_{MN} +\de h_{MN}+\pp$ and taking into account the gauge
(\ref{psiguage2}), one obtains the first-order variations
corresponding to our iteration scheme:
\begin{align}\label{elemvars3}
 \un R_{MN}=- \frac{1}{2}\, \Box\,
\un h_{MN}\, ,\qquad\qquad  \un R =- \frac{1}{2} \, \Box\, \un h
\, , \qquad\qquad \un \hsp G_{MN}=-\frac{1}{2}\,\Box\,\un \hsp
\psi_{MN}\,.
\end{align}
In what follows the first-order  Einstein equation (\ref{emden2})
reads
 \begin{equation}\label{Einm}
\Box \, \un\hsp \psi^{MN}=-\vk \nul T^{MN}\, ,\qquad \qquad \Box\,
\un h^{MN}=-\vk \left(\nul T^{MN}-\eta^{MN}\;\frac{\nul
T}{D-2}\right),\,
 \end{equation}
where $\nul T \equiv \eta_{LR}\nul T^{LR}$\hsp.

Substituting the zeroth-order matter part (\ref{Tzero}) one
obtains $\un h^{MN}$ as a sum
 \begin{equation}\label{sum}
\un h^{MN}=\un h_{\rm m}^{MN}+\un h_{\rm m'}^{MN}
 \end{equation}
due to  linearity of the first order, where each term represents a
solution  of (\ref{Einm}) with  source by the corresponding
particle separately.

Furthermore, the first order of (\ref{emden0b}) reads
\begin{align}\label{F1}
\partial_N \un F^{MN} =-\nul \tilde{J}^{M}
\end{align}
with source given by (\ref{vecq}).

Impose the flat Lorentz gauge for all orders\footnote{Take into
account, it differs from the originally covariant $\nabla_M
A^M=0$.}
\begin{align}\label{Agauge}
\partial_M\,{}^{k}\!A^{M} =0\,, \qquad\qquad ^{k}F_{MN} \equiv {}^{k}\!A^{N,M}-{}^{k}\!A^{M,N}
\end{align}
to derive
\begin{align}\label{A1}
\Box\un A^{M} =\nul \tilde{J}^{M}
\end{align}
as also a d'Alembert equation.

Now consider the first-order equations of motion for two
particles: making use of (\ref{eom_part}), one derives the
electromagnetic part of a force, acting on the charge as
\begin{align}\label{eom_part1}
m \, \un \ztt^M_{\rm em}=e\, \un F^{MN}\smhsp u_N\,.
\end{align}

Whereas $\un F^{MN}$ is produced by the same particle $m$, and one
has to consider $\un F^{MN}$ as \emph{external} field and  omit
the self-action of fields in this order\footnote{The account of
self-action in coordinate representation leads to the
renormalization of mass, radiation and radiation reaction
phenomena \cite{Dirac,Rohrlich,Teitelboim,GaSp} but these effects
are proportional to $\ztt^M$ and its derivatives, and do not
appear in the first order of PT, because of $\nul \ztt=0$   found
above. The appearance of self-action terms in higher orders will
be discussed below.}.

In what follows,  to  first order, both particles move along the
geodesics created by the gravitational field produced by the
partner particle, that we denote schematically
\begin{align}\label{gg'}
g_{MN}= \eta_{MN} + \vk \un h^{MN}_{\rm m'} + \mco(\vk^2)\,,
\qquad \qquad g'_{MN}= \eta_{MN} + \vk \un h^{MN}_{\rm
m}+\mco(\vk^2) \,.
\end{align}
Thus only the gravitational part of force survives \footnote{We
remind that this phenomenon is a direct consequence of that only
one particle is charged in the model-at-hand.} and the total
first-order EoMs for the acceleration (\ref{eom_part}) represent a
motion in the external linearized gravitational field and read
\begin{align}
\label{z1m} \un \ztt^{M}=-\vk \left(\un h^{ML,R}_{\rm m'}-
 \frac{1}{2}\, \un h^{LR,M}_{\rm m'}\right)u_{L}u_{R}\, , \qquad\qquad
 \un \zpp^{M}=-\vk  \left( \un h^{ML,R}_{\rm m}-
 \frac{1}{2}\,\un h^{LR,M}_{\rm m}\right) u'_{L} u'_{R}\, .
\end{align}
For a more complete  derivation of this gravitational part see
\cite{GKST4}. It justifies our model as ''radiation under
gravity-mediated collisions''.

\vspace{0.5em}

\textbf{Second order equation for
$\boldsymbol{A}\mathbf{-}$radiation.} The solution of  linear
equation (\ref{A1}) is the field generated by an uniformly moving
charge and represents the boosted Coulomb field. Hence it does not
contribute to radiation. In four dimensions it explicitly follows
from the Larmor formula for the electromagnetic radiation by an
accelerated charge. In arbitrary dimension it implicitly follows
from the Equivalence principle. We will discuss this more
thoroughly later.

The second order of our scheme leads to the radiation. For the
vector emission in the bremsstrahlung process it is enough to
consider only the correction to electromagnetic field $\de A^M$
and  its source.

Taking the next order of (\ref{emden0b}) together  with the
Lorentz gauge fixing, one obtains
 \begin{equation} \label{Forder1}
\Box \,\de A^{M}=j^M(x)\, ,\qquad  \qquad j^M(x) \equiv \rho^M
(x)+ \sigma^M(x)\, ,
 \end{equation}
where
\begin{align}\label{rholoc}
\rho^M(x) \equiv \un \tilde{J}^M (x)= e \int \left(\vp \un
\dot{z}^M-u^M \un z^N \partial_N \right)\,
\delta^D\!\left(x-\!\nul z\right) \, d\tau
\end{align}
and
\begin{equation}\label{son}
\sigma^M(x)=- \vk \;\partial_N \!\nhsp\left(\un h^{M}{}_{\! L} \un
F^{ LN }+ \un h^{N}{}_{\! \nhsp L} \un F^{ML} - \frac{1}{2}\, \un
h\un F^{MN} \right) ,
\end{equation}
respectively.

We will refer to the first term as the \emph{local} term since it
is fixed on the trajectory of particle $m$, while the second term
will be referred to as the \emph{non-local} current\footnote{Note
that there is some ambiguity with regard to the definition of the
local and the non-local part: indeed, if   both sides of
(\ref{emden0a}) are not multiplied by the factor $\sqrt{|g|}$ and
if vary it instead (\ref{emden0b}), the variation of this factor
will remain in the RHS and will be identified as local.
Nevertheless, for the source of 2nd-order field one needs the sum
of $\rho^M$ and $\sigma^N$ and, of course when such a factor
disappears from one term, it resurrects in another hand side
variation -- hence the sum is insensitive to such algebraical
transformations.}, as it comes from the left-hand side of
(\ref{emden0b}) and represents the non-linear terms of the vector
field with gravity.

A note to be added: in fact, we use the perturbation theory only
over the gravitational coupling $\vk$. This is achieved by the
fact that only the gravitational force acts on the particles up to
the first order. Because of  this fact, both terms in
(\ref{rholoc}) are proportional to $\zt^M$ and $z^M$,
respectively, and thus contain  $\vk$ as a pre-factor.

\subsection{The radiation formula}\label{radiation_formula}
Here  we  highlight the basic steps to derive the momentum
radiated in the form of an electromagnetic field. A flat space
world tube $W$ with a boundary of two space-like hypersurfaces,
$\Sigma_{\pm\infty}$ defined at $t\to\pm\infty$, as well as a
time-like cylindrical surface $C$ located at infinite distance is
considered. Spatially, both particles are located within the
volume or order $b^{\hsp D-1}$, due to the small scattering angle,
while with respect to time the  process is restricted by the
characteristic time of collision, where both fields in the source
(\ref{son}) are of equal significance. Thus one considers the
source of emission to be restricted by the characteristic
space-time volume $\mathcal{V}$. Integrating the flux through the
two hypersurfaces with the time-positive normals, we write the
emitted momentum $P^M$, using the flat-space background concept:
 \begin{equation}
 \label{flux1}
P^M=\int\limits_{\Sigma_{+\infty}}T^{MN}\,
dS^{+}_N-\int\limits_{\Sigma_{-\infty}}T^{MN}\,
dS^{+}_N=\int\limits_{\partial W} T^{MN}\,d S_N=\int\limits_W
\partial^N T_{MN}\,\dx=-\int\limits_W
F^{MN} J_{N} \, \dx\, ,
\end{equation}
where $T^{MN}$ and $J^M$ are   flat analogues of (\ref{totalTf})
and (\ref{emden0}), respectively. Here one uses the Gau{\ss}'s
theorem and the Maxwell equations and  implies the cancelation of
the surface integral over $C$ due to  the fact that it corresponds
to the retarded moment $t\to -\infty$ of emission, where the
motion was free.

Performing a Fourier-transformation\footnote{Our convention on the
Fourier-transforms is
$$
\varphi\hsp(x)=\frac{1}{(2 \pi)^D}\int \varphi\hsp(k) \, e^{-i kx}
\dk \, , \qquad\qquad \varphi\hsp(k)= \int \varphi \hsp(x) \,e^{i
kx} \dx\,.
$$
}, substituting $F^{MN}$ by its retarded solution via the Green's
function and making use of current transversality ($k \cdot j =0$)
with the fact that $j^M(x)$ is a real-valued function, we obtain
\begin{align}\label{flux3}
P^M =\frac{i}{(2 \pi)^D}\int  \dk\, k^M \,G_{\rm ret} (k)\, j^*
(k)\cdot j (k)  \, ,
\end{align}
where $G_{\rm ret}(k)=-\mathcal{P}\left(1/k^2\right)+i \pi \,{\rm
sgn}(k^0)\,\delta(k^2)$. The real part
$-\mathcal{P}\left(1/k^2\right)$ does not contribute to the
integral due to imparity   over time integration. Finally,
transforming the integral into positive values of $k^0$ and
integrating over $|k|$ with $\delta(k^2)$, one finally obtains
\begin{align}\label{flux4}
P^M =-\frac{1}{2(2 \pi)^{D-1}}\int \frac{k^M}{| \mathbf{k}|}  \,
j^*_N(k)\hsp j_L(k) \,\eta^{NL}\, d^{D-1}\mathbf{k}\,,
\end{align}
where $\mathbf{k}$ is an absolute value of $(D-1)$-dimensional
spatial part of $k^M$. Taking into account the transversality and
the on-shell condition $k^2=0$ of the emitted wave, one can
replace the Minkowski metric in $\eta^{MN}$ by
\begin{align}\label{gdddd}
\Delta^{MN} \equiv
\left({}_{g}\Pi^{M}{}_{\!L}\right)\left({}_{k'}\Pi^{LN
}\right)=\eta^{MN}+\frac{k^{M}k^{N}-2\hsp  (k
g)\,k^{(\!M}g^{N)}}{(kg)^2} \, ,
\end{align}
with any time-like unit vector $g$, where
\mbox{$\,{}_{g}\Pi=1-g\otimes g$} and \mbox{$\,{}_{k'}\Pi=1+\,
k'\otimes k'/(kg)^2$} are projectors onto subspaces transverse to
$g$ and \mbox{$k'\equiv\,{}_{g}\Pi k = k-(kg)\hsp g$},
respectively. Since $k'\cdot g=0$, the projectors $\,{}_{g}\Pi$
and $\,{}_{k'}\Pi$ commute. Their product $\Delta^{MN}$ is then a
symmetric projector onto the subspace $M_{k,g}$, perpendicular to
$k$ and $g.$ By construction, the projector $\Delta$ is idempotent
($\Delta^2=\Delta$), thus  it acts on $M_{k,g}$ as the unit
operator. In what follows, we will conveniently choose $g_M=u'_M$
and calculate the flux in the Lorentz frame (referred to as the
Lab frame further) with $u'_M=(1,0, \ldots, 0)$.

We arbitrarily choose the orthonormal basis $\{\varepsilon_i^M\}$
on $M_{k,g}$ and set the resolution of identity $$ \Delta^{MN}
=-\sum\limits_i  \varepsilon_i^M \hsp \varepsilon_i^N\, ,
\qquad\qquad
 \varepsilon_i^M \hsp \varepsilon_{jM}=-\delta_{ij}\,, \qquad\qquad i=1,2...,D-2\,.$$

Finally, setting $M=0$ for the energy, the radiation formula reads
\begin{align}\label{4momentum}
E_{\rad}= \frac{1}{2\left(2\pi\right)^{D-1}}\sum_i
\int\limits^\infty_0 \omega^{D-2}\, d \omega
\int\limits_{S^{D-2}}\!\! d \Omega \;\vert  J \cdot \varepsilon_i
\vert ^2\
\end{align}
as sum over polarizations, where $\omega \equiv k^0$ while $d
\Omega$ stands for  the measure on unit  sphere $S^{D-2}$.

\vspace{0.5em}

\textbf{Polarization vectors.}  Polarization vectors are mutually
orthogonal and satisfy $ \varepsilon_i\cdot k  = \varepsilon_i
\cdot u' =0$. It is convenient to choose the first $D-4$ vectors $
\varepsilon_\alpha$ to be orthogonal to the collision space
(\{scattering plane\} $\times$ \{time\}), defined by the linear
shell of $u^M$, $u'^M$ and $b^{\smhsp M}$. Thereby they satisfy
the relations $ \varepsilon_{\alpha} \cdot k =
\varepsilon_{\alpha} \cdot u'= \varepsilon_{\alpha} \cdot u=
\varepsilon_{\alpha} \cdot b=0$, where $ \alpha=3,\pp, D-2 $.
Choosing the $D$-dimensional unit antisymmetric tensor to be
$\epsilon^{0xyz3,...(D-2)}=1$, we define the remaining two
polarization vectors as
\begin{align}\label{1st_pol}
 \varepsilon_1^M=N^{-1}\left[\left(ku\right)
u'^M-\left(ku'\right)u^M+\left( u\cdot u' -\frac{ k\cdot u }{
k\cdot u' }\right)k^M\right]
\end{align}
and
\begin{align}\label{2nd_pol}
 \varepsilon_2^M=N^{-1}
\epsilon^{MM_1M_2...M_{D-1}}\,u_{M_1}u'_{M_2}k_{M_3}\,\varepsilon_{3M_4}\pp
\varepsilon_{(D-2)M_{D-1}}\,,
\end{align}
respectively, where $N$ is a normalization constant given by
\begin{align}
N^2=-\left[\left(ku'\right)u-\left(ku\right)u'\vp\right]^2.
\end{align}
By construction, it is easy to verify that $\varepsilon_1 \cdot u'
= \varepsilon_2 \cdot u= \varepsilon_2 \cdot u'=0$ and $
\varepsilon_1 \cdot k= \varepsilon_2 \cdot k= \varepsilon_1 \cdot
\varepsilon_2= 0$.

\begin{figure}
\begin{center}
\includegraphics[angle=0,width=13cm]{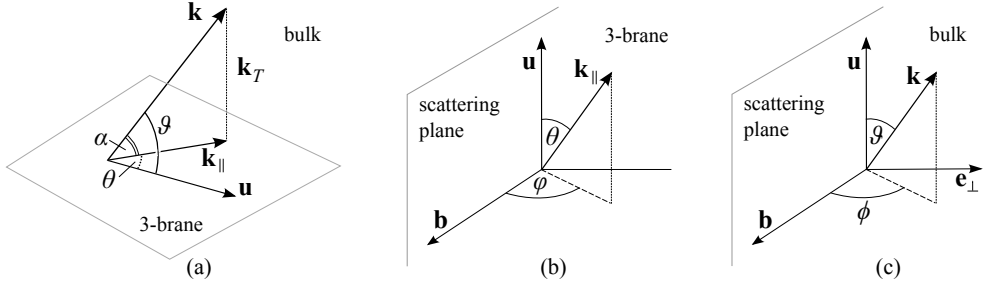}
\caption{The angles in the lab frame used in the text. Vector
$\mathbf{k}$ is split onto brane and bulk components as
$\mathbf{k}=(\mathbf{k_{\parallel}}, \mathbf{k}_T)$.}
\label{branepic}
\end{center}
\end{figure}

Introducing the angles according to Fig.\,\ref{branepic} (for
additional info see Appendix \ref{formulae}), the normalization
factor reads $N=\gamma v \sin \vartheta (ku')$ and  the following
products do not vanish: $(e_1 b)$, $(e_1 u)$ and $(e_2 b)$,
respectively. The values of these contractions are given by
 \begin{align}\label{e_products}
   \varepsilon_1 \cdot b =-b\,\cos \vartheta\cos\nhsp\phi \,, \qquad\qquad
\varepsilon_1\cdot  u =\gamma v \sin\vartheta\, ,  \qquad \qquad
 \varepsilon_2 \cdot b  =-b\,\sin\nhsp\phi\, .
\end{align}
For the derivation, see \cite{GKST4}. Thus the ''bulk''
polarizations do not contribute into radiation; thereby in
addition, one can introduce chiral polarization vectors in a usual
way as
 \begin{align}\label{e_chiral}
\varepsilon_{\pm}^M = \frac{ \varepsilon_1^M \pm i \varepsilon_2^M}{\sqrt{2}}\;.
\end{align}


\vspace{0.8em}

To summarize this Section: the formula for emitted radiation
(\ref{4momentum}) and the appropriate polarization states of
massless $D$-dimensional photon are derived, and for the
problem-at-hand only two polarizations given in the covariant form
(\ref{1st_pol}, \ref{2nd_pol}), contribute into the total emitted
energy, as it is proper in four dimensions.

The source of the emitted field is to be computed within the
iteration scheme based on the perturbation theory over
gravitational constant $\vk$.

Notice, $J^M$ in (\ref{4momentum}) represents the \emph{total}
source of the total $A^M$ as a solution in \emph{flat} space-time,
and thus in our iteration scheme it is given by the series
 \begin{align}\label{J_series}
J^M(k)=\nul J^M (k)+ j^M(k) +\pp \,.
\end{align}
Here the $\nul J^M$ given by (\ref{vecq}) is a source of boosted
Coulomb field, and its square does not contribute to the
radiation. It will be shown below that the contribution of product
$\nul J^{*} \cdot j +\nul J \cdot j^{*}  $  also vanishes, and
$\sum |j \cdot \varepsilon_i|^2$ becomes the first surviving order
which contributes to the total emitted energy.

Thereby $j^M(k)$ (\ref{Forder1}) as well as its constituents
becomes of particular significance and we concentrate on its
evaluation. Looking at $\sigma^M(k)$ (\ref{son}), it is enough to
restrict ourselves on the first-order perturbation of the
gravitational field $\un h^{MN}=\un h_{\rm m}^{MN}+\un h_{\rm
m'}^{MN}$. Thus in order to simplify notations, we keep $h^{MN}$
as a simplified notation of $\un h^{MN}_{\rm m}$ and denote,
respectively, $h'^{MN} \equiv \un h^{MN}_{\rm m'} $.

\section{The radiation amplitudes}\label{radiation_amplitudes}

The first-order fields discussed above  in the momentum  space are
given by
 \begin{align}\label{1h'mn}
 &h_{MN}(q)=
 \frac{2\pi\hsp \vk \hsp m}{q^2}\, \e^{iqb}\,\delta(qu)\nhsp \left(\nhsp \nhsp u_{M}u_{N} -
\frac{1}{d+2}\,\eta_{MN} \nhsp \nhsp \right)\, &\qquad \qquad &
\un A^M(q)=-
\frac{2\pi\,e}{q^2}\, \e^{iqb}\,\delta(qu)\, u^M\,, \\
&h'_{MN}(q)=
 \frac{2\pi \vk m'}{q^2}\, \delta(qu')\left(u'_{M}u'_{N} -
\frac{1}{d+2}\,\eta_{MN} \right)\, &\qquad \qquad& \un
F^{MN}(q)=i\,\frac{2\pi\,e}{q^2}\, \e^{iqb}\,\delta(qu)\nhsp
\left[\smhsp q^M u^N\nhsp -q^N u^M\vp \right]\smhsp . \nn
\end{align}
Respectively,  the Fourier-transform of $\nul\nhsp \tilde{J}
(x)=\nul {J} (x)$ reads
 \begin{align}\label{J0mom}
\nul\nhsp  {J}^M(q)=2\pi\,e\, \e^{iqb}\,\delta(qu)\, u^M\,.
\end{align}

Now we  proceed to compute the two parts of the radiation
amplitude.
\subsection{Local amplitude}
\label{local} The Fourier transform of (\ref{rholoc}) reads
\begin{align}
\label{loc_varn} \rho^{M} (k) =e\,\e^{i (kb)} \!\int \!\left[\vp
{}^{1}\dot{z}^M+ i \hsp(k \un{z})\,u^M  \right] \e^{i(ku)\tau}\,
d\tau\,.
\end{align}
The first order correction to the trajectory is computed in
\cite{GKST3} and we  quote  that result here.
\begin{align}\label{sc6}
 \un z^{M}(\tau)=-\frac{ i m' \varkappa_D^2}{(2
\pi)^{D-1} }  \int \dq\, \frac{\delta(qu')}{q^2 (qu)}\,
\e^{-iqb}\left(\e^{-i(qu)\tau}-1\right) \left[\gamma
u'^{M}-\frac{1}{d+2}\,u^{M}-\frac{ \gamma_*^2}{2 \hsp (qu)}
\,q^{M}\right],
\end{align}
where $\gamma_*^2\equiv \gamma^2-{(d+2)}^{-1}$. We drop all the
terms containing ${u^\prime}^M$ since they are transverse to the
polarization vectors and thus will not contribute to the
radiation. After integrating with respect to $\tau$ we obtain
\begin{equation}\label{rho78}
\rho^{\smhsp M}\nhsp(k)=-\frac{e m^\prime \vk^2 \,\e^{i
(kb)}}{\left(2 \pi \right)^{D-2} } \left[\, \gamma \hsp I
\left({u'}^M-\frac{  k u^\prime   }{  ku  }\,
  \hsp u^M \right)-\frac{\gamma_*^2 }{2 \hsp ( ku )}\, {I}^M+\frac{\gamma_*^2\, k \cdot {I}}{2  \hsp ( ku)^2}
  \,u^M\,
  \right]\, ,
\end{equation}
where the integrals $I$ and $I^{M}$ are defined by
\begin{align}\label{Iints0}
I = \int \frac{ \delta(pu')\, \delta(ku-pu)\,\e^{-i(pb)}}{p^2}\:
\ddp \,,\qquad \qquad I^{M} = \int \frac{ \delta(pu')\,
\delta(ku-pu)\, \e^{- i(pb)}}{p^2} \;p^{\, M} \, \ddp \,.
\end{align}
These integrals have been computed in \cite{GKST2} in terms of
Macdonald functions (modified Bessel functions of 3rd kind):
\begin{align}\label{Iints}
I  =-\frac{(2\pi)^{d/2+1}}{\gamma v\,b^{\hsp d}}\,
{\hat{K}}_{d/2}(z)\, ,\qquad \qquad  I^{M}
 =-\frac{(2\pi)^{d/2+1} }{\gamma v \hsp b^{\hsp d+2}}\left(b\hsp z\hsp {\hat{K}}_{d/2}(z)\,\frac{\gamma
u'^{M}-u^{M}}{\gamma v}+ i \hat {K}_{d/2+1}(z)\, b^{\smhsp
M}\right),
\end{align}
with
\begin{equation} \label{z1}
 z\equiv \frac{(ku) b}{\gamma v}\,, \qquad\qquad z'\equiv \frac{(ku')b}{\gamma v}  \,,
 \end{equation}
where we use the more economic,  non-conventional notation
$\hat{K}_{\nu}(x)\equiv x^{\nu}{K}_{\nu}(x)$,
 in order to simplify the explanation of estimates making use of slowly altering at $[\smhsp 0,\mco(1)\smhsp]$ function.

Substituting (\ref{Iints}) into (\ref{rho78}) one obtains
\begin{equation}\label{rhon(k)}
\rho^{\smhsp M}\nhsp(k)=\frac{\lambda \, \e^{i (kb)}}{  v} \left[
\left(1- \frac{\gamma_*^2}{2 \hsp \gamma^2
v^2}\right)\!\!\left(u'^M- \frac{z'}{z}\,u^M \right)
\hat{K}_{d/2}(z)+\frac{i \gamma_*^2}{2\hsp \gamma^2 v \smhsp
z}\left( \frac{(kb)}{\gamma v z}\, u^M-\frac{b^{\smhsp
M}}{b}\right)\hat{K}_{d/2+1}(z) \right] ,
\end{equation}
with
\begin{equation}\label{lambda}
\lambda \equiv \frac{e\hsp  m'\smhsp \vk^2}{\left(2
\pi\right)^{d/2+1}b^{\hsp d}}\;.
\end{equation}

Here we have restored the dependence on $u'^M$ in order to make
obvious the conservation of the current (Subsection
\ref{conservation}).

The local current $\rho^{M}(k)$ contains Macdonald functions
$K_{\nu}(z)$ and, combined with the volume factor $\omega^{\smhsp
d+2}\sin^{d+1}\nhsp \vartheta$, gives dominant contribution in the
region $\omega \sim \gamma^2\!/\smhsp b$, $\vartheta \sim
\gamma^{-1}$ (i.e. $z \sim 1$), as   was argued in \cite{GKST2}
and will be discussed later in Subsection \ref{vec-vec}. In this
region for the usage below we expand $\rho^M(k)$ in powers of
$\gamma$:
\begin{equation}\label{rhonexp}
\rho^{\smhsp M}\nhsp(k)=\nhsp \frac{\lambda \, \e^{i (kb)}}{2 }
\nhsp \left[-  \frac{z'}{z}\, \hat{K}_{d/2}(z)\,u^M - i \nhsp
\left(\nhsp \frac{z' \sin \vartheta \cos \phi}{z}\, u^M\nhsp
+\nhsp \frac{b^{\smhsp M}}{b} \nhsp \right)\nhsp
\frac{\hat{K}_{d/2+1}(z)}{z}  +
\frac{d+1}{d+2}\,\frac{z'}{\gamma^2 z}\, \hat{K}_{d/2}(z)\,u^M +
  \mco( \gamma^{-2})  \right]\! ,
\end{equation}
where the first term in the parenthesis is of order
$\mco(\gamma)$, the square-bracket-term has order $\mco(1)$, while
the last term is of order $\mco( \gamma^{-1})$ and the rest
represents all subleading terms.

\subsection{Non-local (stress) amplitude}

The Fourier transform of (\ref{son}) is given by
\begin{align}\label{Clinton05}
 \s^M\nhsp(k)=\frac{\vk^2\hsp  e \hsp m^\prime\,
\e^{i\left(kb\right)}}{ (2 \pi )^2 }& \left[(ku^\prime)
 \left((k u^\prime )\,u^M-(k u)\, u'^M \vp \right) J+
 \left(\gamma \left(k u'\right)-\frac{\left(ku\right)}{d+2} \right)J^{M}+ \right. \nn \\
& \quad +\left. \left(\frac{u^M}{d+2}-\gamma u'^M \right) k \cdot
J +\left(\vp (k u)\, u'^M -(k u')\, u^M\right)u' \cdot J
\vphantom{\frac{a}{a}}\right] \,,
\end{align}
where\footnote{We denote these double-propagator Fourier integrals
as $J$ and $J^M$, the same letter as vectorial source introduced
in the Section\,\ref{setup}, in order to keep notation and contact
with \cite{GKST4}, so we hope, this will not bring a reader to
some misleading.}
\begin{align}\label{Jints}
J(k) \equiv \int  \frac{\delta(pu')\, \delta(ku-pu)\,
\e^{-i(pb)}}{ p^{\hsp 2}\,  (k-p)^2  }  \: \ddp \, \,,
\qquad\qquad J^M(k) \equiv \int \frac{\delta(pu')\,
\delta(ku-pu)\, \e^{-i(pb)}}{ p^{\hsp 2}\, (k-p)^2}\, p^M  \, \ddp
\,,
 \end{align}
which have been computed in \cite{GKST4} as   integrals over
Feynman parameter $x$. We keep (\ref{Clinton05}) for the proof of
gauge conservation and further suppress terms proportional to
${u'}^M$, as they do not contribute to the radiation. From this
definition (\ref{Jints}) it follows that $u' \cdot J=0$, thus
$\s^M(k)$ reads:
\begin{align}
\label{sigmaf} \s^M\nhsp(k)=\frac{\lambda \,
\e^{i\left(kb\right)}}{2 \gamma v} \int\limits_0^1 dx \,\e^{-i
(kb )\,x} &
 \left\lbrace \left[\frac{\beta-x \xi^2}{d+2}-\left(\gamma z^\prime - \frac{z}{d+2}\right)
 \left( (1-x )\,z+\gamma z^\prime x\vp\right)+   \gamma^2v^2 {z^\prime}^2\right] \hat{K}_{d/2-1}(\zeta)\, u^M + \right. \nn \\
&\,\, \left. +\hsp i\hsp \left[\frac{ (kb
)}{d+2}\,u^M+\left(\gamma^2v z^\prime-\frac{\gamma z
v}{d+2}\right)\!
 \frac{b^{\smhsp M}}{b}\right]\hat{K}_{d/2}(\zeta)\right\rbrace ,
 \end{align}
with  $$\xi^2 \equiv 2\hsp \ga \smhsp z \smhsp z' - z^2 -
{z'}^2\,, \qquad\qquad \beta \equiv \ga \smhsp z \smhsp z' - z^2\,
, \qquad  \qquad \zeta^2( x) \equiv {z'}^2 x^2+2 \hsp \gamma
\smhsp z \smhsp z'\smhsp x \left( 1-x \right)+z^2
\left(1-x\right)^2 .$$ The non-local amplitude has now been
written in terms of   three scalar integrals of the following
type:
\begin{equation}\label{wedf}
J_{(\sigma,\tau)}\equiv \int\limits_0^1 x^\sigma \, \e^{-i\left( k
b \right)x}\, \hat{K}_{d/2+\tau} (\zeta)\, dx\,, \qquad\qquad
(\sigma,\tau)=(0,-1), (0,0) \,\text{ and }\, (1,-1) \,.
\end{equation}
These integrals have been studied in details in \cite{GKST4}:
introducing parameter $\varrho\sim \omega \smhsp b \hsp
\vartheta$, (\ref{wedf}) is expanded as series over $1/\varrho$.
Thus in the high-frequency region (or $z-$region, for brevity)
$\omega \sim \gamma^2\!/\smhsp b$, $\vartheta \sim \gamma^{-1}$
the dominant contribution  comes from small $x=0 \pp \mco(
\gamma^{-2})$ and all integrals (\ref{wedf}) are to be expanded in
terms of Macdonald functions with argument $\zeta(0)=z$, with
expansion parameter $1/\gamma$. In the large-angle region (or
$z'-$region) $\omega \sim \gamma /\smhsp b$, $\vartheta \sim 1$
the dominant contribution comes from the values of $x$ near 1: $x=
1-\mco( \gamma^{-2})\pp 1$ and all such integrals are to be
expanded in terms of Macdonald functions with argument
$\zeta(1)=z'$.

In the transition region ($\omega \sim \gamma/\smhsp b$,
$\vartheta \sim \gamma^{-1}$) the exponential in (\ref{sigmaf})
does not oscillate rapidly and the whole domain $x=[0,1]$
contributes equally. The series with Macdonald functions
$K_{\nu}(z)$ and $K_{\nu'}(z') $ is also valid (see Appendix
\ref{DI}) but converges very slow since no small factor is
available. Finally, in the ultimate region ($\omega \sim
\gamma^2\!/\smhsp b$, $\vartheta \sim 1$) the whole integral is
exponentially suppressed by $\mco (\e^{\gamma} )$.

Next we  consider more thoroughly the high-frequency behavior of
local and non-local amplitudes.

\subsection{Destructive interference}

We now proceed to demonstrate the cancelation of the two leading
orders of $\sigma^M$ and $\rho^M$ in powers of $\gamma$ in the
$z-$region, which leads to the strong damp of the amplitude by
$\mco(\gamma^2)$ and the emitted energy by four orders of
$\gamma$. We will refer to this effect as destructive
interference. The same effect for gravity was described in
\cite{GKST3,GKST4} by means of the same representation via
Macdonald functions. In another representation  it appeared in
\cite{ggm} dealing with only four dimensions.

We follow \cite{GKST4} and sketch the procedure for showing this:
in the $z-$region ($z\sim 1, z' \sim \gamma$) the integral $J_{(
1,-1)}$ is suppressed by two orders of $ \gamma $ with respect to
the $J_{( 0, -1)}$ and $J_{(0,0)}$  as it was implicitly mentioned
in the previous Subsection and proved in \cite[eqn.(3.28)]{GKST4}.
We now keep only the terms that will give us the first three
orders of (\ref{sigmaf}):
\begin{align}\label{jkl}
 \s^M\nhsp(k)\approx\frac{\lambda \, \e^{i (kb )}z'\gamma}{2
} \int\limits_0^1 dx \, \e^{-i  (kb )\,x} \left[  {z'}
\hat{K}_{d/2-1} (\zeta )\, u^M + \frac{i  }{b}\,
\hat{K}_{d/2}\left(\zeta\right)\,b^{\smhsp M}-\left(
\frac{d+1}{d+2}\, \frac{z}{\gamma} +\frac{{z'}}{\gamma^2} +  z'
x\right) \hat{K}_{d/2-1} (\zeta )\,u^M\right].
 \end{align}
Finally we substitute the approximation \cite[eqn.(B.10)]{GKST4},
appropriately simplified here neglecting the exponentially
suppressed Macdonalds $K_{ \nu'}(z')$ ($a \equiv  {z}/{\sin
\vartheta}$):
\begin{align}\label{refer_int}
 \,\,\, \int\limits^1_0 \!dx \, \e^{-i (kb)\hsp x}\hat{K}_{ \nu-1}(\zeta)\nsmhsp &
\approx \frac{ \beta}{a^2 \coa^2} \left( \hat{K}_{ \nu}(z)-i
\frac{ (kb)}{\beta}\,\hat{K}_{ \nu+1}(z)  -\frac{ 2\nu+1 }{a^2} \,
\hat{K}_{ \nu+1}(z)   + \frac{ \sin^2\! \nhsp \phi }{a^2}\,
\hat{K}_{ \nu+2}(z) \right)+\pp
\\
 & \nsmhsp \approx \nsmhsp \frac{\hat{K}_{ \nu }(z)}{\gamma z
z'}\nsmhsp - \nsmhsp i \frac{ (kb )\, \hat{K}_{ \nu+1}(z)}{
(\gamma z z )^2}\nsmhsp +\nsmhsp \frac{1\nsmhsp - \nsmhsp\gamma^2
\psi}{ \gamma^3 z'z }
  \,\hat{K}_{ \nu }(z)
\nsmhsp  - \nsmhsp  \frac{\sin^2\!\vartheta}{\gamma z^3 z'} \left[
( 2 \nu\nsmhsp +\nsmhsp 1 )  \hat{K}_{ \nu+1}(z) \nsmhsp-
\nsmhsp\sin^2\!\nhsp \phi \,\hat{K}_{ \nu+2}(z) \right].\nn
 \end{align}
For $J_{( 1,-1)}$-type integral \cite[eqn.(3.28)]{GKST4} we retain
only the leading terms:
\begin{align}
\int\limits^1_0 dx \, x \, \e^{-i  (kb )\hsp x}\hat{K}_{ \nu-1}
 (\zeta )\approx -\frac{1}{\left(\ga z'\right)^2}\,\hat{K}_{ \nu
} (z ) -\frac{\left( 2\nu+1\right)}{\left(\gamma z z'\right)^2}\,
\hat{K}_{ \nu+1}(z)+ \frac{1}{\left(\gamma z z'\right)^2}\,
\hat{K}_{ \nu+2}(z)\,.\nn
\end{align}
Thus upon substitution of the latter two into  (\ref{jkl})  and
 retaining the first three orders, one obtains:
\begin{align}
\sigma^M\approx \frac{\lambda \, \e^{i \left(kb\right)}}{2\gamma}
& \left[ \frac{\gamma z'}{z}\,\hat{K}_{d/2}\left(z\right) u^M+i
\left(\gamma \frac{b^{\smhsp M}}{b} -\frac{\left(kb\right)}{z} u^M
\right)\frac{\hat{K}_{d/2+1}(z)}{z}\,   - \frac{d+1}{d+2}\,
 \hat{K}_{d/2} (z
)\,u^M -\right. \\& \; \left.
 -(d+1)\left(1-\frac{\sin^2\!\vartheta}{\psi
}\right)\nhsp\frac{\hat{K}_{d/2+1} (z )}{z^2}\, u^M +\left( \!
\left( \frac{ \sin^2 \!\vartheta \sin^2 \!\nhsp\phi}{\psi}
-1\right) u^M+\frac{ (kb )}{z'}\frac{b^{\smhsp
M}}{b}\right)\nhsp\frac{\hat{K}_{d/2+2} (z) }{z^2}\right].  \nn
\end{align}
The first two orders of this expression exactly cancel with the
first two orders of (\ref{rhonexp}), leaving us with
\begin{align}\label{j_zed}
j^M\approx \frac{\lambda \, \e^{i \left(kb\right)}}{2\gamma}\left[
\frac{d+1}{d+2}\,\left( \frac{1}{\gamma^2 \psi}-1\right)
\hat{K}_{d/2}(z)\,u^M-(d+1)\left(1 -\frac{\sin^2\!\vartheta}{\psi
}\right)\nhsp\frac{\hat{K}_{d/2+1} (z )}{z^2}\, u^M\right. +\nn
\\ \left. +\left( \! \left( \frac{ \sin^2 \!\vartheta \sin^2
\!\nhsp\phi}{\psi} -1\right) u^M+\frac{ (kb )}{z'}\frac{b^{\smhsp
M}}{b}\right)\nhsp\frac{\hat{K}_{d/2+2} (z) }{z^2} \right].
\end{align}
We note that even though the current will finally be projected on
the two polarization vectors, this will not change our conclusion,
as the contractions (\ref{e_products}) add no powers of $\gamma$
\emph{at the region of interest}.

\subsection{The total radiation amplitude}

In order to compute the total radiation energy, we will need to
study the following three regions. The $z$-type radiation emitted
for angles $\vartheta \sim 1/\gamma$ and $\omega\sim
\gamma^2\!/\smhsp b$, the region with frequency $\omega \sim
\gamma/\smhsp b$ again for small angles and finally the radiation
at angles $\vartheta \sim 1$ at medium-frequencies.

\vspace{0.5em}

\textbf{High frequency  regime.} The radiation amplitude in
$z-$regime after the destructive interference was derived in the
previous Subsection. Projecting (\ref{j_zed}) on (\ref{e_chiral}),
the chiral amplitudes $j_{\pm} \equiv j \cdot \varepsilon_{\pm}$
read:
\begin{align}\label{tau1hf}
j_{\pm}\nhsp(k) \approx  \frac{\lambda \, \e^{i  (kb )} \sin
\vartheta}{2\sqrt{2} } & \left[\frac{d+1}{d+2}\,
  \frac{1-\gamma^2 \psi}{\gamma^2 \psi}\,
\hat{K}_{d/2} (z ) - \frac{d+1}{z^2}   \left( \frac{ \sin^2\!
\vartheta}{
\psi} -1 \right) \hat{K}_{d/2+1} (z ) +\right. \nn \\
& \;\;\left.+\frac{\sin^2 \nhsp \!\phi }{z^2}\left(
 \frac{\sin^2 \! \vartheta }{\psi} -1
\right) \hat{K}_{d/2+2} (z )
 \pm i\,\frac{  \sin 2  \phi }{ 2\, z^2}
 \, \hat{K}_{d/2+2} (z ) \right].
\end{align}
All terms in the parenthesis (\ref{tau1hf})  are of order $
{\mco(1)}$ (in $\lambda=1$ units) within $z-$regime, hence the
whole amplitude goes like $\mco( \gamma^{-1})$ due to the common
pre-factor $\sin \vartheta$.

\vspace{0.5em}

\textbf{Large angle  regime.} In this region  of the parameters
($z'-$regime)  $z$ is of order $ \mco( \gamma)$, so the Macdonald
functions that have $z$ as their argument are exponentially
suppressed. Thus we ignore the local part of the current and
consider only  the non-local part. To repeat, in this regime the
main contribution of the integrals with respect to $x$ comes from
the area near $x=1$, and the integrals $J_{0,\tau}-J_{1,\tau}$,
which are of the form $1-x$, are suppressed by a factor of $\mco (
 \gamma^{-2})$ with respect to both $J_{0,\tau}$ and $J_{1,\tau}$.
We rewrite (\ref{sigmaf}) in a way where we are expanding both
with respect to $\gamma$ but also with respect to $1-x$. Taking
also into account that $u^M$ is perpendicular to the second
projection, while it gives us an order of $\gamma$ when projected
on the first polarization, while $b^{\smhsp M}$ gives no
additional powers of $\gamma$ when projected on either
polarization, we write the two leading orders:
\begin{align}\label{j_pr}
j^M\nhsp(k)\approx\frac{\lambda \, \e^{i\left(kb\right)}}{2 \gamma
}
 \int\limits_0^1 dx \, \e^{-i  (kb )\hsp x} & \left\lbrace \left[ \! \vph \right.\right. \! \! \left(\gamma^2
{z'}^2-\frac{d+1}{d+2}\,
  \gamma z z'\right) (1-x )+\frac{{z'}^2}{d+2} \left. \vph \right] \hat{K}_{d/2-1} (\zeta
)\,
   u^M  \nn \\ & \,\,\left. +\, i\hsp \left[\frac{ (kb )}{d+2}\, u^M+\left(\gamma^2 z'-\frac{\gamma z}{d+2}\right)\frac{b^{\smhsp M}}{b}\right]
    \hat{K}_{d/2}\left(\zeta\right)\right\rbrace \, .
 \end{align}
Since no destructive interference is expected, we retain only the
leading terms of integrals, and set $x=1$ inside the integrand of
(\ref{j_pr}). These integrals have been computed in \cite{GKST4}
and give, to the leading order,
\begin{equation}\label{j_pr2}
\int\limits^1_0 \e^{i \left(kb\right)x}\,\hat{K}_\tau (\zeta )\,
dx \approx \, \frac{\e^{-i\left(kb\right)}}{{z'}^2 \gamma^2
\psi}\, \hat{K}_{\tau+1 }  (z' )\, .
\end{equation}
Eventually, the entire first line in (\ref{j_pr}) turns out to be
subleading with respect to the second one, and, upon substitution
(\ref{j_pr2}) $j^M$ reads:
\begin{align}
j^M\nhsp(k)\approx\frac{\lambda\,i}{2 \gamma \psi}
 \left[\frac{ (kb )}{\gamma^2 {z'}^2 (d+2 )}\,
u^M+\left(\frac{1}{z'}-
  \frac{z}{\gamma z' (d+2 )}\right)\frac{b^{\smhsp M}}{b}\right] \hat{K}_{d/2+1} (z' )  \, .
 \end{align}
Finally projecting on $\varepsilon_{\pm}$ (\ref{e_chiral}) the two
significant radiation amplitudes in $z'-$region are given by
\begin{align}\label{j_pr4}
j_{\pm}\nhsp(k)\approx\frac{\lambda\,i}{2 \sqrt{2} \hsp \gamma
\hsp \psi}
 \left[-\frac{ \sin^2 \nhsp\vartheta  \, \cos\nhsp\phi}{{z'} (d+2 )}\,
 +\left(\frac{ \psi}{d+2}-\frac{1}{z'}
  \right) \cos \vartheta\cos\nhsp\phi \pm  i   \left(\frac{ \psi}{d+2}-\frac{1}{z'}
  \right) \sin\nhsp\phi \right] \hat{K}_{d/2+1} (z' )  \, .
 \end{align}
In what follows, the amplitudes are of order $\mco (\gamma^{-1})$.

 \vspace{0.5em}

\textbf{Transition regime.} In this region, the projection of the
current on the polarization vectors will once more not add any
powers of $\gamma$. We have $z \sim 1/\gamma$ and $z' \sim 1$.
Looking at expressions (\ref{rhon(k)}) and (\ref{sigmaf}) we see
that they are of the same order $\mco(\gamma)$ in any dimension in
units $\lambda=1$.

\subsection{Conservation of current and validity of gauge fixation}\label{conservation}

In the above analysis, the following gauges were fixed:
\begin{itemize}
    \item the affine parametrization of the trajectories along the
worldlines of the scattered particles:
\begin{align}\label{affine_gauge}
g_{MN} \zt^M \zt^N = g_{MN} \zp^M \!\zp^N =1\, ;
\end{align}
\item the de\,Donder gauge on the gravitational field:
\begin{align}\label{deDonder_gauge}
\partial_M \,{}^{k} \psi^{MN} =0, \qquad\qquad k=1,2,\pp\, ;
\end{align}
\item the Lorentz gauge on the vector field:
\begin{align}\label{Lorentz_gauge}
\partial_M \,{}^{k} A^{M} =0, \qquad\qquad k=1,2,\pp\, .
\end{align}
\end{itemize}
To verify self-consistency of our scheme (at least to the lowest
orders of interest), we show it explicitly.

To zeroth order, (\ref{affine_gauge}) degenerates into $u^2=1$ and
${u'}^2=1$ which is trivially satisfied.

In the first order, variation of (\ref{affine_gauge}) reads
\begin{align}\label{affine_gauge1}
\vk h'_{MN}(\nul z)\,  u^M u ^N +2 \,( \un \zt  \cdot u)=0 \, ,
\qquad\qquad \vk h_{MN}(\nul z') \, {u'}^M {u'}^N +2\, ( \un \zp
\cdot u')=0\, ,
\end{align}
respectively. From (\ref{1h'mn}) the value of $h'_{MN}(x)$ at the
location of $m-$particle $x=\nul z (\tau)$ reads
 \begin{align}\label{kkll1}
 h_{MN}(\nul z)=
 \frac{ \vk \hsp m'}{(2\pi)^{D-1} }\int \dq \frac{ \delta(qu')}{ q^2}\left( u'_{M}u'_{N} -
\frac{1}{d+2}\,\eta_{MN} \right)\, \e^{- i q z_0} .
\end{align}
Contracting it with $u^M \hsp u^N$ and substituting $\nul
z^M(\tau)= u^M \hsp \tau +b^{\smhsp M}$ one obtains
 \begin{align}\label{kkll1d}
 h_{MN}(\nul z)\,u^M \hsp u^N =
 \frac{ \vk \hsp m'}{(2\pi)^{D-1} }\int \dq \frac{ \delta(qu')}{ q^2}\left( \gamma^2 -
\frac{1}{d+2} \right)\, \e^{- i q \hsp \cdot (u\tau + b)}.
\end{align}

Differentiating (\ref{sc6}) and contracting with $u^M$ one obtains
\begin{align}\label{sc6x}
 \left(\un \zt(\tau)\cdot u\right) =-\frac{   m' \varkappa_D^2 \,\gamma_*^2}{2 (2
\pi)^{D-1} }  \int \dq\, \frac{\delta(qu')}{q^2  }\, \e^{-iqb} \,
\e^{-i\smhsp (qu)\smhsp \tau}\,.
\end{align}
Multiplying it by 2 and combining with (\ref{kkll1d}) one gets the
cancelation and thereby verifies (\ref{affine_gauge1}) to the
first order. The gauge on the trajectory of  $m'-$particle is
checked similarly.

Next,  proceeding to the de\,Donder gauge on $\un \psi_{MN}$: one
rewrites (\ref{1h'mn}):
 \begin{align}
 \un \psi^{MN}(q)=
 \frac{2\pi\, \vk \hsp m}{q^2}\, \e^{iqb}\,\delta(qu)\,  u^{M}u^{N}
\,. \nn
\end{align}
The divergence in   Fourier space reads
 \begin{align}
 q_N \un \psi^{MN}(q)=
  \frac{2\pi  \, \vk \hsp m}{q^2}\, \e^{iqb}\,(qu)\,\delta(qu) \,
u^{M}  =0 \,, \nn
\end{align}
by virtue of distributional identity $x\,\delta(x)=0$.

The divergence of $\un A^M$ (the first order of
(\ref{Lorentz_gauge})) vanishes due to the same reason:
 \begin{align}\label{1h'mn1}
  q_M \un A^M(q)=-
\frac{2\pi\,e}{q^2}\, \e^{iqb}\,\delta(qu)\, (qu)=0\,.
\end{align}
Let also verify  the gauge on $\de A^M$: in the momentum space
$$\de A^M (k) = -\frac{\de j^M(k)}{k^2}\,,$$ where $j^M(k)$ is the
full Fourier-transform taken off-shell $k^2=0$ and with no terms
neglected due to polarizations. Thus Lorentz gauge of $\de A^M$ is
equivalent to $k \cdot j=0$.

The constituents of $j^M(k)$ are given by
(\ref{rho78},\ref{rhon(k)}) and (\ref{Clinton05}). Projecting both
on $k^M$ one concludes $k\cdot \rho(k)=0$ and $k\cdot
\sigma(k)=0$. Thus both
$$\partial_M \hsp\rho^M(x)=0\, ,\qquad\qquad  \partial_M \hsp \sigma^M(x)=0$$
conserve  separately, as well as their sum.

Finally, one has to point out, that the conservation of $\de A^M$
on flat background  represents the same effect as conservation of
$J^{M}$ (\ref{emden0}) (continuity equation) in the curved
background:
\begin{align}
 \nabla_M J^M(x)=0\,.
\end{align}
Explicitly the latter reads
\begin{align}\label{scrack}
 \nabla_M {J^M} = \partial_M J^M+\Gamma^N_{N,M}\hsp  J^M \,.
\end{align}
The zeroth-order variation coincides with the conservation of
$\nul J^M =\nul \tilde{J}^M$ discussed above. The first-order
variation of (\ref{scrack}) is given by
\begin{align}\label{scrack1}
 \un\left[\nabla_M {J^M}\right] = \partial_M \un J^M+\un \hsp\Gamma^N_{N,M} \nul J^M \,.
\end{align}
These terms read
\begin{align}
& \partial_M\un {J^M}  = e \, \partial_M \int \left(\un
\dot{z}^M-u^M \un z^N
\partial_N
-\frac{\vk}{2}\, h \, u^M \right)  \delta^D\!\left(x-\nhsp \nhsp
\nul z\right)\, d\tau\, \nn \\
 &\un\hsp \Gamma^N_{N,M} \nul
J^M=\frac{e\hsp\vk }{2}\, h_{,M} \int u^M \, \delta^D\!
\left(x-\nhsp \nhsp \nul z\right)\,  d\tau\, ,
\end{align}
thus their sum equals
\begin{align}
  e \, \int \left(\un
\dot{z}^M-u^M \un z^N
\partial_N \right)\partial_M \,  \delta^D\!\left(x-\nhsp \nhsp \nul
z\right)\, d\tau= e \int d \left(\vp \un z^M \, \partial_M\,
 \delta^D\! \left(x-\nhsp \nhsp \nul
z\right)\right)=0
\end{align}
as a total derivative. The latter represents the proof in
coordinate-space of the property $\partial_M \hsp \un
\tilde{J}^M(x)=
\partial_M \hsp\rho^M(x)=0 $,
discussed above.

Thus the iteration scheme we use is compatible with the gauge we
fix, and gives the apparent way to compute radiation amplitude
and, eventually, the flux of emitted momentum.

\section{The emitted energy}\label{emitted_energy}
In order to compute the emitted energy, we take the zeroth
component of the emitted momentum (\ref{4momentum}):
\begin{equation}\label{emenrg}
E= \frac{1}{2\left(2\pi\right)^{d+3}}\sum_i \int\limits^\infty_0
\omega^{d+2} d \omega \int d \Omega \; \vert j_{\hsp i} (k) \vert
^2,
\end{equation}

First we summarize the radiation amplitudes derived in the
previous Section and overview the corresponding contributions to
the total flux. In Table I we present the energy emitted in the
several relevant regimes of frequency and angle, where the
estimates of contribution to the total emitted energy are deduced
from (\ref{emenrg}) with the estimate of $j_{\hsp i} (k)$ and the
 characteristic value of $\vartheta$ and $ \omega$ following immediately from the corresponding
Table's entry.
\begin{center}
\begin{table}[h]
\label{energytab} {
\begin{tabular}{|>{\tc}c|c|c|c|c|}
  \hline \backslashbox{$\,\,\vartheta$}{$\,\od$} & $\od\lesssim 1/\smhsp b$  & $ \od \sim \gamma/\smhsp b$ & $\od \sim \gamma^2\!/\smhsp b $& $\od \gg \gamma^2\!/\smhsp b$
  \\\hline
  $ \gamma^{-1} $ & $\begin{array}{c}  \rule{0cm}{1.3em} j \sim \mathrm{Im}\rho\\ E_{\rad}\sim \gamma^2 \\
\text{\footnotesize  subleading by}  \\\text{\footnotesize  (phase
space)}
\end{array}$ &
  $\begin{array}{l} \rule{0cm}{1.3em}  j \sim \mco(\gamma),  \text{\footnotesize{} from } \rho \text{\footnotesize{} and }
 \sigma\\  x\in[0,1],\;\text{\footnotesize{} medium regime}  \\
\text{ \footnotesize no destructive interference
}\\
 E_{\rad} \sim \gamma^3
  \\ \end{array}$  &
   $\begin{array}{l}   \rule{0cm}{1.3em}  j \sim \mco(\gamma^{-1}),  \text{\footnotesize{}  from }  \rho+\sigma(z) \\
 x\in[0,\mco(\gamma^{-2})],\; z-\text{\footnotesize{}regime}
\\ \text{\footnotesize destructive interference}
 \\
E_{\rad} \sim \gamma^{d+2}
   \end{array}$
   & $\begin{array}{c}  \text{\footnotesize  negligible radiation}\\ \text{\footnotesize exponential fall-off}\\ \end{array}$
  \\[18pt]   \hline
  1 & $\begin{array}{c}  j \sim \rho \\E_{\rad} \sim 1 \\ \text{\footnotesize (phase space)} \end{array} $ &
   $\begin{array}{l}   \rule{0cm}{1.3em}  j \sim \mco(\gamma^{-1}),   \text{\footnotesize{} from }
\sigma(z')\\
 x\in[1-\mco(\gamma^{-2}),1] \\ z'-\text{\footnotesize
regime}, \,
 E_{\rad} \sim \gamma^{d+1}\end{array} $&
   $\begin{array}{c}  \text{\footnotesize negligible radiation}\\ \text{\footnotesize exponential fall-off} \\ \end{array}$&
    $\begin{array}{c}  \text{\footnotesize  negligible radiation}\\ \text{\footnotesize exponential fall-off} \\  \end{array} $\\[ 13pt] \hline
    \end{tabular}
} \caption{The behavior of radiation amplitudes and  contribution
to the emitted energy of each of the several characteristic
regions of angle and frequency. The values are normalized as
$\lambda=b=1$.}
\end{table}
\end{center}

Now we illustrate qualitatively the effects described above and
based on the derivation in previous Section.

\begin{figure}
\begin{center}
\includegraphics[angle=0,width=12cm]{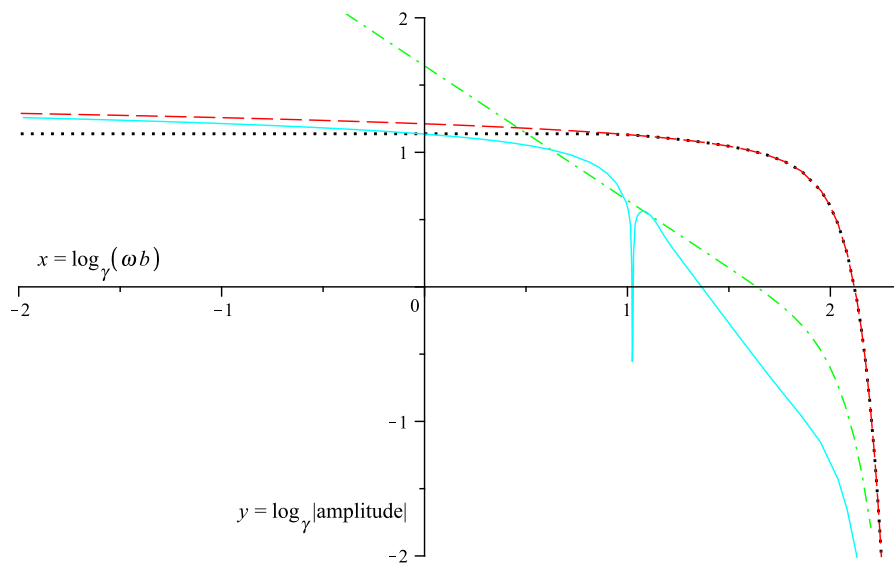}
\caption{Radiation amplitudes of first polarization for $d=0$ and
$\gamma=10^3$
 in logarithmic scales $x=\log_{\hsp\gamma} \omega b$
and $y=\log_{\hsp\gamma} |{\rm amplitude}|$, evaluated at
$\vartheta =1/\gamma,\, \phi=\pi/4$. The plots are given for
$-{\rm Re}\, \rho(k)$ (red, dashed), ${\rm Im} \,\rho(k)$ (green,
dot-dashed), ${\rm Re}\, \sigma$ (black, dotted) and ${\rm Re}\,
j$ (cyan, solid). At $x \approx 1 $ the curve $\log_{\gamma}|{\rm
Re}\, j|$ in logarithmic scale has discontinuity $y=-\infty$
related with the fact that corresponding original amplitude ${\rm
Re}\, j$ changes its sign.}
 \label{ampl_plot}
\end{center}
\end{figure}

 On Fig.\ref{ampl_plot} we plot a characteristic
picture of the behavior of local and non-local amplitudes and
their sum (the radiation amplitude) for the case $d=0$ at
characteristic value of $\vartheta$ and some common value of
$\phi$.

The qualitative features deserving attention  are the following:

\begin{itemize}
    \item At $\omega\to 0$ ${\rm Im} \,\rho(k)$ goes like
$1/\omega$ and dominates in total $j$, in Fig.\,\ref{ampl_plot} it
corresponds to the asymptote with tangent $-1$ on green
(dot-dashed) curve. This property is valid for all $d\geqslant 0$
and will be of usage further, when the zero-frequency limit is to
be computed;

\item At  $x\to -\infty$ ($\omega \to 0$) $|{\rm Re}\,\sigma| \ll |{\rm
Re}\,\rho|$ hence ${\rm Re}\,j \approx {\rm Re}\, \rho$. At this
limit $\omega\to 0$  $|{\rm Re}\,\sigma|$  goes like $\omega^0$
(black, dotted line in Fig.\,\ref{ampl_plot}) for $d=0$, like
$\omega^1$ for $d=1$ and like $\omega^2$ for $d\geq 2$, as it
follows directly from (\ref{sigmaf}) and behavior of hatted
Macdonald functions.

\item At $x>2$ each curve has rapid fall-off at $y=-\infty$, corresponding to the strong  exponential
decay of an amplitude at $\omega \gtrsim \gamma^2\!/\smhsp b$;

\item At $x>1$ ${\rm Re}\,\sigma \approx -{\rm
Re}\,\rho$, so their sum (difference of absolute values in the
plot) ${\rm Re}\,j$ (cyan, solid) is much smaller. At $x\approx 2$
the difference of ${\rm Re}\,j$ and ${\rm Re}\, \rho$ is $\Delta y
\approx 2$, so $j$ is damped by $\gamma^2$ with respect to ${\rm
Re}\, \rho$. This illustrates the destructive interference at
$\gamma^2\!/\smhsp b >\omega \gg \gamma/\smhsp b$, that can be
rewritten as
$$j(\omega) \sim j(\omega_0)\,\frac{ \omega_0^2 }{\omega^2}\, , \qquad\qquad \omega_0 \sim \frac{\gamma}{b}\,;$$

\item In the region $x=(1,2)$  $\log_{\gamma}|{\rm
Re}\, j|$ represents straight-line piece with tangent $-2$, what
corresponds to the destructive interference region
$\omega=(\gamma/\smhsp b, \gamma^2\!/\smhsp b)$. Thus the
radiation amplitude itself goes like $\omega^{-2}$ at this region.
Being averaged over angles (with some average angle
$\bar{\vartheta} =\mco(\gamma^{-1})$), the same is valid for the
frequency distribution. For higher dimensions the corresponding
behavior of the latter $ {dE}/{d\omega}\equiv F$ is
\begin{equation}\label{rrryyy}
  F(\omega) \sim (\omega\bar{\vartheta})^{d+2} j^2(\omega)
  \sim   \frac{ j^2(\omega_0)\, \omega_0^4 \,}{\gamma^{d+2}}\, \omega^{d-2}\sim
\gamma^{4-d}\,\omega^{d-2}, \qquad\qquad \gamma/\smhsp b < \omega
< \gamma^2\!/\smhsp b
\end{equation}
in this region;

\item $|{\rm Im}\, \rho|$ is subleading with respect to  $|{\rm
Re}\,\sigma|$ but larger than $|{\rm Re}\,j|$ (at $x>1$) on this
plot. It is damped by $|{\rm Im}\,\sigma|$ not presented here, so
their sum $|{\rm Im}\,j|$ becomes much smaller than $|{\rm
Re}\,j|$.
\end{itemize}

Thus in fact, we have two radiation amplitudes instead of a single
one in \cite{GKST3}, with obvious identification $f \to e, f' \to
e'$. In other words our primary problem now is to derive the final
overall coefficient.

\subsection{Total radiated energy}\label{totalraden}

As can be seen from Table I, the dominant radiation comes from
different regimes depending on the number of extra dimensions,
$d$. Indeed, as it follows from (\ref{rrryyy}),
\begin{equation}\label{yyyrrr1}
E\sim \int\limits_{\sim \hsp\gamma/\smhsp b}^{\sim
\hsp\gamma^2\!/\smhsp b} \frac{dE}{d\omega}\,
 d\omega\sim  \frac{1}{\gamma^{d-4}}  \int\limits_{\omega_0}^{\gamma \omega_0} \omega^{d-2} \,d
\omega\, ,
\end{equation}
so the dominant contribution comes from the upper limit  $\omega
\sim \gamma^2\!/\smhsp b\hsp$ for $d \geq 2$, from the lower limit
$\omega_0 \sim \gamma/\smhsp b$ for $d=0$ and from the whole
domain  for $d \geq 1$, respectively.

According to this  argument, we need to consider separately the
cases where the number of extra dimensions are $d=0$, $d=1$ and
$d\geq 2$. We start with studying the $d \geq 2$ case.

\vspace{0.5em}

$\boldsymbol{d} \mathbf{ \geq 2.}$ In this case, as can be seen in
the table, the radiation with frequency in the area of $\omega
\sim \gamma^2\!/\smhsp b$ dominates. In the case of interest here,
$R \gg b$, we can replace the summation by integration and use the
uncompactified formula for the emitted energy.

 The next step is to
substitute the expression we have already found for
(\ref{e_chiral}), which will give the dominant contribution in
this case. We notice that when squaring the two amplitudes we will
have products of the Macdonald functions. In order to perform the
integration over $\omega$, we will change variable to $z$ and the
radiated energy will take the following form:
\begin{equation}\label{split_fr_anf}
\frac{dE}{d\Omega}=\frac{\lambda^2 \sin^{d+3} \vartheta}{8 \left(2
\pi\right)^{d+3}b^{\hsp d+3} \psi^{d+3}} \sum_{a,b=0}^2
C_{ab}^{\dm} D_{ab}^{\dm}(\vartheta, \phi)\, ,
\end{equation}
with \footnote{We omit overall pre-factors $v\simeq 1$ where it is
unambiguous.} $C_{ab}^{\dm} \equiv \int
\hat{K}_{d/2+a}(z)\hat{K}_{d/2+b}(z)\,z^{d+2\left(\delta_{0a}+\delta_{0b}-1\right)}\,dz
$. We are now left with the integration over $\omega$. The
expressions for $j_\pm(k)$ (\ref{tau1hf}) are accurate for high
frequencies, however it has been shown \cite{GKST3} that for
$d\geq 2$ it is possible to expand the integration domain $z
=(\sim 1/\gamma, \infty)$ up to $z =(0, \infty)$, with the
relative error $\mco(\gamma^{-1})$. Computing $C_{ab}^{\dm}$ with
help of \cite{Proudn}
\begin{equation}\label{freq_ints}
\int\limits_0^\infty K_\mu (z )K_\nu (z )\,z^{\alpha-1}\hsp
dz=\frac{2^{\alpha-3}
\Gamma\left(\frac{\alpha+\mu+\nu}{2}\right)\Gamma\left(\frac{\alpha+\mu-\nu}{2}\right)
\Gamma\left(\frac{\alpha-\mu+\nu}{2}\right)\Gamma\left(\frac{\alpha-\mu-\nu}{2}\right)}{\Gamma\left(\alpha\right)}\,,
\qquad\qquad \alpha> \mu+\nu
\end{equation}
 and summing up the contributions of two chiral
polarizations, the angular part reads
\begin{align}\label{ang_ints}
\nonumber D_{00}^{\dm}&=\left(\frac{d+1}{d+2}\right)^{\!\nhsp 2}
\left(\frac{1}{\gamma^4 \psi^2}-\frac{2}{\gamma^2
\psi}+1\right)\,, \quad
\quad &D&_{11}^{\dm}=\left(d+1\right)^2 \left(\frac{\sin^2\! \vartheta}{\psi}-1\right)^{\nhsp \!2}\, , \\
\nonumber D_{22}^{\dm}&=\sin^4\!\nhsp \phi \left(\frac{\sin^2\!
\vartheta}{\psi}-1\right)^{\!2}+\frac{\sin^2 \!2\phi}{4}\; , \quad
\quad
&D&_{01}^{\dm}=- \frac{\left(d+1\right)^2}{d+2} \left(\frac{\sin^2\!  \vartheta}{\psi}-1\right) \frac{1 - \gamma^2 \psi}{\gamma^2 \psi}\, ,\\
  D_{02}^{\dm}&= \frac{d+1}{d+2} \left(\frac{\sin^2 \!
\vartheta}{\psi}-1\right) \frac{1 - \gamma^2 \psi}{\gamma^2
\psi}\, \sin^2 \!\nhsp \phi \; , \quad \quad &D&_{12}^{\dm}=-
\left(d+1\right)\left(\frac{\sin^2\!
\vartheta}{\psi}-1\right)^{\nhsp\!2}\sin^2\!\nhsp\phi\; .
\end{align}
By virtue of summation, we can integrate each
$D_{ab}^{\dm}(\vartheta,\phi)$ separately. The integration over
the $\phi$ is trivial using the following relations
\begin{equation}\label{phi_ints}
\int\limits_{S^{d+1}}\! d \smhsp\Omega_{d+1}=\Omega_{d+1}\, ,
\qquad\qquad \int\limits_{S^{d+1}}\!\sin^2\! \nhsp \phi \; d\smhsp
\Omega_{d+1}=\frac{1}{2}\, \Omega_{d+1}\, ,  \qquad\qquad
\int\limits_{S^{d+1}} \!\sin^4\! \nhsp \phi \; d\smhsp
\Omega_{d+1}=\frac{3}{8}\; \Omega_{d+1}\,.
\end{equation}
with the volume of unit sphere of dimensionality $n-1$ (in
Euclidean $\mathbb{R}^{n}$) given by
\begin{equation}
\Omega_{n-1}=\frac{2\, \pi^{n/2}}{\Gamma(n/2)}\,.
\end{equation}
 Making use of
\begin{equation}
\int\limits_0^\pi \frac{\sin^n \vartheta}{\psi^m}\,
d\vartheta=\frac{2^{m-1}\Gamma\left(\frac{n+1}{2}\right)
\Gamma\left(m-\frac{n+1}{2}\right)}{\Gamma\left(m\right)}\;
\gamma^{2m-n-1}\, ,
\end{equation}
(valid for $2m >n+1$, for derivation see Appendix \ref{beam_ang}),
we integrate  over $\vartheta$ to end up with the expression
\begin{equation}\label{sum2}
E=\frac{e^2 {m'}^2 \varkappa_D^4 \gamma^{d+2}}{2^{\hsp 2d+8}
\pi^{\smhsp 3d/2+7/2}\Gamma\left(\frac{d+5}{2}\right)}
\sum_{a,b=0}^2 C_{ab}^{\dm} D_{ab}^{\dm}\, ,
\end{equation}
where now
\begin{align}
\nonumber &D_{00}^{\dm}=\left(\frac{d+1}{d+2}\right)^2 , & \qquad
\qquad &D_{11}^{\dm}=\left(d+1\right)^2 , & \qquad \qquad
&D_{22}^{\dm}=\frac{d+6}{8}\;,  \\ \nonumber &D_{01}^{\dm}=
\frac{\left(d+1\right)^2}{d+2}\;, & \qquad   \qquad
&D_{02}^{\dm}=-\frac{d+1}{d+2}\;, & \qquad
&D_{12}^{\dm}=-\frac{d+1}{2} \, .
\end{align}
 and  summing up in  (\ref{sum2}), we arrive at the
following expression:
\begin{equation}\label{E_final}
E\approx C_d \frac{\left(e m' \vk^2\right)^2}{b^{\hsp
3d+3}}\:\gamma^{d+2}\, .
\end{equation}

We give here the values of $C_d$ for several values of the number
of extra dimensions:  $C_2=4.39 \cdot10^{-6}$, $C_3=1.12 \cdot
10^{-6}$, $C_4=5.63 \cdot 10^{-7}$, $C_5=4.35 \cdot10^{-7}$ and
$C_6=4.63 \cdot 10^{-7}$, respectively.

\vspace{0.5em}

$\boldsymbol{d=1.}$ We now focus our attention to the cases
$d=0,1$. Here we also can use the high-frequency approximation as
for  $d \geq 2$, but it does not represent the main contribution
now. On the other hand, in the transition region $\omega \sim
\omega_0$ the phase of an exponential in the integrand is of order
$\mco(1)$, thereby the integrand does not strongly oscillate and
can be easily computed numerically. So we revert to numerical
methods.

The radiated energy will mostly come from the small angle regime,
i.e. $\theta \lesssim 1/ \gamma$. As mentioned, in 5D the
frequency distribution of the emitted energy
 falls as $1/\omega$ in the regime between $\mco (\gamma/\smhsp b)$
and $\mco (\gamma^2\!/\smhsp b)$. Thus the dependence on $\gamma$
following from \ref{yyyrrr1}, is determined by
$$
E\sim  \gamma^{3}  \int\limits_{\omega_0}^{\gamma \omega_0} \frac{
d \omega}{\omega} \sim  \gamma^3 \hsp \ln \gamma \,. $$  We have
computed this result numerically \footnote{Numerical computation
is performed for following values of $\gamma$: $10^3$, $5\cdot
10^3$, $10^4$, $5\cdot 10^4$, $10^5$. The relative error in
90\%-level of confidence probability  is 5\%.} to deduce:
\begin{equation}\label{E_final_1}
E = C_1 \frac{\left(e m' \varkappa_5^2\right)^2}{b^{\hsp 6}}\:
\gamma^3 \hsp \ln \gamma\, , \qquad\qquad C_1=1.34\cdot 10^{-4}\,
.
\end{equation}

\vspace{0.5em}

$\boldsymbol{d=0.}$ As can be seen from the tables, the radiation
mainly comes from the transition regime ($\theta \lesssim
1/\gamma$ and $\omega \sim \gamma/\smhsp b$). As it follows from
(\ref{rrryyy}), at higher frequencies the frequency-distribution
curve decays as $1/\omega^2$, and according to (\ref{yyyrrr1}),
the estimate of emitted energy reads:
$$
E\sim \gamma^{4} \int\limits_{\omega_0}^{\gamma \omega_0}  \frac{d
\omega}{\omega^2} \sim \frac{ \gamma^{4}}{\omega_0}\sim \gamma^{3}
\, ,
$$
in agreement with the Table I.

Hence we once more use numerical methods to compute the energy:
\begin{equation}\label{E4D}
E \approx C_0 \frac{\left(e m' \varkappa_4^2\right)^2}{b^{\hsp
3}}\: \gamma^3\,, \qquad\qquad  C_0=1.36 \cdot 10^{-4}\, .
\end{equation}
The frequency distribution in four dimensions is given in
Fig.\,\ref{ADDplot}.

\vspace{0.5em}

\begin{figure}
\begin{center}\noindent
\subfigure[]{\raisebox{1pt}{\includegraphics[width=8.45cm]{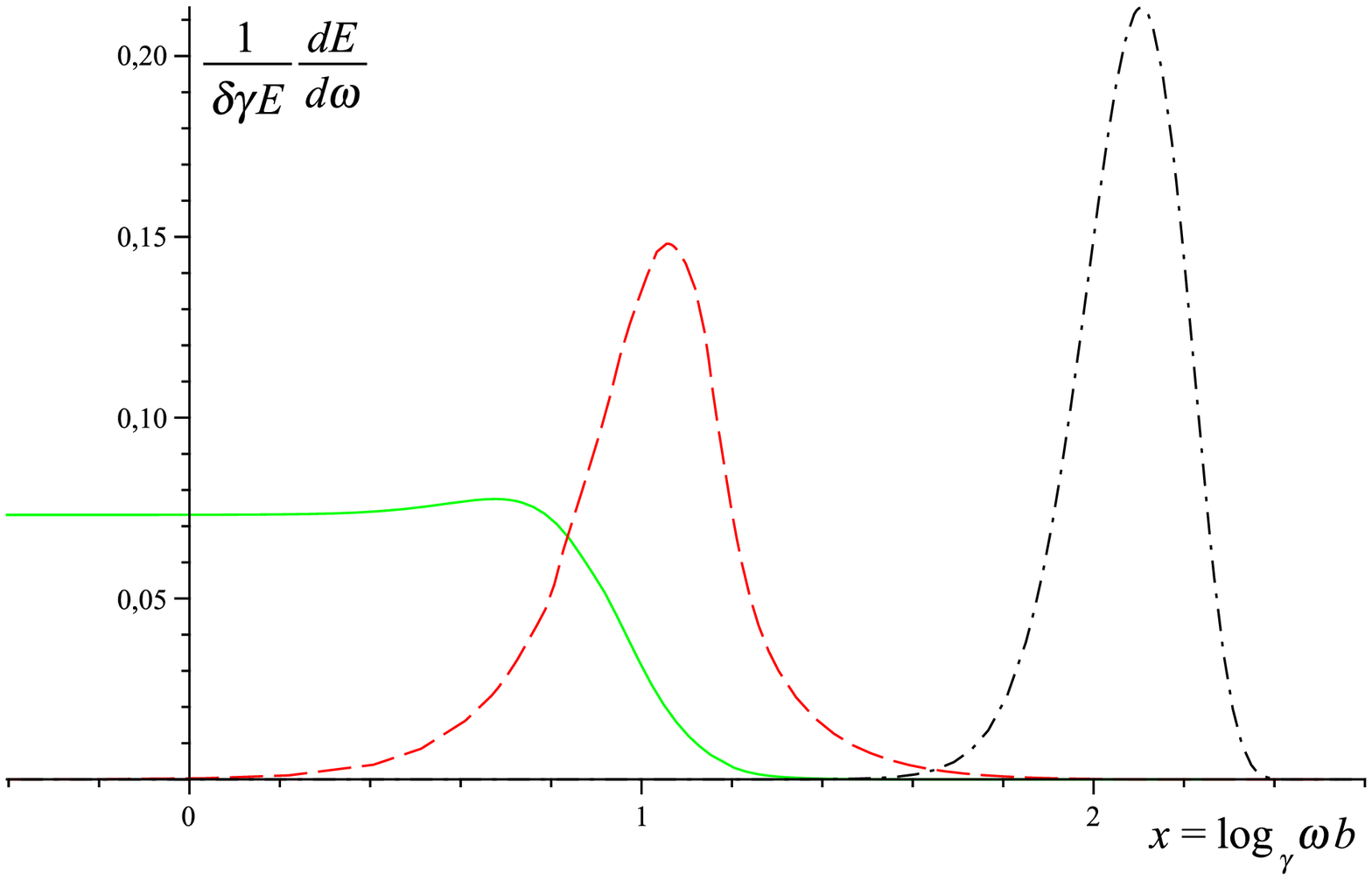}\label{4Ddist_subfig1}}}
\subfigure[]{\includegraphics[width=8.45cm]{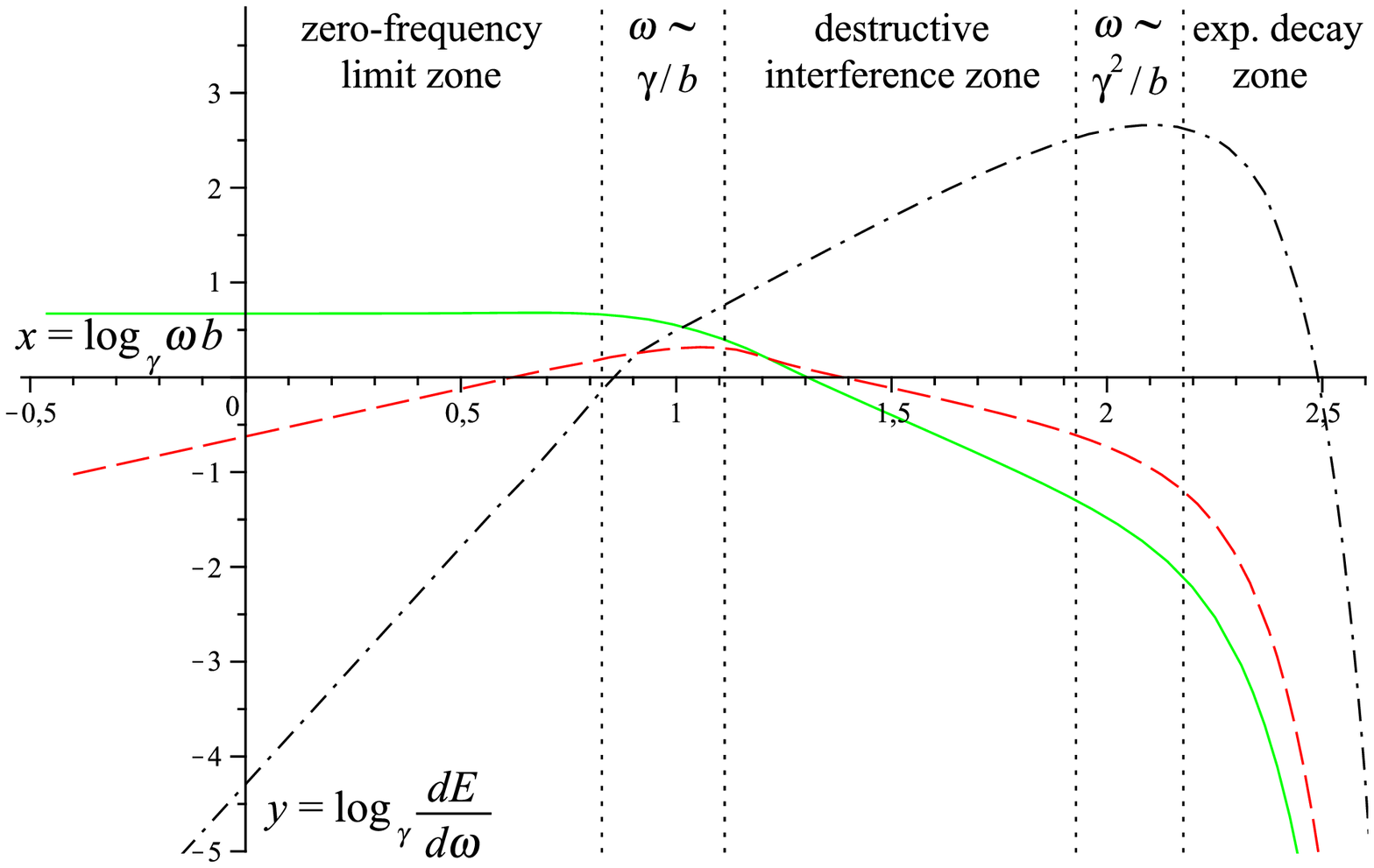}\label{4Ddist_subfig2}}
\end{center}
\caption{Frequency distribution of emitted energy in linear (a),
normalized by a factor $ \delta=
\Gamma^4\!\nhsp\left(\frac{d+1}{2}\right)$, and logarithmic (b)
$y$-scale as function of ${\rm log}_{\gamma}(\omega b)$ for
$\gamma=10^3$. The dimensions are: $d=0$ (green, solid), $d=1$
(red, dashed) and $d=5$ (black, dot-dashed).} \label{4Ddist}
\end{figure}

\textbf{Spectral-angular characteristics.} The frequency
distribution curves in logarithmic $x-$scale are presented in
Fig.\,\ref{4Ddist}: in linear scale of $dE/d\omega$ (a) and, to
illustrate the rate of growth/fall, in logarithmic $y-$scale (b).
Curves at the destructive-interference region \mbox{$x\in (1,2)$}
on the subfigure (b) represent straight lines with integer
tangents $d-2$, confirming the general idea (\ref{rrryyy}), while
at low frequencies ($x<0$) any curve has an asymptote with integer
tangent $d$, to confirm an idea (\ref{rhonexp1}).

The angular distribution $dE/d\vartheta$ curves are presented on
Fig.\,\ref{th_distr}.

\subsection{The ADD bremsstrahlung}

Among the higher-dimensional scenarios the models with direct
Kaluza-Klein modes, where the bulk represents compactification on
a torus $T^d$,  are of particular history and significance. Here
the transformation between $D-$dimensional couplings and their
four-dimensional colleagues can be established directly, via the
dimensional reduction of an action.

The $D-$dimensional propagator is split on the corresponding tower
of KK modes:
$$\frac{1}{q_M q^M} \to\frac{1}{V} \sum_{l\in \mathbb{Z}^d} \frac{1}{q_{\mu}q^{\mu}-l^2/R^2}\, \qquad\qquad \mu=0 \pp 3\,,$$
where $R$ stands for the compactification radius and $V=(2\pi
R)^d$ is a volume of extra dimensions.

Thus, concerning our computation, the momentum integrals $I$,
$I^M$, $J$ and $J^M$ introduced above, arise as a sum over
integer-valued momentum inside the argument of the Macdonald
functions. The summand represents (\ref{Iints}) with $d=0$ and the
argument of the Macdonald functions
  $z_l = \left({z^2+ l^2 b^{\hsp 2}/R^2}\right)^{1/2}$,
 both divided by a normalizing factor $V$. Thus  upon the transfer from summation to integration
according to the Euler -- Maclaurin rule
\begin{equation}\label{sum2int}
    \frac{1}{V} \sum_{l\in \mathbb{Z}^d}\hat{K}_{
\lambda}\!\left(\sqrt{z^2+l^2b^{\hsp 2}/R^2}\right) \to
\frac{1}{V} \int \hat{K}_{ \lambda}\!\left(\sqrt{z^2+l^2\hsp
b^{\hsp 2}/R^2}\right)\,d^{\hsp d}\smhsp l
 = \frac{1}{(2\pi )^{d/2} b^{\hsp d}} \,\hat{K}_{\lambda+d/2}(z)
\end{equation}
(for derivation see \cite{GKST2}) in the final result one restores
the expression (\ref{Iints}) with ''actual'' $d$.

{\noindent
\begin{figure}
\subfigure[ \;  Angular distribution $dE/d\vartheta$ of the
emitted energy ($\gamma=10^3$) for $d=0$ (green, solid), $d=1$
(red, dashed) and $d=5$ (black, dot-dashed), normalized by the
total emitted energy $E$.
 ]{\raisebox{0pt}{\includegraphics[width=8.0cm]{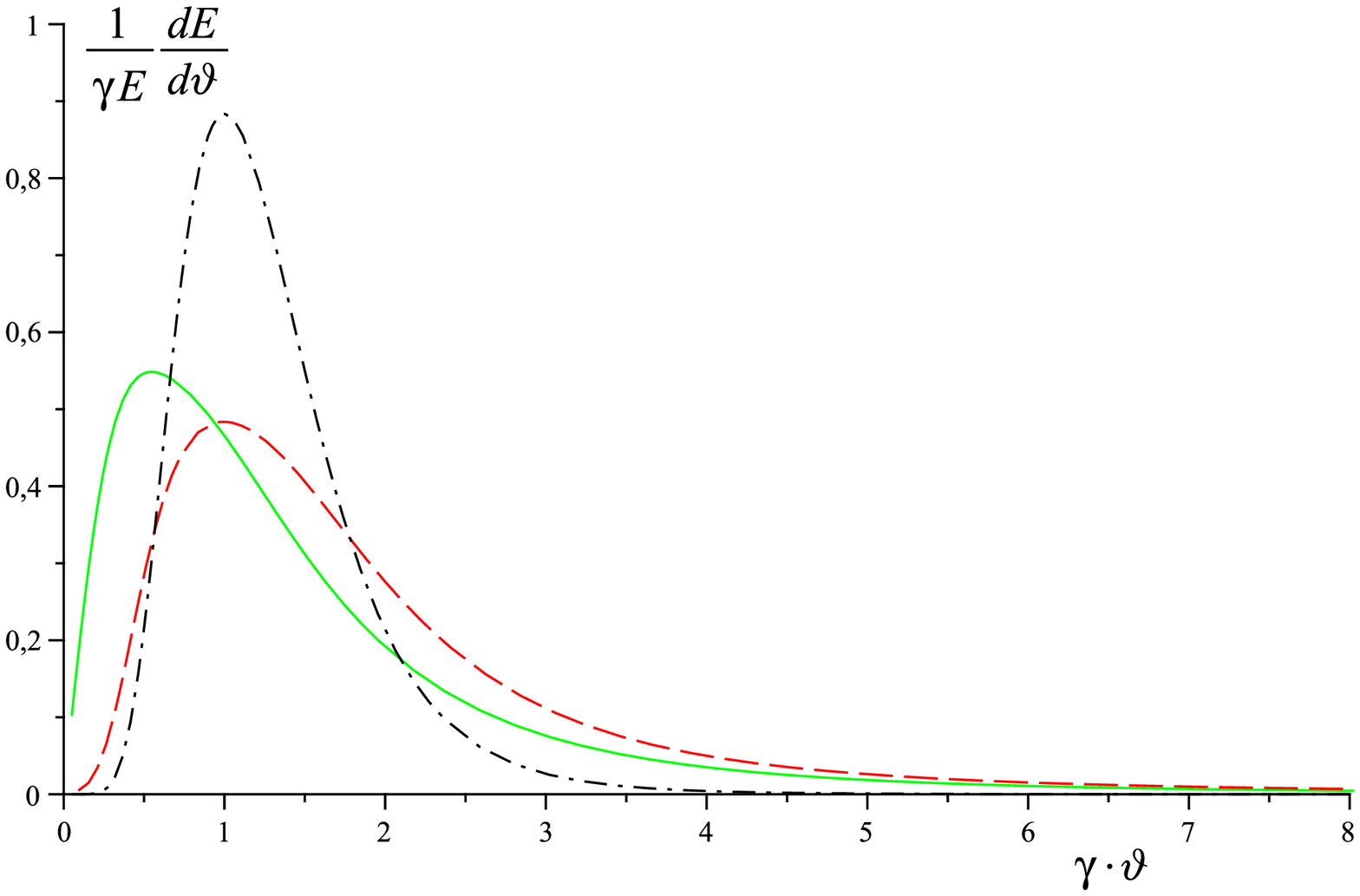}\label{th_distr}}}\hspace{0.8cm}
\subfigure[ \;Frequency distribution plots for ADD-bremsstrahlung
for $d=0$ (green, solid), $d=2$  (red, dashed) and $d=5$ (black,
dot-dashed) ($\gamma=10^3$), normalized by the ZFL factor
$\Delta=\Gamma^2(d/2+1) /(3\cdot 2^{5}
\pi^{d+4})$.]{\raisebox{0pt}{\includegraphics[width=8.0cm]{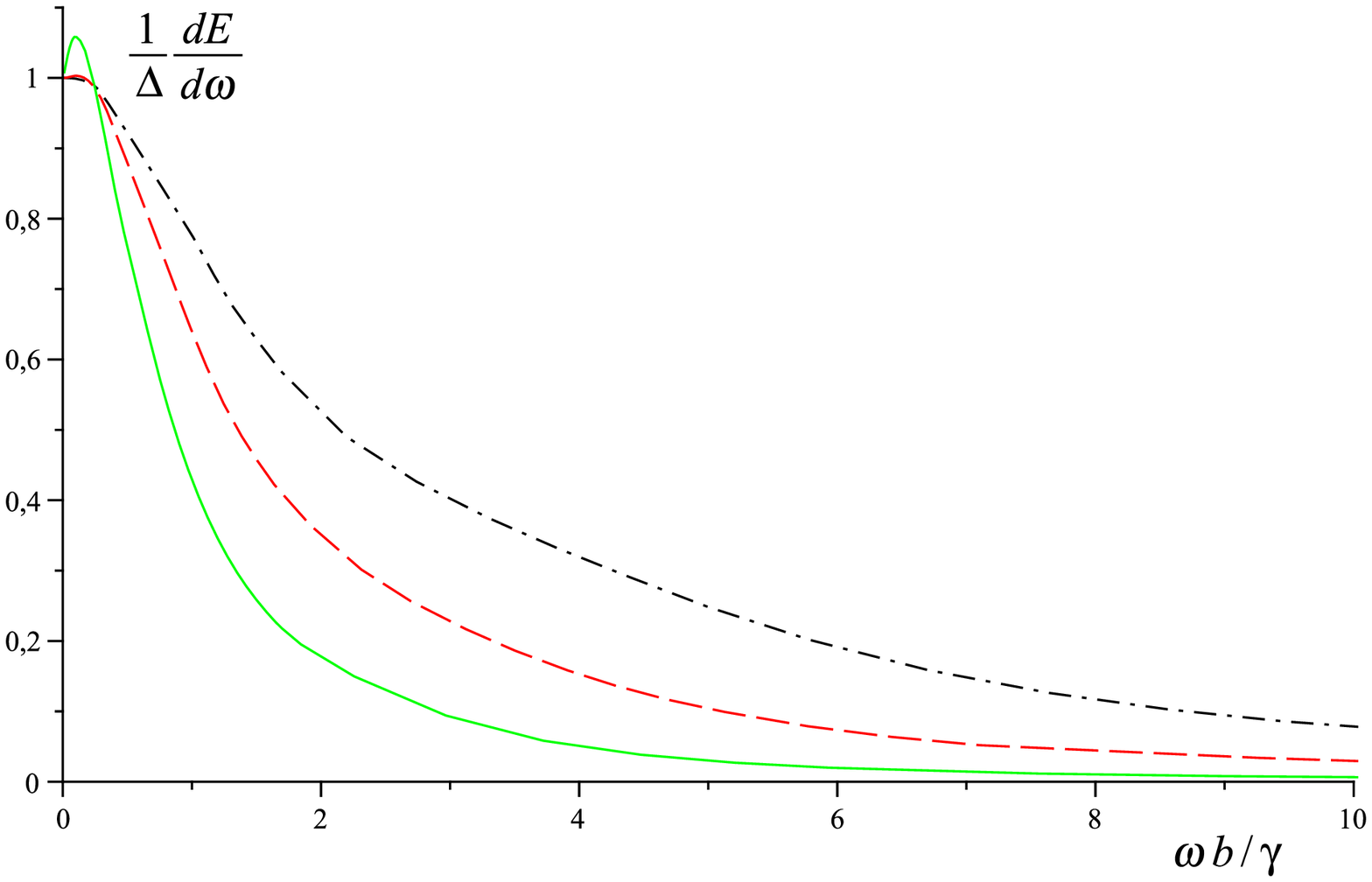}\label{ADDplot}}}
\caption{Angular and frequency distributions.} \label{gggggt}
\end{figure}
}

Apart from the features  common to higher-dimensional models, the
ADD scenario has some particular properties:
\begin{itemize}
    \item the SM fields  and
massive particles live on the 3-brane, while gravity is
essentially higher-dimensional;
\item ADD is initially proposed as linearized theory of gravity.
\end{itemize}

Thus in order to evaluate electromagnetic bremsstrahlung by
gravity-mediated collisions we can not apply some special cases
among those derived before: indeed, $D=4$ does not allow for
gravity to propagate in the bulk, while $D>4$ does allow for the
vector field to live in the bulk.

Meanwhile, the linearized action for gravitational part
 \begin{equation}\label{FP}
 S_g=\int\left[ -\frac{1}{4} h^{MN} \Box h_{MN}
+\frac{1}{4} \,h \,\Box h -\frac{1}{2} \, h^{MN}\, h_{, MN}\,
+\frac{1}{2}\, h^{MN} \, h^P_{N, MP}  \right]\dx \,,
 \end{equation}
and corresponding spin-zero (spin-one) field lead  to the
essentially same picture after KK-summation, as initially
$D$-dimensional gravity with $D-$dimensionally massless photon
(graviton), as it was shown in \cite{GKST3,GKST4}.

In what follows   we have to take a $D-$dimensional source $j^M$
and substitute it into the radiation formula (\ref{4momentum}) for
$d=0$, where we vanish the bulk components $M=4\pp D-1$. Thus the
photon wave vector is parametrized by $k^M =(k^{\hsp\mu}, 0,\pp,
0)$, with
\begin{equation}
\label{kamu} k^{\hsp\mu}=\omega\left(1,
  \sin\theta\cos\ffi,
\sin\theta\sin\ffi,\cos\theta\right)\, .
\end{equation}

Thereby, two KK propagators, corresponding to the interaction in a
source, sit inside the $D-$dimensional amplitudes  $j$ and $j^*$,
while a third propagator from the Green's function in
(\ref{flux3}) appears with normalization factor. Meanwhile, the
model allows for the emitted photon to propagate only along the
brane, that implies only zeroth emitted mode. Thus the sum
degenerates into a single term while the normalizing factor
survives. Eventually, the formula for the emitted energy via the
electromagnetic field in ADD reads
\begin{equation}
\label{DEadd}   E_{\rm ADD}=-\frac{1}{16\pi^3  V}
\int\limits_{0}^{\infty} \omega^{2}\, d \omega
\int\limits_{S^{2}}\!\! d\Omega\, j^{*}_{\mu} (k)\, j^{\mu}(k)
=\frac{1}{16\pi^{3}V}\sum_{i=1,2} \int\limits_{0}^{\infty}
\omega^{2}\, d \omega \int\limits_{S^{2}}\!\! d\Omega\, \left
|j_{\mu} (k)\,\varepsilon^{\mu}_{i}\smph \right|^2   \,.
\end{equation}
In other words, we take the four-dimensional formula for radiation
(normalized by $V$) and put a $D-$dimensional source projected on
the four-dimensional sector: $\ds \left.j^{\mu}(k^{\nu})=j^M(k)\,
\delta^{M}_{\mu}\right|_{k^i=0}$.

\vspace{0.2em}
 Thus we use the four-dimensional coordinate
system (Fig.\,\ref{branepic}, b) (with angles $\theta,\varphi$)
for parametrization of the emitted photon  and keep
$D-$dimensional angles $\vartheta,\phi$ (Fig.\,\ref{branepic}, c)
for   the parametrization of interaction graviton.

The on-shell condition now reads $k_{\mu} k^{\hsp \mu}=0$; taking
into account, that basis vectors $u^M$, ${u'}^M$ and $b^{\smhsp
M}$ do not contain bulk components, it is enough to take
higher-dimensional amplitudes $ \rho(k)$ and $\sigma(k)$ and two
polarization vectors (\ref{1st_pol}) and (\ref{2nd_pol})
\begin{align}\label{1st_pola}
 \varepsilon_1^{\hsp\mu} =\frac{1}{\gamma v z' \sin\theta}\left[z\hsp
u'^\mu-z' u^\mu+\left( \gamma-\frac{z}{z'}\right)
\frac{b\,k^\mu}{\gamma v}\right], \qquad\qquad
 \varepsilon_2^{\hsp\mu}= \frac{b}{\gamma^2 v^2 z' \sin\theta}\;
\epsilon^{\smhsp \mu\nu\lambda\rho}\,u_{
 \nu}u'_{ \lambda}k_{ \rho}\,,
\end{align}
where in addition, contractions (\ref{e_products}) hold under
appropriate substitutions $\vartheta \to \theta, \phi\to \varphi$.

To iterate, one takes $\rho(k)$ by (\ref{rhon(k)}) plus
$\sigma(k)$ in the integral representation (\ref{sigmaf}), square
and integrate with measure $\omega^2$. Thus all notes on the
destructive interference are \emph{still valid}. Eventually,
multiplying by $\omega^2$ leads to the same behavior as in four
dimensions, due to the hatted Macdonald function
$\hat{K}_{\nu}(x)$ goes like $\mco(1)$ at the range $x=0\pp
\mco(1)$ for any non-negative index $\nu$. So the four-dimensional
behavior of the frequency distribution is reproduced, with some
numerical corrections. Respectively, we repeat the strategy of
computation in 4D presented above.

Thus the characteristic frequency and angle are given by
\begin{align}\label{charADD}
\omega_{\rm ADD} \sim \omega_0 = \frac{\gamma}{b}\, , \qquad\qquad
\theta \sim \bar{\vartheta}=\frac{1}{\gamma}\, ,
\end{align}
 i.e. one has
beaming in forward direction with respect to the charged
particle's motion. The total emitted energy reads
\begin{align}\label{rhonexp5}
E_{\rm ADD}  =  \bar{C}_d\,\frac{ (e m' \vk^2)^2}{b^{\hsp 2d+3} V
} \, \gamma^3\,,
\end{align}
with coefficient $\bar{C}_d$ to be defined numerically. The
results of numerical computation (overall coefficients
$\bar{C}_d$) are listed here: $\bar{C}_1=4.90  \cdot 10^{-5}$, $
\bar{C}_2=2.54
 \cdot 10^{-5}$, $\bar{C}_3=1.77  \cdot 10^{-5}$, $\bar{C}_4=1.52   \cdot 10^{-5}$, $\bar{C}_5=1.55  \cdot 10^{-5}$, $
\bar{C}_6=1.85  \cdot 10^{-5} $, while the frequency distribution
plots are shown in  Fig.\,\ref{ADDplot}.

\vspace{0.5em}

\textbf{ZFL of the frequency distribution.} Given that the stress
part (\ref{sigmaf}) of the radiation amplitude is finite (for
$d=0$) and vanishes for $d>0$ at the limit  $\omega \to +0$, the
zero-frequency limit of $dE_{\rm ADD}/d\omega$ is determined by
the imaginary part of the local amplitude (\ref{rhonexp}): indeed
\begin{align}\label{rhonexp1}
j^{\smhsp \mu}  (k)=- i \, \frac{\lambda \, \e^{i (kb)}}{2}
 \left( \frac{z' \sin \theta \cos \varphi}{z}\, u^{\smhsp \mu}\nhsp
+\nhsp \frac{b^{\hsp \mu}}{b} \right)\nhsp
\frac{\hat{K}_{d/2+1}(z)}{z} \sim \frac{1}{\omega} ,
\end{align}
while the other terms are regular or diverge logarithmically (for
$d=0$) at $\omega \to 0$. Such a behavior in $\omega$  is
reminiscent of the infrared divergence of the corresponding
Feynman  diagrams. However, upon multiplication by $\omega^{2}$
from the measure of integration, it contributes a finite amount to
the radiation loss.

Taking the finite limit of hatted Macdonald
$\hat{K}_n(z)=2^{n-1}\Gamma(n)$ (for $ n
> 0$) and omitting the phase factors
\begin{align}\label{rhonexp2}
 j^{\smhsp \mu}  (k) \simeq \frac{\lambda \,\Gamma(d/2+1) }{2^{1-d/2}
\omega\hsp b \,\psi}
 \left( \frac{  \sin \theta\, \cos \varphi}{ \gamma \psi}\, u^{\smhsp \mu}\nhsp
+\nhsp \frac{b^{\hsp \mu}}{b} \right) ,
\end{align}
with $\psi\equiv 1-v\cos \theta$ now.

Squaring it and substituting into the first formula (\ref{DEadd}),
one has
\begin{align}\label{rhonexp3}
\left.\frac{dE_{\rm ADD}}{d\omega}\right|_{\omega=0} = \frac{ (e
m' \vk^2)^2 \,\Gamma^2(d/2+1) }{ 2^{8} \pi^{d+5} b^{\hsp 2d+2} V }
\,\int d\theta\, d\varphi\;\frac{\sin \theta}{\psi^2}
 \left(1- \frac{  \sin^2\! \theta\, \cos^2 \!\varphi}{ \gamma^2 \psi^2} \right) ,
\end{align}
Consecutively integrating over $\varphi$ with help of
(\ref{phi_ints}), and over $\theta$ via (\ref{jj2q}), the ZFL in
ADD bremsstrahlung reads
\begin{align}\label{rhonexp4}
\left.\frac{dE_{\rm ADD}}{d\omega}\right|_{\omega=0} = \frac{
 \Gamma^2(d/2+1) }{3\cdot 2^{5} \pi^{d+4} } \, \frac{ (e m'
\vk^2)^2}{b^{\hsp 2d+2} V } \, \gamma^2\,.
\end{align}
Notice, that this formula is still valid in four dimensions.

 Going back and taking into account that destructive
interference suppresses  not only the radiation amplitude at
frequencies $\omega>\mco(\gamma/\smhsp b)$ -- but also the flux,
one concludes that frequency
$$\omega_{\rm ADD} \sim \omega_0 =\frac{\gamma}{b}$$ gives the effective
cut-off for all cases of ADD, as well as to four-dimensional
bremsstrahlung. Thereby the realistic estimate is
\begin{align}\label{rhonexp6}
E_{\rm ADD} \sim \left.\frac{dE_{\rm
ADD}}{d\omega}\right|_{\omega=0}\!\!\times \omega_{\rm ADD} =
\frac{
 \Gamma^2(d/2+1) }{3\cdot 2^{\smhsp 5} \pi^{d+4} } \, \frac{ (e m'
\vk^2)^2}{b^{\hsp 2d+3} V } \, \gamma^3\,.
\end{align}
Such an approach is used by Smarr \cite{Smarr} to estimate
four-dimensional gravitational bremsstrahlung.

Therefore, the vector bremsstrahlung in ADD case repeats the
four-dimensional picture, up to numeric coefficient.

 \subsection{The UED bremsstrahlung and average number of Kaluza-Klein modes}

Through the entire text we implied that (\ref{restr_B}) is
satisfied and one has large number of KK-modes, that allows to
pass from KK-summation to the continuous integration and that
eventually leads to the enhancement of $\gamma-$factor power.

Meanwhile, for the UED models, where the vector field can
propagate through the bulk, the contemporary constraints
\cite{UED} on the size of the extra dimensions, coming from the
experimental data (including the recent ATLAS and CMS
experiments), give the following bound:
\begin{equation}\label{impact_2}
   1/R_{\rm UED} \sim 300-3000\,{\rm GeV}\,, \qquad\qquad R_{\rm UED} \sim 10^{-16}\,\mathrm{cm}\,.
\end{equation}
In this case the inequality $b<R$ (\ref{restr_B}), combined with
$b>r_{\rm cl}$, to have a charge point-like, does not hold. Does
it imply that the whole derivation presented above, fails?

Consider the situation more thoroughly: we first restore the
original KK-summations, before switching to integration. The
 analogue of (\ref{4momentum}) reads:
\begin{equation}\label{E_mom_UED}
E=\frac{1}{16 \pi^{3} V}\sum_{i} \sum_{n\in \mathbb{Z}^d}
\int\limits_{0}^{\infty}\!\! \varpi^{2}\, d \varpi
\int\limits_{S^{2}}\!\! d\Omega\, \left |j_{i}(k)\right|^2 \left.
\vphantom{\sqrt{d}} \right|_{k^0=\sqrt{ \varpi^2+n^2b^2/R^2}}\,
\qquad\qquad \varpi^2=\sum_{a=1}^{3}(k^a)^2,
\end{equation}
with $\varpi=|\mathbf{k}|$ being a continuous frequency in
four-dimensional sector.

The local current is given by  (\ref{rho78}), after the
corresponding change of the integrals $I$ and $I^{M}$ in
(\ref{Iints0}), given in \cite{GKST3}, to:
\begin{align}\label{Iints_UED}
I  =-\frac{2\pi }{\gamma v V}\,\sum_{l\in \mathbb{Z}^d}
{{K}}_{0}(z_l)\, ,\qquad \qquad  I^{M}
 =-\frac{2\pi }{\gamma v \hsp b^{\hsp 2} V}\sum_{l\in \mathbb{Z}^d}\left(b\hsp z\hsp {{K}}_{0}(z_l)\,\frac{\gamma
u'^{M}-u^{M}}{\gamma v}+ i \hat {K}_{1}(z_l)\, b^{\smhsp
M}\right),
\end{align}
respectively, with $z_l^2\equiv z^2+l^2b^{\hsp 2}/R^2$. A similar
summation arises in the stress integrals.

When $b\gg R$, one passes in (\ref{Iints_UED})  to integration
according to (\ref{sum2int}), and the expressions (\ref{Iints})
are restored. The stress amplitude is split into the KK-sum in a
similar way, for more information see \cite{GKST3}.

Such a summation appears inside the amplitude $j^M(k)$ and
corresponds to the KK-compactification of the interaction
graviton. So the effective number of KK-modes of interaction is
determined by the exponential decay of Macdonald function
($l^2b^{\hsp 2}/R^2\lesssim 1$) and reads
\begin{align}\label{Iints_UED_a}
N_{\rm int}\equiv l_{\max}=[L/\smhsp b]+1\,,
\end{align}
independent of the value $0 \leqslant z\lesssim 1$.

In the ADD-case the bound on the compactification radius is
$R_{\rm ADD} \sim 10^{-2}\,{\rm cm}$ (for $d=2$), and
(\ref{restr_B}) is well satisfied, thus  one has a large number of
the interaction KK-modes.

In the case of UED, one has $ R_{\rm UED}<l_C$ and one has to
revisit the computation. The above condition implies that  the
interaction has only \emph{zeroth} KK-mode.

Thus the sum in (\ref{Iints_UED}) degenerates into
\begin{align}\label{Iints_UED_b}
I  =-\frac{2\pi }{\gamma v V}\, K_{0}(z)\, ,\qquad \qquad I^{M}
 =-\frac{2\pi }{\gamma v \hsp b^{\hsp 2} V}\left(b\hsp z\hsp K_{0}(z)\,\frac{\gamma
u'^{M}-u^{M}}{\gamma v}+ i \hat {K}_{1}(z)\, b^{\smhsp M}\right),
\end{align}
plus exponentially-suppressed terms, and the radiation amplitude
represents the expressions derived in Section
\ref{radiation_amplitudes} for $d=0$, but normalized by the factor
$V$.

Therefore, the emission modes are determined by the exponential
decay of Macdonalds $K_0(z)$ and $K_1(z)$. In the original
KK-treatment the argument $z$ becomes dependent upon the number
$n$ of emission KK-quantum as
\begin{equation}\label{z_KK}
    z\equiv \frac{(ku)b}{\gamma v}\simeq \sqrt{\varpi^2 b^2 + n^2
b^2 /R^2}- \varpi b \hsp v \cos \theta
\end{equation}
In the total absence of emission KK-modes, the characteristic
frequency is given by its $d=0-$value $ \varpi \sim \omega_0$
(\ref{charADD}), thus the typical value of $\varpi b $ is at least
$\varpi b \gtrsim \gamma$. Assume that
\begin{equation}\label{bound2}
   b<R \gamma\, ,
\end{equation}
that is reasonable for $R$  given by (\ref{impact_2}) and
$\gamma\sim 10^{14}$. Then the first massive KK-mode is available,
and some number $n<N_{\rm emit}$ of first KK-modes satisfy $n b/R
<\gamma$. In this case one expands the radical in (\ref{z_KK}) to
obtain
\begin{equation}\label{z_KK_2}
  z \approx  \varpi b+ \frac{n^2 b}{2 \hsp \varpi R^2}  - \varpi b \hsp v \cos
\theta=\varpi b  \psi+ \frac{n^2 b}{2 \hsp \varpi R^2}
\end{equation}
Thus the effective number of emission KK-modes
\begin{align}\label{Nemit_UED}
N_{\rm emit}\equiv n_{\max}(\varpi)= \sqrt{\frac{2\varpi
R^2}{b}}\,,
\end{align}
becomes dependent on the frequency. In the most favorable case the
maximal frequency is determined  from the first term of the RHS of
(\ref{z_KK_2}), which should be less than unity independently
\cite{GKST2}: $\varpi \sim \psi^{-1}\!/\smhsp b \sim
\gamma^2\!/\smhsp b$. Thus
\begin{align}\label{Nemit_UED2}
N_{\rm emit}\sim   \frac{ \gamma R}{b} >1 \,,
\end{align}
according to the necessary condition (\ref{bound2}).

In addition, now assume the stronger condition \footnote{We will
return to the validity of this condition in the Subsection
\ref{restrictio}.}:
\begin{align}\label{Nemit_UED3}
 \frac{ \gamma R}{b} \gg 1 \,,
\end{align}
Then $N_{\rm emit}\gg 1$, so the modes are quasi-continuous, and
we combine quasi-continuous momenta with continuous $\varpi$ into
single $\omega$, shift angles $(\theta, \varphi)\to (\vartheta,
\phi)$ and we return to the case (\ref{emenrg}), where we
integrate the square of radiation amplitude with volume measure
\begin{equation}\label{mensura}
     \mathcal{V}_d
=\frac{1}{2(2\pi)^{d+3}}\:\omega^{d+2}\,\sin^{d+1}\!\vartheta\:
d\omega\, d\vartheta\,d\Omega_{d+1}\,.
\end{equation}

Given that the hatted Macdonald function $\hat{K}_{\nu}(z)$ alters
slowly with the change of  index $\nu\geq 0$, the integration
should be performed along the same lines as in
Subsection\,\ref{totalraden}. Namely, for $d\geq 2$ the
high-frequency regime dominates, and for the radiation amplitudes
one has instead  (\ref{tau1hf}) and (\ref{lambda}), the following
one:
\begin{align}\label{tau1hf_gg0}
j_{\pm}\nhsp(k) \approx  \frac{\lambda_0 \, \e^{i  (kb )} \sin
\vartheta}{2\sqrt{2} } & \left[\frac{d+1}{d+2}\,
  \frac{1-\gamma^2 \psi}{\gamma^2 \psi}\,{K}_{0} (z ) - \frac{1}{z^2}   \left( \frac{ \sin^2\!
\vartheta}{
\psi} -1 \right) \hat{K}_{1} (z ) +\right. \nn \\
& \;\;\left.+\frac{\sin^2 \nhsp \!\phi }{z^2}\left(
 \frac{\sin^2 \! \vartheta }{\psi} -1
\right) \hat{K}_{2} (z )
 \pm i\,\frac{  \sin 2  \phi }{ 2\, z^2}
 \, \hat{K}_{2} (z ) \right],
\end{align}
with\footnote{The numeric coefficient before $\hat{K}_1(z)$ is
related with the index of Macdonald function in the series
(\ref{refer_int}) and corresponds to the same expression as in
(\ref{tau1hf}), with $d=0$ is fixed. The numeric coefficient
before ${K}_0(z)$ is coming from the $D$-dimensional $h'_{MN}$ and
keeps $d-$dependence inside itself.} $\lambda_0 \equiv e\hsp
m'\smhsp \vk^2/2 \pi V$.

Again, we split the integrals on frequency and angular parts, as
in (\ref{split_fr_anf}):
\begin{equation}\label{cabdab_UED}
\frac{dE}{d\Omega}=\frac{\lambda_0^2 \sin^{d+3}\! \vartheta}{8
\left(2 \pi\right)^{d+3}b^{\hsp d+3} \psi^{d+3}} \sum_{a,b=0}^2
\tilde{C}_{ab}^{(d\smhsp)} \tilde{D}_{ab}^{\dm}(\vartheta, \phi)\,
,
\end{equation}
where $\tilde{C}_{ab}^{\dm} \equiv \int
\hat{K}_{a}(z)\hat{K}_{b}(z)\,z^{d+2\left(\delta_{0a}+\delta_{0b}-1\right)}\,dz\hsp
$. As before, these integrals are to be computed with help of
(\ref{freq_ints}).

Comparing (\ref{tau1hf_gg0}) with (\ref{tau1hf}), one concludes
that the angular coefficient functions $\tilde{D}_{ab}^{\dm}$ have
the corresponding changes with respect to those ones
$D_{ab}^{\dm}$ given in (\ref{ang_ints}):
$$\tilde{D}_{01}^{\dm}=\frac{{D}_{01}^{\dm}}{d+1}\, ,\qquad\qquad \tilde{D}_{11}^{\dm}=\frac{{D}_{11}^{\dm}}{(d+1)^2}\, , \qquad\qquad
\tilde{D}_{12}^{\dm}=\frac{{D}_{12}^{\dm}}{d+1}\, . $$

The same relations exist for the integrated over all angles
constants. Combining them all and substituting to
(\ref{cabdab_UED}), one obtains the energy loss
\begin{equation}\label{E_final_UED}
E_{\rm UED}\approx \tilde{C}_d \frac{\left(e m' \vk^2\right)^2
}{V^2b^{\hsp d+3}}\:\gamma^{d+2}\, .
\end{equation}

The values of $\tilde{C}_d$ for small values of the number of
extra dimensions are listed as:
 $\tilde{C}_2=7.8 \cdot 10^{-6}$,
$\tilde{C}_3=1.5\cdot 10^{-6}$, $\tilde{C}_4=4.5\cdot 10^{-7}$,
$\tilde{C}_5=1.7\cdot 10^{-7}$.

\vspace{0.5em}

$\boldsymbol{d=1.}$ Repeating the same arguments, we compute the
total radiation numerically:
\begin{equation}\label{E_final_1_UED}
E = \tilde{C}_1 \frac{\left(e m' \varkappa_5^2\right)^2}{V^2
b^{\hsp 4}}\: \gamma^3 \hsp \ln \gamma\, , \qquad\qquad
\tilde{C}_1=2.82 \cdot 10^{-5}\, .
\end{equation}

The spectral characteristics in UED bremsstrahlung are the same as
in higher-dimensional case (Subsection \ref{totalraden}), while
the angular characteristics are similar to all cases considered
above.

\vspace{0.5em} \textbf{A summary.} In Table II we summarize the
ultimate cases of an ultrarelativistic bremsstrahlung from the
viewpoint of average numbers of the Kaluza-Klein modes excited in
the bremsstrahlung process.

\begin{center}
\begin{table}[h]
\label{casestab}
\begin{tabular}{|c||c l|l|}
\hline  \backslashbox{$N_{\rm emit}$}{$\rule{0cm}{1.1em}N_{\rm
int}$} &    & \phantom{hhhh}
 $ N_{\rm int} \lesssim 1$
 &  \phantom{hhh} $
N_{\rm int} \gg 1$
\\  \hline  $ N_{\rm emit} \lesssim 1$  & $\begin{array}{rc} \rule{0cm}{1.3em} \text{\footnotesize space-time model}= & \\
 \text{\footnotesize characteristic frequency}=  & \\ \text{\footnotesize radiation amplitude}= \\  \text{\footnotesize phase volume}= \\
 \text{\footnotesize KK modes}= \\ \text{\footnotesize emitted energy}= \end{array} $
 &  \phantom{h} $\begin{array}{l} \rule{0cm}{1.3em} M_{1,3}
\\
\omega\sim \omega_{\smhsp 0} \\  j=j_{\smhsp 0} \\  \mathcal{V}=\mathcal{V}_0 \\
 N_{\rm int}= N_{\rm emit}=1 \\ \gamma^3 \end{array} $  & \phantom{hi}
$\begin{array}{l}  \rule{0cm}{1.3em}\text{ADD} \\
\omega\sim \omega_{\smhsp 0} \\ j=j_{\smhsp d} \\ \mathcal{V}=\mathcal{V}_0/V \\  N_{\rm emit}=1 \\ \gamma^3/V  \end{array} $ \\[24pt]
  \hline  $ N_{\rm emit} \gg 1$ & $\begin{array}{rc} \rule{0cm}{1.3em} \text{\footnotesize space-time model}= & \\
 \text{\footnotesize characteristic frequency}=  & \\ \text{\footnotesize radiation amplitude}= \\  \text{\footnotesize phase volume}= \\
 \text{\footnotesize KK modes}= \\ \text{\footnotesize emitted energy}= \end{array} $ &
$
 \phantom{h,} \begin{array}{l} \rule{0cm}{1.3em} \text{UED} \\
 \omega\sim \gamma\omega_{\smhsp 0} \\ j=j_{\smhsp 0} /V \\ \mathcal{V}=\mathcal{V}_d \\ N_{\rm int}=1 \\
\gamma^{d+2}\!/V^2\end{array}$
&  \phantom{hi} $\begin{array}{l} \rule{0cm}{1.3em} M_{1,d+3} \\
 \omega\sim \gamma\omega_{\smhsp 0} \\ j=j_{\smhsp d} \\ \mathcal{V}=\mathcal{V}_d \\ N_{\rm emit}=\gamma N_{\rm int}\\ \gamma^{d+2} \end{array} $  \\[24pt] \hline
\end{tabular}
\caption{The qualitative relation between the cases of
gravity-mediated vector bremsstrahlung from viewpoint of number of
KK-modes. The values are normalized as $\lambda=b=e=1$. $N=1$
implies that only the zeroth KK-mode is actual. The measure of the
phase volume integration is defined by (\ref{mensura}).}
\end{table}
\end{center}

\section{Discussion}
According to the computation presented above, we overview possible
effects and give the estimates on them.

\subsection{Scattering of two charges}\label{vec-vec}

When both particles are charged by the vector field $A^M$ then the
direct electromagnetic interaction is expected to be the dominant
force. Then the acceleration (and, being integrated, the
trajectory deflection) represents (to first order of PT) the sum
of two contributions of electromagnetic and gravitational nature,
respectively. In turn, these addenda to trajectory may lead to
radiation via vector and tensor fields. We do not consider
gravitational waves in this work, and thus focus here to the pure
vector bremsstrahlung.

A similar approach (i.e. bremsstrahlung without accounting for
gravity) was considered in \cite{GKST2} for the scalar
bremsstrahlung, so it is not necessary to reproduce that
computation in details. Instead of the detailed computation, we
highlight the main steps and overview the results.

Making use of  perturbation theory over $e$ and considering
(\ref{eom_part}) on the flat background with $F_{MN}$
(\ref{1h'mn}) generated by charge $e'$, the acceleration on
trajectory reads:
\begin{align}\label{acce_em}
\un\hsp \ddot{z}^{M}_{\rm em}(\tau)=i\, \frac{e'e}{(2\pi)^{d+3}m}
\int \dq \frac{ \delta(qu')}{q^2}\;e^{-iqb}e^{-i(qu)\tau}\left[
\vp \gamma \hsp q^{M}-(qu)\,u'^{M}\right].
\end{align}
The scattering angle, computed along the same lines as in
\cite{GKST1}, is given by
\begin{align}\label{angle_em}
\alpha_{\rm em} \sim \frac{e\,e'}{m\,\gamma\, b^{\hsp d+1}} \sim
\left[\frac{\sqrt{r_{\rm cl}r'_{\rm cl}}}{b} \,
\frac{\sqrt{m'}}{\sqrt{m }}\right]^{d+1}\frac{1}{\gamma}<
\frac{(m'/m)^{\frac{d+1}{2}} }{\gamma}\,.\end{align}
 Performing
the perturbation-theory scheme (with the obvious restriction
$b>r_{\rm cl}$), one obtains the following second-order source
valid in \emph{all} frequency regimes:
\begin{align}\label{j2}
j^{M}  (k)\sim  i \, \e^{i (kb)}\,\frac{e^2 \, e'}{m \gamma
b^{\hsp  d}}\,
 \left( \frac{  \sin \vartheta \cos \phi}{ \gamma \psi}\, u^{M}\nhsp
+\nhsp \frac{b^{\smhsp M}}{b} \right)\nhsp
\frac{\hat{K}_{d/2+1}(z)}{z}\,.
\end{align}
It is produced by the fast particle, while the corresponding terms
due to the target and the interference give subleading in $\gamma$
contribution.

As  was mentioned above, such an argument of the Macdonald
function leads to the dominance of  $z-$region  in the entire
spectrum. Thus in the Lab frame the characteristic
spectral-angular values are:
\begin{align}\label{em_char}
\omega_{\rm em} \sim \frac{\gamma^2}{b}\,, \qquad\qquad
\vartheta_{\rm em} \sim \frac{1}{\gamma}\, ,
\end{align}
On the other hand we see that such a behavior at low frequencies
leads to the finite ZFL of frequency distribution, which for the
case of non-compactified extra dimensions reads
\begin{align}\label{ZFL_DD}
\left(\frac{1}{\omega^d}\frac{dE}{d\omega}\right)_{\!\omega=0}
\sim \frac{(e^2 \, e')^2 }{  b^{\hsp 2d+2}}\,\gamma^{-d}\,.
\end{align}
Here no process which drastically changes the amplitude (like
destructive interference) occurs in the whole frequency domain
$\omega \in [0,\omega_{\rm em}]$, and one applies
ZFL-approximation with maximal frequency given by (\ref{em_char}):
\begin{align}\label{total_em}
E_{\rm em} \sim
\left(\frac{1}{\omega^d}\frac{dE}{d\omega}\right)_{\!\omega=0}
\!\! \times \omega_{\rm em}^{d+1}\sim \frac{ e^4\, {e'}^2}{m^2
b^{\hsp 3(d+1)}} \,
 \gamma^{d+2}\, .
\end{align}
Roughly speaking, the total emitted energy carried by the vector
field is twice that of the scalar situation due to the two
polarization states, after making the identifications $f\to e,
f^{\hsp\prime} \to e'$, respectively. Therefore most of emitted
waves are beamed into the cone with characteristic angle
$1/\gamma$.

The efficiency is given by
\begin{align} \label{sc12}
\epsilon_{\rm em} \sim  \left(\gamma\, \frac{r_{\rm cl}^3}{b^{\hsp
3}}\right)^{\!\!1+d}.
\end{align}

Taking into account that when interacting  gravitationally, the
charge emits $E_{\rad} \sim \gamma^3$ in four dimensions, while
only $E_{\rm em} \sim \gamma^2$ in Coulomb-field collision,
 it seems intriguing to derive that value  of $\gamma$,  for which the two contributions  become
comparable.

\vspace{0.5em}

\textbf{Correction to gravity-mediated vector bremsstrahlung.} The
acceleration of both particles in the first order of PT represents
the sum of gravitational and Lorentz-force parts. The
electromagnetic part causes $e^2\,e'-$contribution to the vector
current  and leads to the pure electromagnetic bremsstrahlung
reviewed above in this Subsection.

The appearance of a second charge $e'$ (with mass $m'$) adds some
terms to the radiation amplitudes: namely, local (\ref{rhonexp})
and non-local (\ref{Clinton05}) parts will acquire addenda
$\rho'(k)$ and $\sigma'(k)$, based on the integrals (\ref{Iints0})
and (\ref{Jints}) where primed and unprimed quantities are
mutually interchanged. These terms also can be derived in the same
way in the Lorentz frame associated with $e-$charge (comoving
 frame), and then Lorentz-transformed into the Lab frame. With $e$
and $e'$ to be of the same order, in the comoving frame the
emission is dominant due to these new terms, and governed by
Macdonald function $K_{\nu}(z') $. Hence in this frame the
emission is beamed inside the cone
$\vartheta^{\smhsp\prime}\lesssim 1/\gamma$ with respect to
$\mathbf{u}'$. Being transformed to the Lab frame, these terms
remain to be $K_{\nu}(z') $ since $z'$ is a Lorentz-scalar
(\ref{z1}). Thus these addenda are not important in higher
frequencies and represent subleading, by an order of $\gamma$
terms (with respect to the terms we keep) due to the Lorentz
transformation, with a corresponding interchange of primed and
unprimed couplings in (\ref{E_final}).

The conservation of these terms is easily verified using the same
strategy as for the basic terms. The self-action terms appearing
here, are discussed in Appendix\,\hsp{}\ref{self}.

\subsection{Coherence length}

In this subsection we consider qualitatively the effects arising
in the bremsstrahlung process, and  the spectrum of emitted waves,
from the viewpoint of coherence length, coming from consideration
of the particle's  equation of motion in the presence of external
field.

While accelerating, the  particle emits radiation. Its spectral
characteristics are translated from the corresponding temporal
ones, related with the duration of accelerated motion, and with
the value of acceleration and type of external force.

Apart from the formulae for the total energy loss on radiation in
the coordinate and momentum representations given in
subsection\,\ref{radiation_formula}, the intensity of
electromagnetic emission can be characterized by the square of the
incomplete Fourier-transform of $A^{M}(x)$ considered as an
integral over the particle's classical trajectory $z^{M}(\tau)$:

$$A^{M}(\omega, \mathbf{r})\sim \frac{e}{\rho} \int u^{M}(\tau)\, \e^{i(\omega t-\mathbf{k}\mathbf{z})}\,d \tau\,,
\qquad\qquad \rho \equiv |\mathbf{r}-\mathbf{z}|\,.$$

Being squared, the combination $|A^{M}(\omega, \mathbf{r})|^2$
contains a double integral over $\tau_1 \tau_2$ with $\e^{i  k
\hsp \cdot \Delta z}$ in the integrand.

 Expanding $\Delta
z^{M}=u^M+\ddot{z}^M \tau+ \dddot{z}^M \tau^2/2+\pp$, where $ \tau
\equiv \tau_2-\tau_1$, the difference in the phases of the two
waves emitted by a charge in the same direction $\mathbf{n}$ at
close moments $\tau_1$ and $\tau_2$ of proper time, is determined
by
$$\Delta \varphi = k \cdot \Delta z =\omega\left[\rule{0cm}{1em} t -
 \mathbf{n} \,\Delta \mathbf{z}(t) \right]\, , \qquad\qquad  t \equiv \tau_2-\tau_1\,.$$
In addition, in ultrarelativistic motion the transverse component
of the  force acts much more effectively than the longitudinal
one. Because of this, one can transit from $D-$dimensional
expansions to their spatial sector, and the latter equation can be
rewritten as
$$\Delta \varphi =  \omega t \left(1-\mathbf{n}\mathbf{v} -
\frac{t}{2}\,\mathbf{n}\mathbf{\dot{v}}+ \frac{t^2}{6}\, \mathbf{\dot{v}}^2+\pp\right)$$

Thus to the leading order $\Delta \varphi \approx
 \omega t  \smhsp (1-\mathbf{n}\mathbf{v}) =\omega t\smhsp ( 1-v\cos \vartheta) = \omega t\smhsp \psi\,.$
When $\Delta \varphi$ becomes of order $\mco(1)$, the waves with
antiphase are present in the spectrum, so they annihilate and
decoherence happens.

Thus the maximal duration of coherence is given by
\begin{equation}\label{coh}
 t_{\rm coh}\sim \frac{1}{\omega\smhsp \psi}\,, \qquad\qquad
   \tau_{\rm coh}\sim \frac{t_{\rm coh}}{ \gamma} \sim \frac{1}{\omega\smhsp \gamma\smhsp \psi} \,.
\end{equation}

 Let us consider
the wave formed  within the coherence length (during the coherence
time) and emitted in the angle $\vartheta$ with respect to
$\mathbf{u}$. The characteristic duration of this signal in the
Lab frame is determined by the difference of distances covered by
two waves, emitted at the start and finish of the coherence
interval and received far from the particle's location. Computing
it, one obtains $t_{\smhsp \rm Lab}=(1-v\cos \vartheta)\, t_{\rm
coh}=\psi \hsp t_{\rm coh}$. Going back to all cases of
bremsstrahlung, most of the emitted radiation is beamed inside the
cone $\vartheta \lesssim \bar{\vartheta}=1/\gamma$,
 that is confirmed by the curves in Fig.\,\ref{th_distr}.

Given that at coherence interval the deflection angle should be
$\alpha <\gamma^{-1}$, the Lab-frame duration is estimated as
\begin{align}\label{chart}
t_{\rm Lab} \simeq\frac{ \vartheta^2+\gamma^{-2}}{2}\: t_{\rm coh}
\,.
\end{align}

Finally, using (\ref{coh}) one has:
\begin{align}\label{charw}
\omega_{\rm com} \sim  \frac{1}{t_{\rm coh}}\, , \qquad\qquad
\omega  \sim \frac{1}{t_{\rm coh} \psi }\sim \frac{1}{t_{\rm Lab}}
\sim \gamma^2 \omega_{\rm com}\, .
\end{align}
The frequency in the Lab frame is, thereby, $\gamma^2$ larger than
the frequency in the comoving frame, according to the Doppler
effect.

Therefore we analyze the average time of accelerated motion.

\vspace{0.5em}
 \textbf{Classical electrodynamics.} Expanding (\ref{acce_em}) near $\tau=0$ one deduces
that the  acceleration is determined by the transverse component
$\un\hsp \ddot{z}^{\hsp x}_{\rm em}$ with characteristic value
\begin{align}\label{accex_em}
\un\hsp \ddot{z}^{\hsp x}_{\rm em}(0) \sim \frac{ee'}{m}
\,\frac{\gamma}{b^{\hsp d+2}}\,.
\end{align}
 The duration of the
accelerated motion is characterized by that interval, for which
the trajectory is deflected on an angle, comparable to the total
deflection angle $\alpha_{\rm em}$ given by (\ref{angle_em}):
\begin{align} \label{duration_em}
 \tau_{\rm
em} \sim \frac{b}{\gamma}\, , \qquad\qquad t_{\rm em} \sim b\,.
\end{align}
For details, see \cite{Spir}. Next, consider the radiative part of
the Lorentz-Dirac force in higher dimensions: it is determined by
averaging over angles of the corresponding part of energy-momentum
tensor, the latter reads $T_{\rm em}^{\rm emit} \sim e^2/r^{d+2}$,
where $r$ stands for the retarded Lorentz-invariant distance
parameter  (for construction see \cite{Spir-2009}).

For instance, in four dimensions it represents well-known
(relativistic) Larmor formula for the emission intensity (in the
units we use)
$$ \frac{d {E}_{\rm rad}}{dt} =-\frac{1}{6\pi} \,e^2 \ztt^{\hsp 2}_{\rm em}\,, \qquad\qquad
  \dot{E}_{\rm rad} =-\frac{1}{6\pi} \,e^2 \ztt^{\hsp 2}_{\rm em} \zt^0\,.$$
In even higher dimensions the analogue of the Larmor formula  can
be computed in the a closed form and reads schematically (in the
gauge $\zt^2=1$)
\begin{align} \label{qqq}
\dot{E}_{\rm rad} \sim  e^2  \gamma \left[ B_{\scriptstyle
\underbrace{\scriptstyle (2,2;\smhsp 2 \pp 2)}_{d+2}}\; (\ztt_{\rm
em}^{\hsp 2})^{\smhsp d/2+1}+\pp+    B_{(D/2,D/2)}\left( z_{\rm
em}^{(D/2)} \cdot z_{\rm em}^{(D/2)} \right) \right]\,.
\end{align}
 with some positively
defined form in the parenthesis. Here $B_{(\alpha_{k\pp})}$ is a
constant with list of orders of derivatives, constituting the
corresponding scalar products, while dots represent all
intermediate scalar terms with the same dimensionality of mass
($[m]={\rm cm}^{-1}$).

Taking into account that for higher derivatives
$$ \frac{d^{D/2} }{d\tau^{D/2} }\,\un\hsp z^M_{\rm em} \sim \un{z^x_{\rm em}}^{(D/2)} \sim
 \un \hsp {\ztt^x_{\rm em}} \,\frac{\gamma^{d/2}}{b^{\hsp d/2}}\, ,$$
that follows from (\ref{acce_em}), and substituting
(\ref{accex_em}), one obtains the estimate
\begin{align}\label{accex_n_em}
\un\hsp  {z^x_{\rm em}}^{(D/2)}  \sim \frac{e\smhsp e'}{m}
\,\frac{\gamma^{d/2+1}}{b^{\hsp 3d/2+2}}\,.
\end{align}
Given that all terms in the parenthesis have the same total
dimensionality ${\rm cm}^{-(d+2)}$, and that each derivative adds
$\gamma/\smhsp b$, one concludes that all terms have the same
order of $\gamma-$factor. In what follows, the leading term is
determined by the perturbation theory, and given by the term with
minimal number of scalar products, namely, the last term in
(\ref{qqq})\footnote{According to the affine parametrization, (i)
one can exclude velocity from such scalar products and (ii) terms
with scalar products of the form, for instance $\left(z^{(D/2+1)}
,z^{(D/2-1)} \right)$, are equivalent to the retained
$\left(z^{(D/2)} ,z^{(D/2)} \right)$ by virtue of relation
$$\left(z^{(D/2+1)} ,z^{(D/2-1)} \right)=\frac{d}{d\tau}\left(z^{(D/2)}
,z^{(D/2-1)} \right)-\left(z^{(D/2)} ,z^{(D/2)} \right),$$ where
the full derivative does not contribute to the radiation and can
be dropped. The same concerns the other scalar products
$\left(z^{(D/2+k)} ,z^{(D/2-k)} \right)$.}. From the dimensional
analysis it is easy to see that all other terms contain more than
two first-order kinematical quantities.

Thus the total emitted energy during the whole bremsstrahlung
process is given by
\begin{align}\label{totalo_em}
E_{\rm em} \sim e^2 \left[\un\hsp {z^x_{\rm em}}^{(D/2)}\right]^2
t_{\rm em} \sim  \frac{e^4 {e'}^2}{m^2}
\,\frac{\gamma^{d+2}}{b^{\hsp 3d+3}}\, ,
\end{align}
in agreement with (\ref{total_em}). Thus the estimate of vector
bremsstrahlung as induced emission of a charge in the external
Coulomb field is valid within the same perturbation theory.

Finally,  (\ref{duration_em}) represents the coherence length of
emitted waves in the comoving Lorentz frame -- the characteristic
length of the trajectory, where the signal is formed. Applying the
transformation (\ref{charw})) to (\ref{duration_em}, one obtains
\begin{align}\label{charw_em}
\omega_{\rm em}  \sim \frac{1}{t_{\rm em} \psi } \sim
\frac{\gamma^2}{b}
\end{align}
in the Lab frame, in agreement with (\ref{em_char}).

\vspace{0.5em}

 \textbf{Classical electrodynamics in external curved background.} The deflection angle in a static
gravitational potential in $D$ dimensions is given by \cite{GKST1}
\begin{align} \label{grav_angle}
 \alpha_{\rm gr} \sim \frac{ {r'_{\nhsp g}}^{d+1}}{
 b^{\hsp d+1}}    \ll 1\,,
\end{align}
and, according to the Equivalence principle, does not depend upon
the energy of the scattered particle.

Double-differentiating (\ref{sc6}), one obtains the estimate of
the transverse component of an acceleration caused by the
gravitational force:
\begin{align}\label{accex_gr}
\un\hsp \ddot{z}_{\rm gr}^{x}(0) \sim
\frac{{r'_g}^{d+1}\gamma^2}{b^{\hsp d+2}}\,,
\end{align}
while the characteristic time of acceleration is governed,
essentially, by the same factors as before and reads
\begin{align}
\label{duration_gr}
 \tau_{\rm
gr} \sim \frac{b}{\gamma}\, , \qquad\qquad t_{\rm gr} \sim b\,.
\end{align}
Nevertheless, the dominant contribution into $\ztt^{\hsp 2}(\tau)$
is given by domains $\tau\sim   b/\gamma$ and $\tau\sim  -
b/\gamma$ where
 $|\un\hsp \ddot{z}_{\rm
gr}^{\smhsp 0}|$ reaches its maximum\footnote{In four dimensions
see (\ref{vel_gr_5}) for the components of velocity and its
derivatives.}, despite the fact that at $\tau=0$ it vanishes:
\begin{align}\label{accex_gr1}
\un\hsp \ddot{z}_{\rm gr}^{\smhsp 0}(\pm \tau_{\rm gr}) \sim
\frac{{r'_g}^{d+1}}{b^{\hsp d+2}}\,\gamma^2\,, \qquad\qquad
\un\hsp \dot{z}_{\rm gr}^{\smhsp  0}(\pm \tau_{\rm gr}) \sim
\frac{{r'_g}^{d+1}}{b^{\hsp d+1}}\,\gamma\,\,{}^{*\nhsp
}\footnotemark.
\end{align}
\footnotetext{In what follows the validity of perturbation theory
to this order: $\sup \un\hsp \dot{z}_{\rm gr}^{\smhsp  0} \ll u^0$
if $b\gg r'_g$ holds.}

 If the space-time the had been flat, the direct application of
estimate (\ref{totalo_em}) would lead  to the result
\begin{align}\label{totalo_em_gr}
E_{\rm em/curve} \sim e^2 \left[\un\hsp {z^x_{\rm
gr}}^{(D/2)}\right]^2 t_{\rm gr} \sim   e^2 G_D^2 {m'}^2
\,\frac{\gamma^{d+4}}{b^{\hsp 3d+3}}\, .
\end{align}
However, not only is this result overestimated -- it totally
vanishes due to the following reasoning.

The analogue of Larmor formula in four dimensions in a
\emph{fixed} curved space-time is given by the finite part of
formula by de\,Witt and Brehme \cite{DeWitt:1960fc}, corrected by
Hobbs \cite{Hobbs}\footnote{Here and below the lower-case Greek
indices emphasize the fact, that contraction of indices is
performed in the curved background.}:
\begin{align}\label{deW-B}
f^{\hsp 0}_{\rm em}(\tau)=\frac{ e^2 }{4\pi} \left[ \Pi^{0\nu}\!\!
\left(\frac{2}{3}\, D^2 \zt_{\nu}+\frac{1}{3}\,R_{\nu\lambda}\hsp
\dot z^{\lambda}\right) +
 \dot z^{\nu }(\tau )\!\!\int\limits_{-\infty }^{\tau }\!\!\! \left(\vp v^{\smhsp
0}{}_{\nhsp  \lambda' ; \nu} - v_{\nu  \lambda'}{}^{\nhsp
;0}\right) \dot z^{ \lambda'} (\tau' )\, d\tau'\right] ,
\qquad\qquad \Pi^{\hsp \mu\nu} \equiv g^{\hsp
\mu\nu}-\frac{\zt^{\mu}\zt^{\nu}}{\zt^2}\, ,
\end{align}
Here $v^{\nu \alpha }$ represents the non-local part of the
vectorial Green's function in a curved background in terms of
bi-tensor quantities, evaluated at points $z^{\mu}(\tau)$ and
$z^{\mu'}(\tau')$.

In flat background one has \mbox{$g_{\mu\nu}\to \eta_{MN}
\delta^{M}_{\mu} \delta^{N}_{\nu}$}, \mbox{$ D\dot z^{\mu} \to
\ztt^M \delta_{M}^{\hsp \mu}$}, \mbox{$ D^2\dot z^{\mu} \to
\dddot{z}^M \delta_{M}^{\hsp \mu}$} etc.,  and (\ref{deW-B})
passes into the Lorentz-Dirac equation, there the radiative part
is constituted from the radiation part $\sim \ztt^2 \,\zt^{M} $
and radiation-reaction (''Schott'') part $\sim\dddot z^{M} $.

 The ''Larmor'' part here is given by
\begin{align}\label{larmor}
 \frac{1}{6\pi}\, e^2\,  \Pi^{0\nu} \,
D^2 \zt_{\nu} = \frac{1}{6\pi}\, e^2 \nhsp \left[ D \zt_{\nu}
 D \zt ^{\nu}\,\zt^0 + D^2 \zt^{0} \vp \right]  \,.
\end{align}
But the charge is moving across the geodesics, hence the covariant
acceleration $D\zt^{\hsp \mu}$ and its covariant derivatives
vanish. The local term with Ricci-tensor of the \emph{exact}
metric also vanishes outside the source. Thus in the total-metric
description all radiation effects come from the \emph{tail} term
in (\ref{deW-B}). The same structure of tail term appears in any
dimensionality.

Instead of derivation of tail integral according to the total
metric, we consider the perturbation theory and give a direct
correspondence to reconcile with what we do. In fact, we have been
computing the lower orders of constituents of equation
(\ref{deW-B}).

 In
this case one can expect deflections from the common rule.

First we check that $D\zt^M$ is still zero in the first order:
indeed, as it follows from (\ref{cov_diff}), the flat derivative
$\un \ztt^M$ is given by double derivative of (\ref{sc6}), while
the Christoffel part is given by (\ref{elemvars}) and
(\ref{1h'mn}). Roughly speaking, their sum is
(\ref{affine_gauge1},b) contracted with ${u'}^N$ and thus
vanishes. The next orders do not affect on the order $(r'_g/\smhsp
b)^{2}$ we need. The same concerns the covariant derivatives of
covariant acceleration in higher dimensions.

Now consider the Ricci-term. In the first order of PT the Newton
field coincides with the linearized Schwarzschild metric and thus
still represents Ricci-flat space-time outside the $m'$:
$$R_{ \lambda\rho}\!\left[\eta_{\mu\nu}+\vk \un h_{\mu\nu}^{\rm [PL]}\right]
=\mco \! \nhsp
 \left({(r'_g/r)^{2}}\right)\,, \qquad\qquad R_{ \lambda\rho}\!\left[\eta_{\mu\nu}+\vk \un h_{\mu\nu}^{\rm [S]}\right]
=\mco \! \nhsp
 \left({(r'_g/r)^{2}}\right)\,,
$$
i.e. no terms $\mco(m')$ in both expansions of $R_{\mu\nu}$. The
superscript indices ''[PL]'' (Post-Linear) and ''[S]''
(Schwarzschild) are understood.

Now consider the second-order metric. In out treatment, the
following contributions into $\de h_{MN}$ are expected: $m^2$,
$mm'$ and ${m'}^2$. But from
de\hsp{}Witt\hsp{}--\hsp{}Brehme\hsp{}--\hsp{}Hobbs equation, in
order to keep the field produced by $m'$ as \emph{external}, we
have to retain only the ${m'}^2$-contribution. Throughout the
entire text we have omitted such terms as giving vanishing
contribution to the emitted energy, since on-shell $k^2=0$ these
terms vanish. But off-shell the self-action term ${m'}^2$ is
well-surviving, as it shown in the Appendix \ref{self}. Being
translated back into the coordinate representation, these terms
represent repulsive contribution into $g_{00}$; meanwhile, the
expansion of $g_{00}$ in Schwarzschild metric does not contain
$(r'_g/r)^{2 }-$terms:
$$\vk \de h_{00}^{\rm [PL]} = \mco \left((r'_g/r)^{2(d+1)}\right)\,, \qquad\qquad g_{00}^{\rm  [S]}=1- \frac{r'_g}{r} \,.$$
This fact is reflected into the Ricci-tensor, where non-vanishing
diagonal terms are estimated now:
$$R_{ \lambda \lambda}\!\left[\eta_{\mu\nu}+\vk \un h_{\mu\nu}^{[\rm PL]}+\vk \de h_{\mu\nu}^{[\rm PL]}\right]
=\mco \! \nhsp
 \left({{r'_g}{}^2/r^{4}}\right)\,, \qquad\qquad
R_{ \lambda \lambda}\!\left[\eta_{\mu\nu}+\vk \un h_{\mu\nu}^{\rm [S]}+\vk \de h_{\mu\nu}^{\rm [S]}\right]
=\mco \! \nhsp
 \left({{r'_g}{}^4/r^{6}}\right)\,,
$$
To repeat, the appearance of Ricci-term here is not an excess of
precision, which would take place in the consistent consideration
of the total background as curved.  As it was for vector field in
the Section \ref{radiation_amplitudes}, the delocalization of
gravitational source is a consequence of the flat space-time
description instead of the curved concept.

The analogue of (\ref{deW-B})  in six dimensions is given in
\cite{GaSp}. One can show directly, that radiative part in even
dimensionality coincides with its flat-space analogue, with
obvious generalization of derivatives from common to the
covariant. Thereby on the geodetic motion this part vanishes by
the same reason.

The curved local part (constituted from the single Ricci-term in
four dimensions) comes from the derivative of $\theta(\text{Synge
function})$, accompanying the  $v_{\mu\nu{\smhsp '}}$, and from
the coinciding-point limit of the  covariant expansions of
bi-tensor quantities \cite{christ76}. Given that the
dimensionality of $e^2$ is $[e^2]={\rm cm}^{d}$, the curved local
in $D$ dimensions ($D=\text{even}$) is constituted from
combinations of Ricci- and Riemann tensors with $D^{k} \zt^{\nu}$
of total dimensionality ${\rm cm}^{-(d+2)}$. Among these terms,
taking into account  $\nul \ztt^{\hsp \mu}=0$,  the maximal in
$\gamma$ order has a term of the following type:
\begin{align}
\Pi^{0\nu}R_{\nu\alpha;}{}_{ \underbrace{\scriptstyle
\beta\gamma\delta \pp}_{d}}\, \zt^{\alpha}\zt^{ \beta}\zt^{
 \gamma}\pp \;  \sim\;   R_{\nu\alpha;
\beta\gamma\delta \pp }\, \underbrace{\zt^{0} \zt^{\nu}
\zt^{\alpha}\zt^{ \beta}\zt^{
 \gamma}\pp}_{d+3}\, , \nn
\end{align}
with \emph{positive}  coefficient of proportionality in even $d$,
coming from the construction of curved Green's functions.

 Given that
for Newton field in first non-vanishing order $R_{ \lambda\lambda
}\sim (r'_g)^{2d+2)}/r^{2d+4}$ (for $b\gg r'_g$) and that $\zt^0$
and $\zt^z$ give $\gamma-$factor each, the local curvature term is
of order
\begin{align}\label{yy1}
 \dot{E}_{\rm curv}(\tau)\equiv - f^{\hsp 0}_{\rm curv}(\tau)\; \sim \; - e^2\, R_{\nu\alpha;}{}_{ \underbrace{\scriptstyle
\beta\gamma\delta \pp}_{d}}\, \underbrace{u^{0} u^{\nu}
u^{\alpha}u^{ \beta}u^{
 \gamma}\pp}_{d+3}
\end{align}
Since the metric is static and spherically-symmetric, only the
radial derivatives of Ricci-tensor appear. Finally among $R_{00}$
and $R_{rr}$ the latter is dominant: $$ R_{rr}=- (d+1)(d+2) \,
\frac{{r'_g}^{2(d+1)}}{r^{ 2(d+2)}}+ \mco\!
\left({r'_g}^{3(d+1)}/r^{3d+5}\right)\, , \qquad\qquad
r=\sqrt{b^{\hsp 2}+\gamma^2 v^2 \tau^2}\,.$$ Substituting it into
(\ref{yy1}) and taking care of the sign, one has:
\begin{align}\label{yy2}
 \dot{E}_{\rm curv}(\tau) \; \sim \; - e^2\,  R_{rr;}{}_{ \underbrace{\scriptstyle
rrrr \pp}_{d}}\,\, u^{0}\hsp \underbrace{ u^{z} u^{z}u^{ z}u^{
 z}\pp}_{d+2} \; \sim\; \frac{ {r'_g}^{2(d+1)}}{r^{3 d+4 }}\,
\gamma^{d+3}>0\,.
\end{align}

The characteristic spatial distance, where the curvature alters
significantly across the particle's trajectory, is of order
$\mco(b)$, thus the mean time and mean proper time are given by
(\ref{duration_gr}), in what follows that $r\sim b$ and the
relative contribution reads
\begin{align}\label{yy3}
  {E}_{\rm curv}(\tau)\; \sim \;  \dot{E}_{\rm curv}(\tau) \,\tau_{\rm gr}  \; \sim\; \frac{e^2 \hsp {m'}^2\hsp G_D^2}{b^{\hsp 3d+3}}\,
\gamma^{d+2}\,.
\end{align}
The characteristic times (\ref{duration_gr}) find a reflection in
the characteristic frequencies for this partial process. These
frequencies are given by $\omega \sim \gamma^2\!/\smhsp b$ as a
full analogy with (\ref{charw_em}).

Looking at the Table I one concludes that this sub-process
corresponds to the high-frequency entry, with the proper  estimate
of partial contribution into the total emitted energy.

\vspace{0.5em}

\textbf{A tail.} Next, proceed to the last, tail,  term in
(\ref{deW-B}): it comes from the modification of the self Coulomb
field of a particle, by the weak curved  background:
\begin{align}\label{tail1}
\dot{E}_{\rm tail}(\tau)\equiv -\frac{e^2}{4\pi}
 \dot z^{\nu }(\tau )\!\!\int\limits_{-\infty }^{\tau }\!\! \left(\vp v_{ \smhsp 0 \lambda' ; \nu} - v_{\nu  \lambda' ; 0}\right)
\dot z^{ \lambda' } (\tau' )\, d\tau'
\end{align}

Thus the basic problem is to estimate the tail function in
(\ref{deW-B}) as tensor in Minkowski space-time, for the weak
Newton field. The basic step in four dimensions was made in
\cite{DeWitt-deWitt}, and applied to the non-relativistic motion.
The first order of this expression:
\begin{align}\label{tail2}
\un\dot{E}_{\rm tail}(\tau)= \frac{e^2}{4\pi}
 u^{\nu }  u^{ \lambda' }  \!\int\limits_{-\infty }^{\tau }\!\! \left[\vp \un v_{\nu  \lambda', 0}\left(\nul z(\tau),\nul z(\tau')\right)
-\un v_{\smhsp 0\lambda' , \nu}\left(\nul z(\tau),\nul
z(\tau')\right)\right] \, d\tau'
\end{align}
represents the full derivative over $\tau$ and, being integrated
further from $\tau=-\infty$ to $\tau=+\infty$, vanishes. A more
detailed derivation is to be given in \cite{Spir2}. The
second-order (${m'}^2$) is given by six terms
\begin{align}\label{tail3}
\frac{4\pi}{e^2}\,\de\dot{E}_{\rm tail}(\tau) = & \;
 u^{\nu } \!\int\limits_{-\infty }^{\tau }\!\! \left(\vp  \un v_{\nu  \lambda', 0}-\un v_{\smhsp 0\lambda' , \nu}
\right)\! \un \zt^{ \lambda' }\nhsp \nhsp (\tau')  \: d\tau' +
 \un \zt^{\nu }(\tau) \, u^{ \lambda' }  \!\int\limits_{-\infty }^{\tau }\!\! \left(\vp \un v_{\nu  \lambda', 0} - \un v_{\smhsp 0\lambda' , \nu}
 \right) \,
d\tau'+\nn \\  & +
 u^{\nu }  u^{ \lambda' }  \un z^{\sigma}\!(\tau) \!\int\limits_{-\infty }^{\tau }\!\! \left( \vp \un \smhsp v_{\nu  \lambda', 0 \sigma}-
\un\smhsp v_{\smhsp 0\lambda' , \nu \sigma} \right) \, d\tau'+
 u^{\nu }  u^{ \lambda' }  \!\int\limits_{-\infty }^{\tau }\!\! \left(\vp  \un \smhsp v_{\nu  \lambda', 0 \sigma'}-
\un\smhsp v_{\smhsp 0\lambda' , \nu \sigma'} \right) \! \un
z'^{\sigma'}\!\! (\tau')\: d\tau' +\nn \\ & +
 u^{\nu }  u^{ \lambda' }  \!\int\limits_{-\infty }^{\tau }\!\! \left(\vp  \de \smhsp v_{\nu  \lambda', 0} - \de\smhsp v_{\smhsp 0\lambda' ,
\nu} \right) \, d\tau' -
 u^{\nu }  u^{ \lambda' }  \!\int\limits_{-\infty }^{\tau }\!\! \left(\vp \un\smhsp \Gamma^{\hsp 0}_{\sigma \nu} \un v^{\sigma}{}_{\nhsp \lambda'}
+ \un \smhsp \Gamma^{\smhsp\sigma}_{\nu\hsp 0}
 \un v_{ \sigma  \lambda' }\right) \,
d\tau'
\end{align}
where the integrals are to be evaluated on the \emph{unperturbed }
trajectory. The first line represents the variation of $\zt^{\nu
}\zt^{\lambda'}$, the second one is a first term of Taylor
expansion of $\un v_{\mu\nu',\lambda}$ while the third line is
constituted from second-order $v_{\mu\nu',\lambda}$ and
$\Gamma-$terms from covariant differentiation of $v_{\mu\nu'}$,
respectively.

Direct application of the PT gives $\un v_{\mu\nu'}$ as some
combination of the second-order derivatives of generic integral
\begin{equation}\label{uuui8}
I(x, x')=\int \delta^{(d/2)}\!\left((x'-x'')^2\right)
\delta^{(d/2)}\!\left((x-x'')^2\right)\frac{\dx''}{{r''}^{d+1}}\,,
\qquad\qquad x''=(t'',\mathbf{r}'')\, ,
\end{equation}
which can be interpreted as a matrix element of Newtonian
potential from initial state \mbox{$|\hsp \mathrm{in} \rangle =
{}^{D}\nhsp G |x\rangle$} to the final \mbox{$|\hsp\mathrm{out}
\rangle = {}^{D}\nhsp G|x\rangle$}, with \mbox{${}^{D}\nhsp G$} is
a Green's function in flat $D-$dimensional space-time.

In particular, the consistent account of the non-relativistic
limit leads to the Smith\hsp{}--\hsp{}Will force in higher
dimensions\footnote{In fact, Smith and Will \cite{Smith-Will} have
shown that the four-dimensional result by de\hsp{}Witt and
de\hsp{}Witt for newtonian (weak) field \cite{DeWitt-deWitt} is
still exact in the total Schwarzschild metric even for the case of
strong field.}. The discussion of all terms in (\ref{tail3}) and
all derivatives of (\ref{uuui8}) goes beyond our primary goal
here. We will highlight here the four-dimensional estimate, with
generalization to be done in forthcoming publication: the integral
$I(x, x')$ in (\ref{uuui8}) is computed in \cite{DeWitt-deWitt}
and reads
\begin{equation}\label{uuui9}
I(x, x')=\frac{1}{|\mathbf{r}-\mathbf{r}'|}\left[
\theta(r+r'-t+t')\,\ln\frac{r+r'+|\mathbf{r}-\mathbf{r}'|}{r+r'-|\mathbf{r}-\mathbf{r}'|}
+
\theta(t-t'-r-r')\,\ln\frac{t-t'+|\mathbf{r}-\mathbf{r}'|}{t-t'-|\mathbf{r}-\mathbf{r}'|}\right].
\end{equation}

The third-order derivatives over $t$ and $z$ have maximal value
only if one keeps $\theta(t-t'-r-r')$ and differentiates the
logarithm, otherwise for $\tau,t'\gg b/\gamma v$
$\delta^{(k)}(t-t'-r-r')$ contains $\gamma$ inside an argument and
$\gamma$  goes to denominator.

Thereby
\begin{equation}\label{uuui10}
v_{\mu\nu',\lambda}(x, x')\sim r'_g
\theta(t-t'-r-r')\,\frac{(x-x')_{\mu}(x-x')_{\nu'}(u_{\lambda}/\gamma)}{\left[(x-x')^2\right]^3}.
\end{equation}
For $x$ and $x'$ are taken on the unperturbed trajectory,
$(x-x')_{\mu}=u_{\mu}\,(\tau-\tau')$ contains $\gamma$ (for
$\mu=0,z$), while $(x-x')^2=(\tau-\tau')^2$ -- does not, thus the
typical term reads
\begin{equation}\label{uuui11}
v_{\mu\nu',\lambda}(x, x')\sim r'_g\,
\frac{\theta(t-t'-r-r')}{\gamma}\,\frac{u_{\mu}u_{\nu'}u_{\lambda}}{(\tau-\tau')^4}\sim
\gamma^2\,\frac{\theta(t-t'-r-r')}{(\tau-\tau')^4}.
\end{equation}

The solution for $\un z^0$ coming from (\ref{sc6}) is given by
\begin{align}\label{vel_gr_5}
 \un \zt^{0}(\tau)=   \frac{\, m' \varkappa_4^2\, }{8\,\pi^2  } \frac{ \gamma}{ \sqrt{b^{\hsp 2}+(\gamma v
\tau)^2}}\, , \qquad\qquad \un \zt^{z}(\tau)=- \frac{d+4}{2\hsp
(d+1)}\,\un \zt^{0}(\tau)
\end{align}

According to $\theta(t-t'-r-r')$,  $t-t' =\gamma(\tau-\tau')\equiv
\gamma \xi$ is larger  than $r+r'>2b\hsp .$ Thus $\xi> 2b/\gamma$.
Substituting $r=\sqrt{b^{\hsp 2}+\gamma^2 v^2 \tau^2}$ and
$r'=\sqrt{b^{\hsp 2}+\gamma^2 v^2 (\tau-\xi)^2}$, such an argument
of Heaviside function has a solution only if $\tau \xi
>b^{\hsp 2}$. Taking into account the double $\tau\xi$-integration
and that integration ranges of both $\xi$ and $\tau$ are equally
important, one expects the domination from the range
\begin{equation}\label{uuui12}
|\tau | \sim \xi \sim b\,.
\end{equation}

Therefore the typical term of the total energy associated with a
tail, reads
\begin{equation}\label{uuui13}
\de{E}_{\rm tail} \sim e^2 (r'_g)^2  \gamma^4 \int\limits_{-\infty
}^{ \infty} d \tau \int\limits_{b^{\smhsp 2}\!\nhsp\smhsp /\tau
}^{ \infty} \frac{d\xi }{\xi^4}
   \frac{1}{
\sqrt{b^{\hsp 2}+(\gamma v \tau)^2}}
\end{equation}
Substituting the estimate (\ref{uuui12}), one obtains finally
\begin{equation}\label{uuui14}
\de{E}_{\rm tail} \sim e^2 (r'_g)^2  \gamma^4  \, \frac{ \tau \xi
}{\xi^4}
   \frac{1}{
\sqrt{b^{\hsp 2}+(\gamma v \tau)^2}}\!\left.\vph\right|_{\tau \sim
\xi \sim b}\sim \frac{e^2 (r'_g)^2}{b^{\hsp 3}}\, \gamma^3\,,
\end{equation}
in agreement with (\ref{E4D})\footnote{From the consideration made
above we can say nothing about a sign of this expression. The main
goal of this subsection is to qualitatively explain the spectral
characteristic of this process arising do to the tail. However,
giving the direct correspondence to the positively-defined
expression in the text, we hope that a consistent accounting of
all terms in (\ref{tail3}) will lead to the conclusion concerning
the sign.}.

Thus, despite the rapid decrease of $\un \zt^M$ at
$\tau>b/\gamma$, the main contribution comes from $\tau\sim b$ due
to the fact that $\un v$ alters slowly.

According to (\ref{uuui12}), the characteristic duration in the
comoving and in the Lab frames, due to the Doppler effect, are
given by
\begin{equation}\label{uuui15}
\tau_{\rm tail}=b =\gamma \tau_{\rm em} \, \qquad\qquad t_{\rm
tail }=\gamma\hsp \tau_{\rm tail}=\gamma\hsp  b\, \qquad\qquad
t_{\rm Lab,tail}\sim \frac{ t_{\rm em}}{\gamma^2} \sim \frac{b}{
\gamma}\,,
\end{equation}
respectively, while applying the same deduction as in
(\ref{charw}) one obtains the characteristic frequencies of this
tail effect:
\begin{align}\label{charw1}
\omega_{\rm com,tail} \sim  \frac{1}{t_{\rm tail}} \sim
\frac{1}{\gamma\hsp b}\, , \qquad\qquad \omega_{\rm tail} \sim
\frac{1}{t_{\rm Lab,tail}} \sim \frac{\gamma}{b}=\omega_0\, ,
\end{align}
in agreement with (\ref{charADD}), taken for $d=0$.

Thus we arrive at the conclusion, that, at least in four
dimensions, the transition region in the Table I corresponds to
the tail effects of non-linearity in
de\hsp{}Witt\hsp{}--\hsp{}Brehme sense. The generalization into
higher dimensions represents the goal of forthcoming work.

Comparing with the bremsstrahlung by non-gravitational force, we
conclude that in gravity the Lorentz transformation of frequency
is determined not only by simple ultrarelativistic consideration
of Doppler effect, but also by curved geometry and non-linear
effects.

Thus we arrive at the following scheme:
$$\text{tail in curved space-time} \to \text{Ricci-term in } M_{1,D-1}+ \text{tail term for } v \text{ treated perturbatively in } M_{1,D-1}\, .$$

Thereby, to conclude: the contribution coming from a tail in the
curved-space concept reappears as local curvature terms. This
phenomenon is directly related with PT over Minkowski background,
and with ultrarelativistic character of a motion. In our scheme it
represents the same effect as the effective delocalization of the
second-order-field source in the flat space.

The analogy of such a resurrection was proposed by
\cite{DeWitt-deWitt} for the opposite ultimate case of
non-relativistic motion along a bounded orbit, where
originally-tail contribution (with respect to the total metric)
reappeared as non-conservative non-relativistic Larmor energy.

\subsection{Restrictions and possible cut-offs}\label{restrictio}

Here we assume that $m\gamma \gg m'$ and the emitted energy is
determined  by those values obtained in the Section
\ref{emitted_energy}. Thereby the total initial energy is
essentially the energy  of the fast particle:
$\mathcal{E}_0\approx m\gamma\,.$ Our goal here is to set bounds
on the minimal value of an impact parameter $b$ and to confirm the
validity of the classical approach applied above.

The condition on the weakness of gravitational field, $b \gg
r'_g$, has been discussed in (\ref{restr_0}). The condition $b \ll
R$ (\ref{restr_B}) is related with the treatment of  space-time as
higher-dimensional. Additionally, in the ADD model, it is directly
related to the pass from KK-mode-summation to the quasi-continuous
integration. Finally, the condition on the classicality of the
emitted vector field obviously reads
\begin{equation}\label{restr_1}
   b>r_{\rm cl}=(e^2/m)^{d+1}.
\end{equation}

Next consider the conditions which do not follow from the
classical theory but are necessary for the classical result to fit
the quantum one.

The simple quantum-mechanical restrictions
\begin{equation}\label{restr_2}
   \omega_{\max} \ll
 {E}_{\rad} \, , \qquad \qquad E_{\rad} < \mathcal{E}_0\approx
m\gamma
\end{equation}
reflect the fact that the particle can not lose energy more than
it had initially (being free at infinity). The ultimate situations
of hard bremsstrahlung, when the charge emits almost all its
energy, are admissible in  QED \cite{Landau4}. Next, for the
treatment of the emitted photons as classical, we need a large
number of their quanta, which implies the  weak particle-recoil.
For the radiation problem at hand, the weak particle-recoil
condition due to the emission of photons with frequency $\omega$
is satisfied if the momenta of the emitted photons are much
smaller than the momentum transfer of the elastic collision. For
the hard-photon emission with $\omega < \mathcal{E}$ the latter
condition is satisfied if the emission angle $\vartheta$ is less
than the deflection angle $\alpha_{\rm gr}$, while for $\omega \ll
\mathcal{E}$ this condition can be relaxed.

Substituting the characteristic emission angle $\vartheta \sim
\bar{\vartheta}=1/\gamma$ into (\ref{grav_angle}) one obtains
\begin{equation}\label{restr_3}
     b > r'_g \gamma^{\frac{1}{\rule{0cm}{0.43em}d+1}}\,. \,{}^{*}
\footnotemark
\end{equation}
\footnotetext{The latter quantity coincides with the
energy-associated Schwarzschild radius $r'_S$ of $m'$ in the
comoving (with $m$) Lorentz frame and approximately equals $r_S$
(of $m$) in the Lab frame for comparable $m\sim m'$.} This
condition differs from the one, (\ref{restr_A}), given in the
Introduction for gravitational bremsstrahlung. It is stronger than
the weak-field condition (\ref{restr_0}) but weaker than
(\ref{restr_A}).

Indeed, according to the iteration scheme, the ultrarelativistic
charge emits the energy after its  trajectory is gravitationally
perturbed, so we do not need to accounting for the back-reaction
of the gravitational field due to the fast charge, on the
uncharged, target, particle.

Moreover, the experience from analogous computations of the total
energy of synchrotron radiation shows that this condition can be
relaxed and replaced, instead, by the weaker $\om \ll {\mathcal
E}_0$ without restriction on the angles of the emitted photon.
When the emitted energy $E$ is of order ${\mathcal E}_0$, this
condition also guarantees a large number of emitted quanta, and
justifies further the description of radiation with a classical
field.

Estimating  the efficiency of the emitted energy in four
dimensions according to (\ref{E4D}), one gets
\begin{equation}\label{restr_4}
\epsilon_{0} \equiv \frac{E_{\rad}}{\mathcal{E}_0} \sim
\frac{e^2\, {m'}^2 G_4^2}{m b^{\hsp 3}}\,\gamma^2\sim \frac{r_{\rm
cl}}{b}  \left(\frac{\gamma r'_g}{ b}\right)^{\!\nhsp 2}< 1\, ,
\end{equation}
by virtue of restrictions (\ref{restr_1},\ref{restr_3}).

For the ADD bremsstrahlung (\ref{rhonexp5}), with the same
characteristic frequency $\omega \sim \gamma/\smhsp b$, the
efficiency reads
\begin{equation}\label{restr_5}
\epsilon_{\rm ADD}   \sim \frac{e^2\, {m'}^2 G_D^2}{m V b^{\hsp
2d+3}}\,\gamma^2\sim \left(\frac{r_{\rm
cl}}{b}\right)^{\!d+1}\!\left( \frac{ b}{R}\right)^{\! d} \!
\left(\frac{\gamma^{\frac{1}{\rule{0cm}{0.43em}d+1}} r'_g}{
b}\right)^{\!\nhsp 2(d+1)}< 1\, ,
\end{equation}
if one also takes into account (\ref{restr_B}).

In higher dimensions with characteristic frequency $\omega \sim
\gamma^2\!/\smhsp b$ the direct application\footnote{We neglect
here the $\ln \gamma$ in (\ref{E_final_1}).} of the above
estimates gives $\epsilon_{d} \ll \gamma^{d-1}$. Thereby this
might lead to the efficiency catastrophe for $d>1$.

The possible resolutions of this
 paradox may be related with:

\begin{itemize}
\item A small pre-factor, of order of $C_d \sim 10^{-5}$, in (\ref{E_final});
\item Frequency $\omega\sim \gamma^2\!/\smhsp b$ is incompatible with
the requirement $m< M_*$. Thereby one needs a cut-off on the
frequency;
\item The possible Vainshtein limit of the process in a space with compactified
radii;
\item Combination of (\ref{restr_1}) with (\ref{restr_3}) gives $$b>\max \left\{(e^2/m)^{d+1},   r'_g \gamma^{\frac{1}{\rule{0cm}{0.43em}d+1}}
\right\}.\nn $$
\end{itemize}
Let us consider the latter possibility in practice.

For instance, for the scattering of protons on neutrons with
$\gamma= 10^{14}$, available at the LHC, the classical radius of a
proton and $\gamma r'_g$ for neutron are given ($d=0$) by
\begin{equation}\label{rad_class}
    r_{\rm cl}=1.53\cdot 10^{-16}\,\text{cm}, \qquad\qquad \gamma r'_g
= 2.48\cdot 10^{-38}\,\text{cm},
\end{equation}
 respectively, while in higher
dimensions the ratio $r_{\rm cl}/ r'_g \gamma^{ {1}/(d+1)}$ is
even larger. Thus the restriction on $b$ is determined,
essentially, by $r_{\rm cl}$. Moreover, the latter is less than
the actual size of a proton $l_p$ and its Compton wavelength $l_C$
of it: $$l_p=0.84\cdot 10^{-13}\,\text{cm}, \qquad\qquad l_C=2.10
\cdot 10^{-14}\,\text{cm}.$$ The scattering of nuclei present
similar features.

On the other hand, the radiated energy efficiency coming from
(\ref{E_final})   can be presented as
\begin{equation}\label{restr_6}
\epsilon_{ d}   \sim \frac{e^2\, {m'}^2 G_D^2}{m b^{\hsp
3d+3}}\,\gamma^{d+1}\sim \left(\frac{r_{\rm
cl}}{b}\right)^{\!d+1}\! \left(\frac{\sqrt{\gamma}\, r'_g}{
b}\right)^{\!\nhsp 2(d+1)}\, ,
\end{equation}
and, by virtue of $b>r_{\rm cl}> {\gamma}\, r'_g> \sqrt{\gamma}\,
r'_g$, easily becomes smaller than unity. This practically
resolves the efficiency paradox. The same argument makes  the
dominance of gravitational radiation over the electromagnetic,
almost impossible, an issue raised above according to the naive
comparison of the power of $\gamma$.

For the scattering of electrons one takes the Compton length.
Thereby  there is no the efficiency catastrophe in the
problem-at-hand, but one sets the following bound on the value of
the impact parameter:
\begin{equation}\label{impact_1}
   l_C<b\,.
\end{equation}

In UED, from (\ref{E_final_UED}) one obtains:
\begin{equation}\label{eff_UED}
\epsilon_{\rm \smhsp UED}   \sim  \frac{\left(e m' \vk^2\right)^2
}{m V^2 b^{\hsp d+3}}\:\gamma^{d+1} \sim \left(\frac{b}{R}
\right)^{2d}\left(\frac{r_{\rm cl}}{b}\right)^{\!d+1}\!
\left(\frac{\sqrt{\gamma}\, r'_g}{ b}\right)^{\!\nhsp 2(d+1)} \! .
\end{equation}

Taking into account $b>l_C> R_{\rm UED}$ (\ref{impact_2}), one
rewrites (\ref{eff_UED}) as
\begin{equation}\label{eff_UED_1}
\epsilon_{\rm \smhsp  UED} < \left(\frac{b}{  R}
\right)^{2(d+1)}\left(\frac{r_{\rm cl}}{b}\right)^{\!d+1}\!
\left(\frac{\sqrt{\gamma}\, r'_g}{ b}\right)^{\!\nhsp 2(d+1)} \sim
\left(\frac{r_{\rm cl}}{b}\right)^{\!d+1}\!
\left(\frac{\sqrt{\gamma} \, r'_g}{ R}\right)^{\!\nhsp 2(d+1)}
\!\!\ll 1\, ,
\end{equation}
if directly compare $\sqrt{\gamma} \, r'_g \ll  {\gamma} \, r'_g
\ll R_{\rm UED}$ by values (\ref{impact_2}) and (\ref{rad_class}).

Now return to the large-modes condition (\ref{Nemit_UED3}):
substituting $R_{\rm UED}$ by (\ref{impact_2}) and comparing with
(\ref{impact_1}) one concludes that for $\gamma =10^{14}$ the
condition
\begin{align}\label{Nemit_UED3st}
 \gamma R \sim 10^{-2}\,\cm   \gg b >  10^{-14}\,\cm \sim \lambda_C\,,
\end{align}
 is well satisfied and a large number of the emission modes are excited, that gives the enhancement
of the bremsstrahlung radiation.

Going back to the spectrum we see that if $b>1/m$ holds, then $$
\omega_{\max}=\mathcal{E}_0 =m \gamma>
\frac{\gamma}{b}=\omega_0\,.$$ Thus the maximal value of the
frequency lies inside the domain $(\gamma/\smhsp b,
\gamma^2\!/\smhsp b)$, so the part of destructive interference
region, the main point of our computation, can be detected in
practice in all kinds of the extra dimensions and corresponding
gravity models.

Despite the radiation efficiency being tiny, one can expect that
absolute amounts of the emitted radiation, due to the relatively
large $r_{\rm cl}$ with respect to $r_g$, can be determined (for
instance, for heavy nuclei) and can give information on the
(possible) size and number of extra dimensions.

\subsection{Results and conclusions}

A detailed study  of  classical electromagnetic (vector) radiation
emitted in ultra-relativistic collisions of massive point-like
particles was presented. The space-time was assumed to have an
arbitrary number of toroidal or non-compact extra dimensions and
the post-linear approximation scheme of General Relativity was
employed for the computation. The angular and frequency
distributions of radiation, as well as the total emitted energy
were studied in detail up to leading ultra-relativistic order.

Three characteristic frequency regimes ($1/\smhsp b$,
$\gamma/\smhsp b$ and $\gamma^2\!/\smhsp b$) of the emitted
radiation were identified and the characteristics of the dominant
contribution was determined in various dimensions, depending on
the gravity model.

In particular, in any number of dimensions the soft component of
radiation is mainly due to the scattered particles, with
negligible contribution coming from the cubic
graviton-graviton-photon interaction term\footnote{In four
dimensions this is a well-known fact, verified easily also in the
context of Feynman diagram infrared graviton summation.}. In all
cases of bremsstrahlung  most of the emitted waves are beamed (in
the Lab frame) inside a narrow cone with angle $1/\gamma$ and
along the spatial direction of fast-particle's motion.

Among the notable features we would like to mention, are the
following:

\begin{itemize}
\item The radiation amplitude is damped by the factor $(\omega_0/\omega)^2$ at
the frequency region $\gamma/\smhsp b \lesssim \omega \lesssim \gamma^2\!/\smhsp b
$:
$$j(\omega) \sim j(\omega_0)\,\frac{ \omega_0^2 }{\omega^2}\, , \qquad\qquad \omega_0 \sim \frac{\gamma}{b}\,;$$
Thus at $\omega \sim \gamma^2\!/\smhsp b$ the amplitude $j(\omega)
$ is suppressed by  $ \gamma^2$ with respect to $j\left( \mco
(\gamma/\smhsp b)\right)$, that represents the destructive
interference (DI) effect;

\item The frequency distribution goes like
$$\frac{dE_{\rm rad}}{d\omega}\sim \gamma^{4-d}\,\omega^{d-2}$$
inside this  frequency regime. Hence for $d=0$ and in the ADD-case
  most of the radiation has characteristic frequency $\omega \sim
\omega_0$, for $d\geq 1$ the dominant frequency is $\omega \sim
\gamma\smhsp \omega_0$ while in the transition case $d=1$ the
entire domain  $\gamma/\smhsp b \lesssim \omega \lesssim
\gamma^2\!/\smhsp b $ contributes equally to add a logarithm of
$\gamma$ into the total emitted energy;

\item ZFL gives qualitatively adequate result for the ADD bremsstrahlung (where DI happens beyond
$\omega_{\rm ADD}\sim \gamma/\smhsp b$) and for pure
electromagnetic bremsstrahlung (where no DI happens and the
amplitude has the same behavior up to $\omega_{\rm em} \sim
\gamma^2\!/\smhsp b$) in the small-angle region;
\item No efficiency catastrophe for  reasonable values of
 the Lorentz factor and charges;

\item The applicability of perturbation theory is essentially
determined by the Compton length of a charge: $$b\gg l_C\,;$$

\item The coherence length argument gives an adequate explanation
of the frequency-angular characteristics of the radiation
amplitude but does not predict which frequencies will dominate in
spectrum.
\end{itemize}

 However, in contrast to the four-dimensional
case, in any number of extra dimensions $d>0$ the frequency
spectrum of the emitted radiation vanishes as $\omega\to 0$ and
the total emitted energy in soft gravitons is negligible.

Also, contrary to what happens with soft radiation emission, the
cubic graviton-graviton-photon interaction and the scattered
particles themselves are equally important as sources of radiation
with high frequency. In fact it was shown that in any dimension
they lead to partial cancelation ({\it destructive interference})
of the total beamed radiation amplitude in the high frequency
domain, as a result of which the emitted energy in the
$\gamma^2\!/\smhsp b\hsp-$\hsp{}frequency regime is reduced by two
powers of the Lorentz factor $\gamma$ in the Lab frame.

The relevance of the classical analysis to the full {\it quantum
radiation} problem was also discussed. The {\it classicality
conditions}, necessary for the classical treatment to be a good
approximation to the full quantum problem were derived and the
radiation efficiency $\epsilon$, i.e. the fraction of the initial
energy which is emitted in gravitational radiation, was computed
for values of the  parameters within the region of validity of our
classical computation.

Thus one concludes that the gravitational scattering of charges
and corresponding bremsstrahlung, at least classically, is a more
reliable scheme to detect extra dimensions already in contemporary
colliders, though, of course the quantum-field treatment of this
process (at least for the vector field) is necessary and
represents the direct prospect of further study.

Finally, the spectral characteristics are qualitatively discussed
in the context of coordinate-space equation of a charge in
dimensions of the even space-time dimensionality
(Lorentz\hsp{}--\hsp{}Dirac and
de\hsp{}Witt\hsp{}--\hsp{}Brehme\hsp{}--\hsp{}Hobbs types of
equations). The pure vector bremsstrahlung is qualitatively
described by the radiative part of the  higher-dimensional
Lorentz\hsp{}--\hsp{}Dirac equation in flat space. For the vector
bremsstrahlung under the gravity-mediated collision it was found
that the observable competition of frequencies originates from the
different terms of the
de\hsp{}Witt\hsp{}--\hsp{}Brehme\hsp{}--\hsp{}Hobbs equation,
describing the motion of a charge in the fixed external curved
background.

 Thus one concludes that as qualitative argument,
the concept of coherence length is valid and   directly
corresponds to the similar behavior of amplitudes at
ultrarelativistic characteristic frequency regimes $\omega\sim
\gamma/\smhsp b$ and $\omega\sim \gamma^2\!/\smhsp b$.
Nevertheless, as a quantitative argument, coherence length is much
less useful when the total physical process is split into some
sub-processes. coherence length consideration does not predict
which frequency will dominate in the spectrum, since it does not
take into consideration inside itself the possible competition
between the spectral-angular characteristics of a source and
volume factor in the integration measure when the flux is
computed.

However, the implementation of this interpretation and the proper
treatment of this classical computation have to be confirmed by
the corresponding quantum approach. Meanwhile, even low- and
medium-frequency parts of the  spectral distribution, which are
definitely in  agreement with the quantum case, contain some
distinctive features for the possible presence of extra dimensions
to be detected.

\vspace{0.5em}

\textbf{Acknowledgement.} The authors would like to thank Profs.
D.\,Gal'tsov and \,Th.\,Tomaras for useful discussions.

This work is supported by grant 11-02-01371-a of RFBR (PS) and
(YC) in part by grants (FP7-REGPOT-2012-2013-1) no 316165,
PIF-GA-2011-300984, the EU program ''Thales'' MIS 375734 and was
also cofinanced by the European Union (European Social Fund, ESF)
and Greek national funds through the Operational Program
''Education and Lifelong Learning'' of the National Strategic
Reference Framework (NSRF) under ''Funding of proposals that have
received a positive evaluation in the 3rd and 4th Call of ERC
Grant Schemes''.

Finally, PS is grateful to the University of Crete for hospitality
in some stage of this work and to the non-commercial ''Dynasty''
foundation (Russia) for co-funding.

\appendix
\addcontentsline{toc}{section}{Appendices}
\renewcommand{\theequation}{\Alph{section}.\arabic{equation}}

\section{Useful kinematical formulae}\label{formulae}

\subsection{Notations}\label{notazioni}

The angles in the formulae below are defined in
Fig.\,\ref{branepic}.
\begin{align}
&u^\mu\!\equiv \ga(1, 0, 0, v) \,,  \quad  u'\equiv (1, 0, 0,
0)\,, & \qquad\qquad &
\psi\equiv 1-v\cos\vartheta \,, \;\;  \nonumber \\
&z' \equiv \frac{(ku')b}{\ga v} = \frac{\om b}{\ga v  }\,,&
\qquad\qquad & z \equiv\,\frac{(ku)b}{\ga v} = \frac{\om b}{v }\,
\psi=z' \ga \psi\,,
 \nonumber \\
&
\xi^2\equiv 2\ga z z'-z^2-{z'}^2 = (\ga v z' \sin\vartheta)^2\, , & \qquad\qquad & \beta\equiv \ga z z' - z^2  =\ga^2 {z'}^2 \psi (1-\psi)\, , \nonumber \\
&(kb)=- \gamma z' v \sin \vartheta \cos\phi =-\omega b \sin \theta
\cos \varphi \, ,   & \qquad\qquad &  a \equiv  {z}/{\sin
\vartheta}\,. \nonumber
\end{align}

\subsection{Beaming angular integrals}
\label{beam_ang}

 In the main text
the following angular integrals over $\vartheta$ were needed for
integer $m$ and $n$
\begin{align}\label{jj0}
V_{m}^n\equiv \int\limits_0^{\pi}\frac{\sin^n \!\vartheta}{\vph
\left(1-v \cos{\vartheta}\right)^m}\:d\vartheta \,.
\end{align}
Consider small-angle contribution, corresponding to the beaming of
emitted quanta. For $\gamma \gg 1$ and $\vartheta \lesssim
\gamma^{-1}$
  the numerator and denominator go like
\begin{align}\label{jj1}
\sin^n \!\vartheta \simeq  \gamma^{-n}\, , \qquad \qquad\left(1-v
\cos{\vartheta}\right)^m \simeq \gamma^{-2m}\,,
\end{align}
respectively, thus if $2m>n+1$ one expects the dominance of
small-angle region over the other integration domain.

Expanding
\begin{align}\label{jj2}
\sin \vartheta= \vartheta+ \mco(\vartheta^2)\, , \qquad\qquad  1-v
\cos{\vartheta} = \frac{\vartheta^2+\gamma^{-2}}{2}+
\mco(\vartheta^4) \,,
\end{align}
the integral (\ref{jj0}) reads
\begin{align}\label{jj3}
V_{m}^n=2^m \!\! \int\limits_0^{\sim
1/\gamma}\!\!\frac{\vartheta^n}{\vph \left(
\vartheta^2+\gamma^{-2}\right)^m}\:d\vartheta \,.
\end{align}
Rescaling $\vartheta \to \vartheta/\gamma$ leads to
\begin{align}\label{jj4}
V_{m}^n=2^m \gamma^{2m-n-1}\! \int\limits_0^{\sim
1}\!\frac{\vartheta^n}{\vph \left(
\vartheta^2+1\right)^m}\:d\vartheta \,.
\end{align}
This integral (without pre-factor) is of order $\mco(1)$. Due to
the integrand in (\ref{jj4}) falls rapidly at $\vartheta \gg 1 $
one expands the upper-limit to infinity. Indeed, for any $a\gg1$
the contribution
\begin{align}\label{jj5}
 \int\limits_{a}^{\infty} \! \frac{\vartheta^n}{\vph \left(
\vartheta^2+1\right)^m}\: d\vartheta\simeq \int\limits_{a}^{
\infty} \vartheta^{n-2m}\,d\vartheta \sim a^{-(2m-n-1)}\ll 1 \,.
\end{align}
Thus both the initial integral (\ref{jj0}) and modified one
(\ref{jj4}) have subleading contribution from large values of an
integration argument due to the rapid fall of integrands.

Thus
\begin{align}\label{jj6}
V_{m}^n=2^m \gamma^{2m-n-1}\!
\int\limits_0^{\infty}\!\frac{\vartheta^n}{\vph \left(
\vartheta^2+1\right)^m}\:d\vartheta \,.
\end{align}
Introducing new integration variable $y$ according to $1+
\vartheta^2=1/y $, the (\ref{jj6}) is presented as
\begin{align}\label{jj7}
V_{m}^n=2^{m-1} \gamma^{2m-n-1}\! \int\limits_0^{1}
\left(1-y\right)^{\frac{n-1}{\rule{0cm}{0.43em}2}}
y^{\frac{2m-n-3}{\rule{0cm}{0.43em}2}}\, d\vartheta \,,
\end{align}
that is exactly the Euler's Beta-function
$\mathrm{B}\!\left(\frac{n+1}{2}\, ,\frac{2m-n-1}{2} \right)$.
Rewriting it via Gamma-functions, we finally arrive at
\begin{align}\label{jj2q}
V_{m}^{n}=   \frac{2^{m-1}\,  \Gamma\!
 \left(\frac{n+1}{2}\right)\,\Gamma\! \left(
\frac{2m-n-1}{2}\right)}{\Gamma(m)}\, \gamma^{2m -n-1}\,.
\end{align}
In \cite{GKST2}, with another derivation of the above integral via
Legendre functions, it was shown that first correction to the
(\ref{jj2q}) is of relative order $\mco(\gamma^{-2})$.

 In the case
  $2m=n+1$ an expansion of the integral is logarithmic.

\section{Self-action account}\label{self}

We have already discussed the reason we do not consider the self
action as far as radiation is concerned. It is however useful to
show that including the self action leads to a conserved current.

When one includes the self action, the equations of motion change
are of the same form but we should substitute $h$ and $h'$ with
$h+h'$. This produces some extra terms in the local and non-local
currents. We write here the extra terms in the local current:
\begin{align}\label{rho_self1}
\rho^M_{\rm self} (k )=-\frac{e \hsp m \hsp \vk^2 \, \e^{i (kb)}}{
(2 \pi )^2} \int \frac{\delta (qu )\, \delta(ku-qu)}{q^{\hsp
2}}\left[\frac{d+1}{2\left(d+2\right)} \frac{ (kq ) u^M}{ (qu)}-
\frac{d+1}{2\left(d+2\right)} \frac{q^M}{ (qu )}\right] \dq
\end{align}
Making use of delta function and contracting with $k_M$, one
 obtains zero in what immediately follows that the above
expression is a conserved quantity.

Similarly for the non-local part,
\begin{align}\label{sigma_self}
 \sigma^M_{\rm self}(k)=\frac{e \vk^2 m}{\left(2
\pi\right)^2}\int \frac{\delta(qu)\, \delta(ku-qu)\,\e^{-i(q\cdot
b)}}{ q^2 (k-q)^2 } \left[\frac{ k  u }{d+2}\,q^M-\frac{ k u }
{d+2}\,u^M+\frac{d+1}{2\left(d+2\right)} \,\left( k  q \,u^M- k
u\,q^M\right) \right] \dq
\end{align}

Integration of both (\ref{rho_self1}) and (\ref{sigma_self}) over
$q^0$ gives
\begin{align}\label{rho_self2}
\rho^M_{\rm self} (k )\sim \delta(ku) \int \frac{1}{(q^{\hsp
z})^2/\gamma^2+q_{\perp}^2}\left[\frac{d+1}{2\left(d+2\right)}
\frac{ (kq ) u^M}{ (qu)}- \frac{d+1}{2\left(d+2\right)}
\frac{q^M}{ (qu )}\right] dq^z\,{d^{\hsp D-2}\nhsp
\mathbf{q}_{\perp}}\,.
\end{align}
Thus the account of self-terms leads to the terms proportional to
$\delta(ku)$. These terms are analogous to the Fourier-transforms
of Coulomb field which does not contribute to the radiation.

The conservation of additional terms concerned with the appearance
of second charge and the self-action terms is analogous to the
proof presented in the Subsection \ref{conservation}.

\section{An alternative proof of destructive interference}\label{DI}

We provide another proof for destructive interference in the
$z-$region, with $\vartheta < 1/\gamma$. This differs from the
method followed in the main part of a paper, which covered the
full angular range. In angular region discussed here, we show the
destructive interference effect rigorously, by the
integration-by-parts technique.

We begin with (\ref{sigmaf}). First of all we will perform a variable change from $x$ to $\zeta$, where
\begin{equation}
dx = \frac{\zeta(x)}{f(x)} \, d\zeta\, , \qquad\qquad
f(x)=\left(z^2+{z'}^2-2\ga z z'\right)\, x+\ga z z'-z^2 \,.
 \end{equation}
We will also be using the identity
\begin{equation}
\zeta{\hat K}_\nu(\zeta)=-{\hat K}'_{\nu+1}(\zeta)\,.
\end{equation}
Integrating the expression (\ref{sigmaf}) by parts we obtain the
following:
\begin{align}
 & \s^M(k)=\frac{\lambda}{2\gamma v} \left \lbrace \lph \right.\!
\left [\frac{\xi^2}{d+2}+\left( \gamma z' - \frac{z}{d+2}\right)\
\left(\gamma z'-z\right) \right] \! \left(\lph
\right.\frac{\hat{K}_{d/2}\left(z'\right)}{{z'}^2-\gamma z z'} +\!
\int\limits^1_0 \hat{K}_{d/2}(\zeta) \left( \frac{x \,\e^{i
\left(kb\right)\left(1-x\right)}}{f(x)}\right)^{\!\prime}
dx\!\!\left. \lph \right)u^M+
  \nn \\
&+ i\left[\frac{\left(kb\right)}{d+2}\, u^M \nhsp +\,
\left(\gamma^2v z^\prime \nhsp -\nhsp \frac{\gamma z
v}{d+2}\right)\frac{b^{\smhsp M}}{b}\right] \!\! \left(\lph
\right. \frac{\hat{K}_{d/2+1}\left(z\right)}{\gamma z
z'-z^2}\,\e^{i
(kb)}-\frac{\hat{K}_{d/2+1}\left(z'\right)}{{z'}^2-\gamma z z'} \!
+\!  \int\limits^1_0 \!   \hat{K}_{d/2+1}(\zeta) \left(
\frac{\e^{i \left(kb\right)(1-x)}}{f(x)}\right)^{\!\prime}\! dx
\!\! \left. \lph \right) \nn
\\
& +\left[\frac{\beta}{d+2}-\gamma z'+\frac{z^2}{d+2}+\gamma^2 v^2
{z'}^2\right] \! \left( \lph \right.
\frac{\hat{K}_{d/2}\left(z\right)}{\gamma z z'-z^2}\,\e^{i
\left(kb\right)}-\frac{\hat{K}_{d/2}\left(z'\right)}{{z'}^2-\gamma
z z'} +\int\limits^1_0 \hat{K}_{d/2}(\zeta) \left( \frac{\e^{i
\left(kb\right)(1-x)}}{f(x)}\right)^{\!\prime}\! dx \! \! \left.
\lph \right)u^M\!\! \left. \lph\right \rbrace \, . \nn
\end{align}
Further integration by parts gives
\begin{align}
 \s^M &(k) \!=\! \frac{\lambda}{2\gamma v} \!\left\lbrace
\left[\frac{\xi^2}{d+2}\nhsp +\nhsp\left(\nhsp \gamma z'\! -\!
\frac{z}{d+2}\right)\!  (\gamma z'\nhsp-\nhsp z ) \right]\!
\left(\lph \frac{\hat{K}_{d/2} (z')}{{z'}^2\nhsp-\nhsp\gamma z
z'}\!-\! \frac{\hat{\boldsymbol{K}}_{d/2+1}(z)}{\left(\gamma z
z'\!-\!z^2\right)^2}      +
(iq_1+1)\frac{\hat{K}_{d/2+1}(z')}{\left({z'}^2\nhsp-\nhsp\ga z
z'\right)^2}+\! R_1 \! \right)\! u^M   \right. \nn \\
&   +i  \left[\frac{ (kb )}{d+2}\, u^M \!+\!\left(\gamma^2v
z^\prime \! -\! \frac{\gamma z
v}{d+2}\!\right)\!\frac{b^M}{b}\right]\!\!
\left(\frac{\hat{\boldsymbol{K}}_{d/2+1}(z)}{\gamma z z'\!-\!z^2}
-\frac{\hat{K}_{d/2+1}(z')}{{z'}^2\!-\!\gamma z z'}\smhsp + \smhsp
i \,q_0
\frac{\hat{\boldsymbol{K}}_{d/2+2}(z)}{\beta^2}\smhsp-\smhsp i
\,q_1\frac{\hat{K}_{d/2+2}(z')}{\left({z'}^2\!-\!\ga z
z'\right)^2}+\!R \right)\nn
\\
 & + \left.  \left[\frac{\beta+z^2}{d+2}-\gamma z \hsp z'+ \gamma^2 v^2
{z'}^2\right]\! \left(\frac{\hat{\boldsymbol{K}}_{d/2}(z)}{\gamma
z z'-z^2} -\frac{\hat{K}_{d/2}(z')}{{z'}^2-\gamma z z'} -   i
\,q_0 \frac{\hat{\boldsymbol{K}}_{d/2+1}(z)}{\left(\gamma z
z'-z^2\right)^2} +i
\,q_1\frac{\hat{K}_{d/2+1}(z')}{\left({z'}^2-\ga z z'\right)^2}+R
\right)  u^M  \right\rbrace \, . \nn
\end{align}
with notations
\begin{equation}
\hat{\boldsymbol{K}}_{\tau}(z) \equiv \e^{i(kb)}\,
\hat{K}_{\tau}(z)\, , \qquad\qquad
 q_0=(kb)-i\frac{{z}^2+{z'}^2-2\ga
zz'}{\ga zz'-z^2}\,,\qquad\qquad q_1=(kb)-i\frac{{z}^2+{z'}^2-2\ga
zz'}{{z'}^2-\ga zz'}\nn
\end{equation}
and residues
\begin{align} R=\int\limits_0^1 dx \, \hat{K}_{d/2 +1}(\zeta(x)) \left[
\left(\frac{\e^{- i x (kb)}}{f(x)}\right)^{\!\prime}
\frac{1}{f(x)}\right]^\prime\,, \qquad \qquad R_1=\int\limits_0^1
dx \, \hat{K}_{d/2 +1}(\zeta(x))  \left[ \left(x \frac{\e^{- i x
(kb)}}{f(x)}\right)^{\!\prime} \frac{1}{f(x)}\right]^\prime.\nn
\end{align}
Thus after each iteration of integration by parts, Macdonalds of
$z$ come with phase $\e^{i(kb)}$ from boundary $x=0$, while those
ones with argument $z'$ come with phase 1 from boundary $x=1$.

 If keep on integrating by parts, we will
obtain an expansion. In the region that we are interested in, i.e.
the $z-$region we have $z\sim 1,\; z'\sim \ga,$ so that $\;
\coa^2\sim \beta\sim\ga^2 \sim  (\beta-\coa^2), \; q_0\sim
q_1\sim\ga$. From this we see that the expansion parameters are:
$q_0 \beta^{-1}\sim \ga^{-1}\ll 1,\;q_1 (\beta-\coa^2)^{-1}\sim
\ga^{-1} \ll 1$. With this accuracy one can set $q_0=q_1=(kb),\;
\beta=\ga zz'$ the leading part is then:
\begin{equation}
\s^M\nhsp (k)=\frac{\lambda}{2 \gamma}\,\left[ \gamma
\frac{z'}{z}\, \hat{K}_{d/2}(z)\,u^M- i \left(
\frac{\left(kb\right)}{z}-\gamma \frac{b^{\smhsp M}}{b}\right)
\frac{\hat{K}_{d/2+1}(z)}{z}\right] \, ,
\end{equation}
which exactly cancels with the leading part of (\ref{rhon(k)}),
leaving only the subleading terms. The series converges thus
establishing further the effect of destructive interference.

\end{document}